\documentclass[fleqn,usenatbib]{mnras}

\usepackage{graphicx} 
\usepackage{times}
\usepackage{subfig}
\usepackage{amssymb} 
\usepackage{txfonts}
\usepackage{appendix}
\usepackage{multirow}
\usepackage{textcomp}
\usepackage{natbib}
\usepackage{longtable, portland}
\usepackage{supertabular}
\usepackage{rotating}
\usepackage{array}
\usepackage{caption}
\usepackage{gensymb}
\usepackage[T1]{fontenc}
\usepackage{color}
\usepackage{ulem}
\usepackage{ae,aecompl}

\title[Morphologies of FIR green valley AGN and non-AGN]{Star formation in far-IR AGN and non-AGN galaxies in the green valley. II. Morphological analysis}

\author[A. Mahoro et al.]{Antoine Mahoro $^{1,\,2,\,3}$\thanks{E-mail: antoine@saao.ac.za}, Mirjana Povi\'c$^{4,\,5}$, Pheneas Nkundabakura$^{3}$, 
 \newauthor Beatrice Nyiransengiyumva$^{3}$, and Petri V\"ais\"anen$^{1,\,6}$ \\ 
$^{1}$South African Astronomical Observatory, P.O. Box 9 Observatory, Cape Town, South Africa\\
$^{2}$Department of Astronomy, University of Cape Town, Private Bag X3, Rondebosch 7701, South Africa\\
$^{3}$MSPE Department, School of Education, College of Education, University of Rwanda,  P.O. Box 5039, Kigali, Rwanda\\
$^{4}$Ethiopian Space Science and Technology Institute (ESSTI), Entoto Observatory and Research Center (EORC), \\
Astronomy and Astrophysics Research Division, P.O. Box 33679, Addis Ababa, Ethiopia\\
$^{5}$Instituto de Astrof\'isica de Andaluc\'ia (IAA-CSIC), Glorieta de la Astronom\'ia s/n, 18008 Granada, Spain\\
$^{6}$Southern African Large Telescope, P.O. Box 9 Observatory, Cape Town, South Africa}

\date{Accepted XXX. Received YYY; in original form ZZZ}

\pubyear{2019}

\begin{document}
\label{firstpage}
\pagerange{\pageref{firstpage}--\pageref{lastpage}}
\maketitle

\begin{abstract}
This paper studies morphological properties of 103 green valley FIR active and 2609 non-active galaxies presented in \cite{Mahoro2017}. The photometric data from
the COSMOS survey were used, and the morphological parameters, such as  Abraham and Conselice-Bershady concentration indices, Gini, M20 moment
of light, and asymmetry, were analysed taking into account public catalogues. Furthermore, a visual classification of galaxies was performed. We
found that the fraction of peculiar galaxies with clear signs of interactions and mergers
is significantly higher in AGN (38\%) than non-AGN (19\%) green valley galaxies, while non-AGN galaxies from our sample are predominantly spirals (46\%). We found that the largest fraction of our green valley galaxies is located on the main-sequence (MS) of star formation, independently on morphology, which is in contrast with most of previous studies carried out in optical. We also found that FIR AGN green valley galaxies
have significantly higher star formation rates in all analysed morphological types. Therefore, our results suggest that interactions and mergers obtained in the high fraction of FIR AGN contribute significantly to high star formation rates observed in the selected sample, but are not the only mechanism responsible for enhancing star formation, and others such as AGN positive feedback could contribute as well. In future we will study in more details the possibility of AGN positive feedback through the spectroscopic analysis of public and our SALT data. 

\end{abstract}




\section{Introduction}
\label{sec_intro}

Observational studies and large surveys showed that galaxies have bi-modal properties in terms of stellar mass, star formation, colours, luminosities, different morphological parameters, etc. \citep{Kauffmann2003, Baldry2004, Salim2007, Brammer2009, Povic2013a, Walker2013, Schawinski2014, Ge2018}. The region between the two peaks in the bi-modal distribution (the 'blue cloud' and 'red sequence') is referred to as the 'green valley' and it is considered to be the transition area between the late-type and early-type galaxies \citep[][and references therein]{Salim2014}. Studying the properties of green valley galaxies is therefore very important for the better understanding of star formation quenching mechanisms, morphological transformation in galaxies, and galaxy evolution across cosmic time. 

Previous works studied morphological properties of the green valley galaxies and found that they show intermediate values (between blue cloud and red sequence) in terms of concentration (C), asymmetry (A) and smoothness (S) indices \citep{Mendez2011, Pan2013}. \cite{Mendez2011} found also that the green valley galaxies are generally massive $(M_{*}\sim 10^{10.5}M_{\odot})$ disk galaxies; 12\% of all their sample are bulgeless galaxies, and they found that the merging fraction of green valley galaxies is lower than in late-types. Using the morphological diagrams such as Gini vs. A and Gini vs. M$_{20}$ moment of light, it is found that the green valley galaxies are located in the middle of late- and early-types. Different studies suggested that the morphological transformation of galaxies happens in the green valley during different timescales \citep{Schawinski2014, Lee2015, Smethurst2015, Trayford2016, Bremer2018}, being also dependent on morphology \citep{NogueiraCavalcante2018}. Using the MaNGA\footnote{https://www.sdss.org/dr13/manga/} integral field spectroscopy (IFS) data, \cite{Belfiore2018} however suggested that green valley is a quasi-static population, that requires slow quenching process which uniformly affects the entire galaxy. The authors also found suppression of star formation with respect to mass-matched main sequence galaxies at all radii. In addition, in \cite{Belfiore2017} using again IFS data found that $\sim$\,40\% of green valley galaxies have quiescent central regions, while hosting star formation in extended outer discs.    

Previous works have suggested that active galactic nuclei (AGN) can be responsible for stopping the star formation in galaxies \citep{DiMatteo2005, Nandra2007, Leslie2016}, since most of X-ray detected AGN were found in the green valley \citep{Coil2009, Hickox2009, Cardamone2010, Povic2012, Povic2013b}. In addition, optically selected AGN in the Sloan Digital Sky Survey (SDSS), classified using traditional emission-line BPT diagrams \citep{Baldwin1981}, were found to mainly reside below the main-sequence of star formation, moving from the star-forming toward more passive galaxies, suggesting again a role of AGN in quenching star-formation in galaxies \citep{Leslie2016}.

In \cite{Mahoro2017} we studied a sample of green valley active and non-active galaxies selected from the Cosmological Evolution Survey (COSMOS\footnote{http://cosmos.astro.caltech.edu/}) \citep{Scoville2007}, where $\sim$\,90\% of sources in our sample have redshifts 0.2\,$\le$\,z\,$\le$\,1.2. We wanted to study in detail the star-formation properties of both active and non-active galaxies, and to understand better the role of AGN and its feedback on star-formation quenching. We measured the star formation rates (SFRs) using the far-infrared (FIR) \textit{Herschel}/PACS data. We found that FIR selected green valley AGN have still very active star formation, with 82\% being located either on or above the main sequence, showing therefore signs of star formation enhancement rather then its quenching. The obtained result appears to suggest that for X-ray detected AGN with FIR emission surprisingly, there appears to be positive feedback on star formation from the AGN, rather than negative feedback. 

In work presented in this paper we want to go a step further and to study in more detail morphological properties of these galaxies to understand better the obtained results of higher SFRs in AGN. In particular, we would like to see if higher SFRs in AGN could be related with interactions and mergers, or whether AGN positive feedback indeed plays a role, as suggested in \cite{Mahoro2017}. We go through the visual classification of all active and non-active galaxies, and analyse various morphological parameters and classifications obtained from the public catalogues. The paper is organised as follows: data and sample selection are described in Section 2; morphological classifications and main results are explained in Section 3; and finally, in Section 4 we discuss and summarise our results. 

We assume the following cosmological parameters throughout the paper: $\Omega_{m}=0.3,\,\Omega_{\Lambda}=0.7$, with $H_{0}=70$\,km\,s$^{-1}$\,Mpc$ ^{-1}$. All magnitudes given in this paper are in AB system. The stellar masses are given in units of solar masses (M$_{\odot}$), and both SFR and stellar masses assume \cite{Salpeter1955} initial mass function (IMF).

\section{Data and sample selection}
\label{sec_sample_selection}

In this paper we used public photometric data from the COSMOS 2\,deg$^2$ survey, one of the deepest extragalactic surveys with multi-wavelength data available, centered at RA\,(J2000)\,=\,10:00:28.6 and DEC\,(J2000)\,=\,+02:12:21.0 \citep{Scoville2007}. 

For visual morphological classification we used the \textit{Hubble Space Telescope (HST)} images taken with the Advanced Camera for Surveys (ACS), under the cycle 12-13 (July 2003 - June 2005). Third public release of the HST/ACS COSMOS observations were used covering 1.7\,deg$^2$ in F814W band and with a resolution of 0.003"/pixel \citep{Koekemoer2007, Massey2010}. For extracting images of our sample, the COSMOS cutouts tool\footnote{http://irsa.ipac.caltech.edu/data/COSMOS/index\_cutouts.html} was used. 

The selection of sample is fully explained in \citet{Mahoro2017}. The green valley galaxies were selected using the optical U-B rest-frame colour and criteria $\rm{0.8\leq U-B\leq 1.2}$ \citep{Nandra2007, Willmer07}. Optical data were extracted from the catalogue of \cite{Tasca2009}, complete down to the magnitude of I\,=\,23, and based on the ACS photometric catalogue of \citet{Leauthaud2007}. We used X-ray \textit{XMM-Newton} and \textit{Chandra} data for AGN selection \citep{Brusa2007, Civano2012}, and X-ray (2\,-\,10 keV)-to-optical (I band) flux ratio of $\rm{-1\leq\,logF_{x}/F_{0}\leq1}$ \citep{Alexander2001, Bauer2004, Bundy2007, Trump2009}. For photometric redshifts we used two catalogues, \cite{Salvato2011} and \cite{Ilbert2009} for AGN and non-AGN samples, respectively. The FIR \textit{Herschel}/PACS 160, 100, and 24\,$\micron$ \citep{Lutz2011} and Spitzer DR1 24\,$\micron$, \citep{Rieke2004} data were used for measuring IR luminosities and SFRs through the spectral energy distribution (SED) fitting by using the Le Phare\footnote{http://www.cfht.hawaii.edu/~arnouts/lephare.html} code \citep{Arnouts2011, Ilbert2006}. For measuring SFRs in active galaxies, we used \cite{Kirkpatrick2015} templates and we corrected IR luminosity for AGN contribution \citep{Mahoro2017}. The final selected sample contains 103 and 2609 green valley FIR AGN and non-AGN emitters, respectively, with good SFR measurements and with redshifts z\,$\le$\,3 where $\sim$\,90\% of galaxies have redshifts 0.2\,$\leq$\,z\,$\leq$\,1.2. For more details regarding green valley selection, AGN selection, redshift distributions, and SFR measurements, see \citet{Mahoro2017}.

\section{Morphological analysis and results}
In this section we describe our morphological analysis of selected AGN and non-AGN galaxies. We first extracted HST/ACS images of all AGN and non-AGN galaxies and carried out detailed visual morphological classification. We then compared our classification with different non-parametric classifications available for the COSMOS field in the public catalogues. Finally, we analysed the distributions of different morphological parameters of selected FIR AGN and non-AGN samples.    

\subsection{Visual Classification}
\label{sec_visual_class}

\begin{figure*}
\centering
\begin{minipage}[c]{\textwidth} 
\includegraphics[width=2cm]{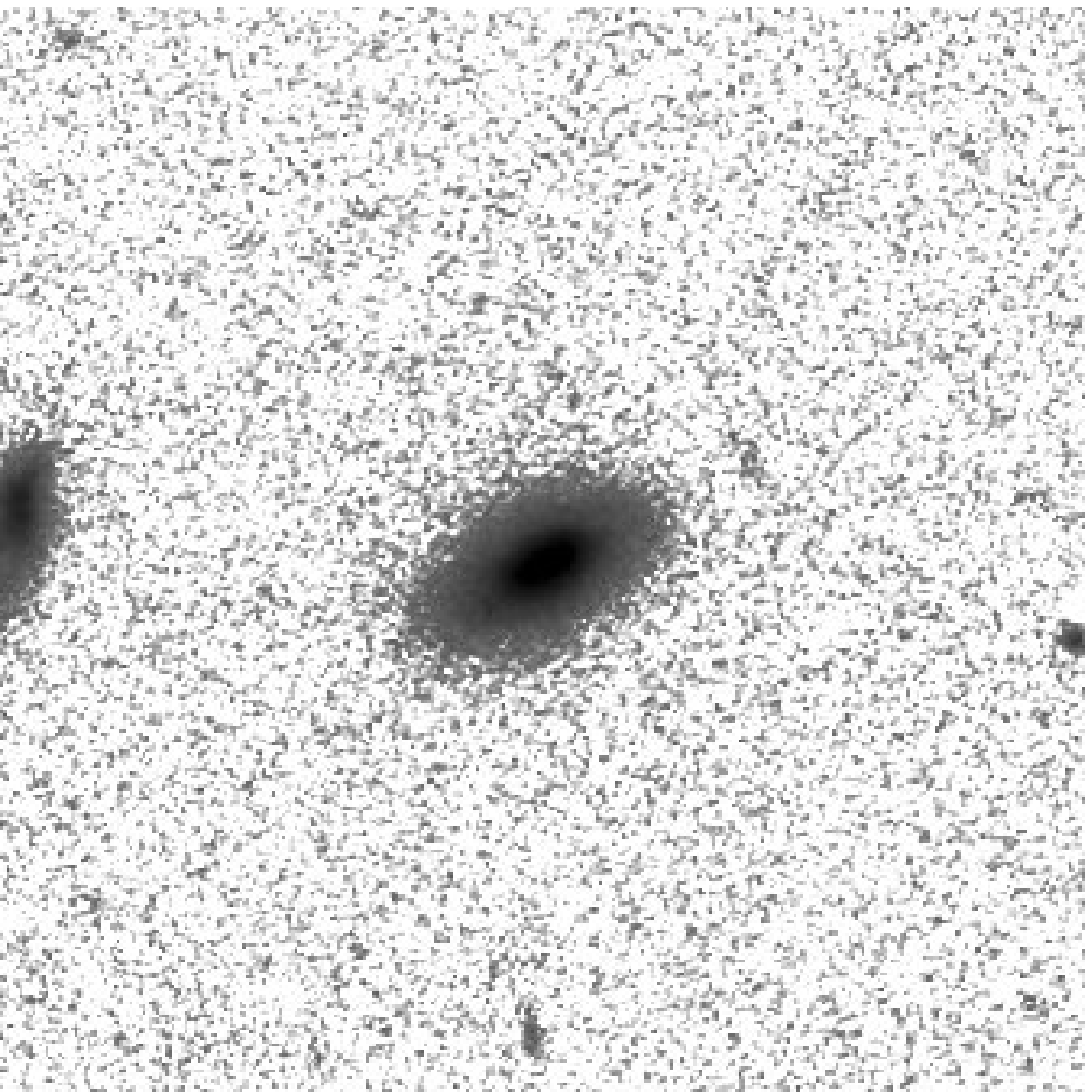}
\includegraphics[width=2cm]{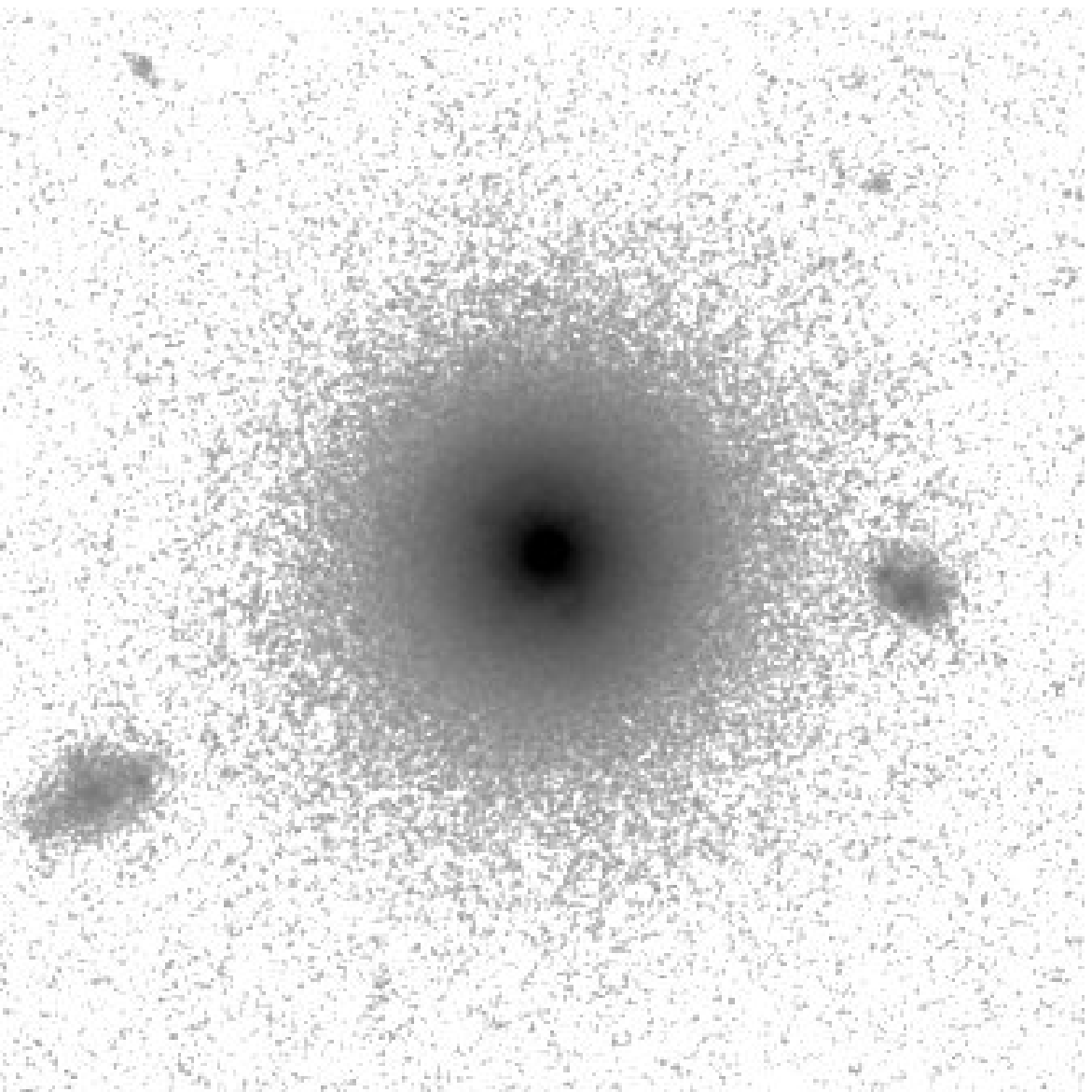}
\includegraphics[width=2cm]{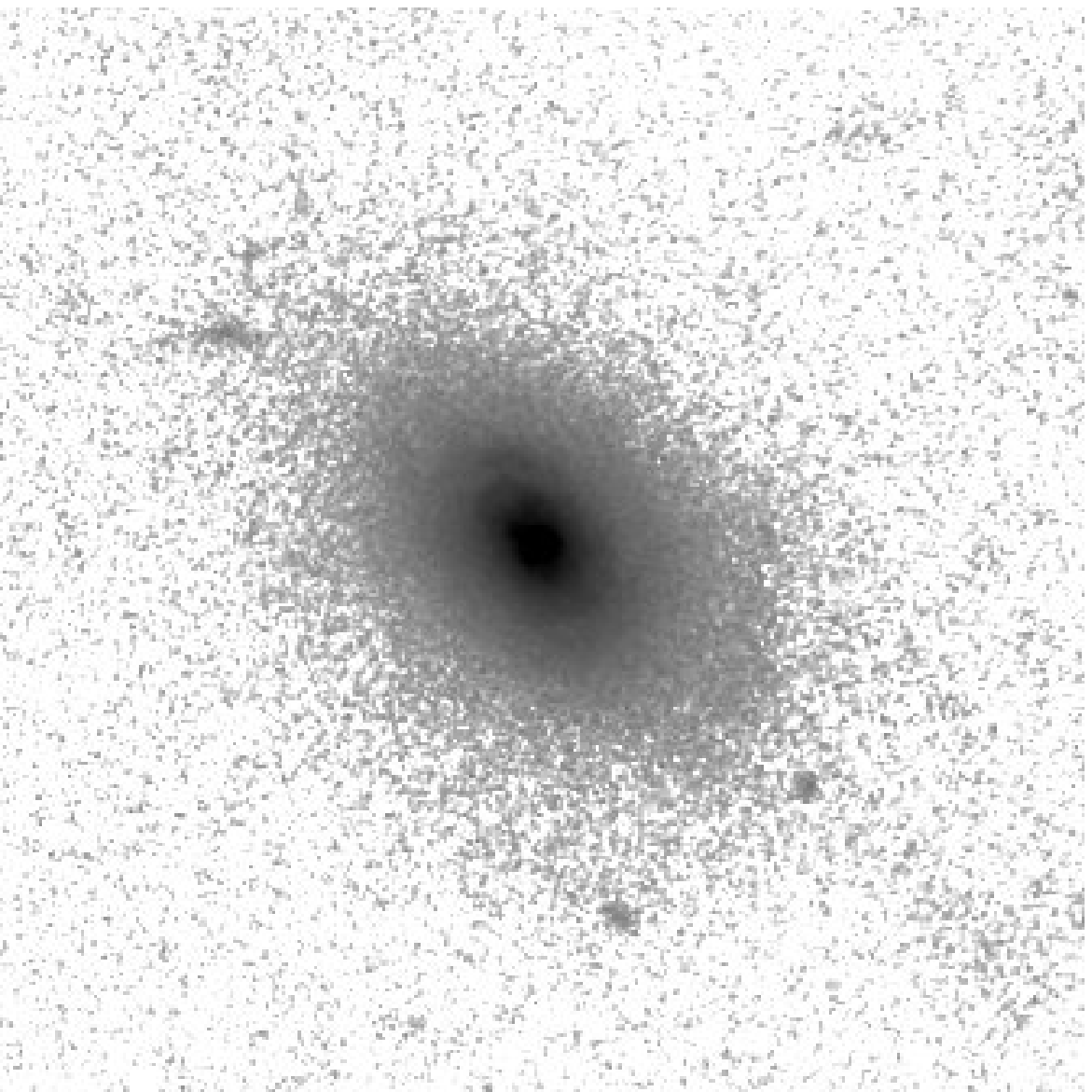}
\includegraphics[width=2cm]{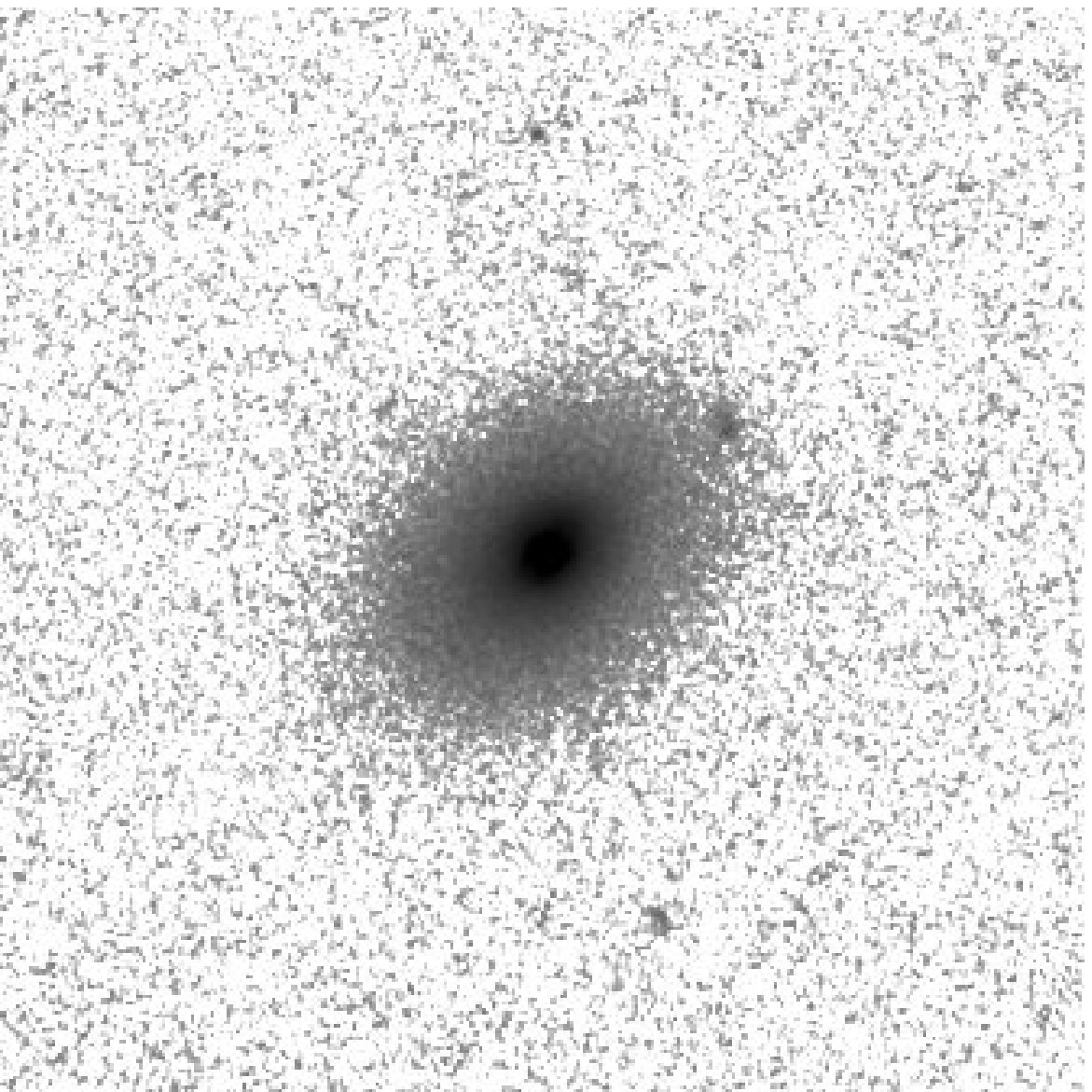}
\includegraphics[width=2cm]{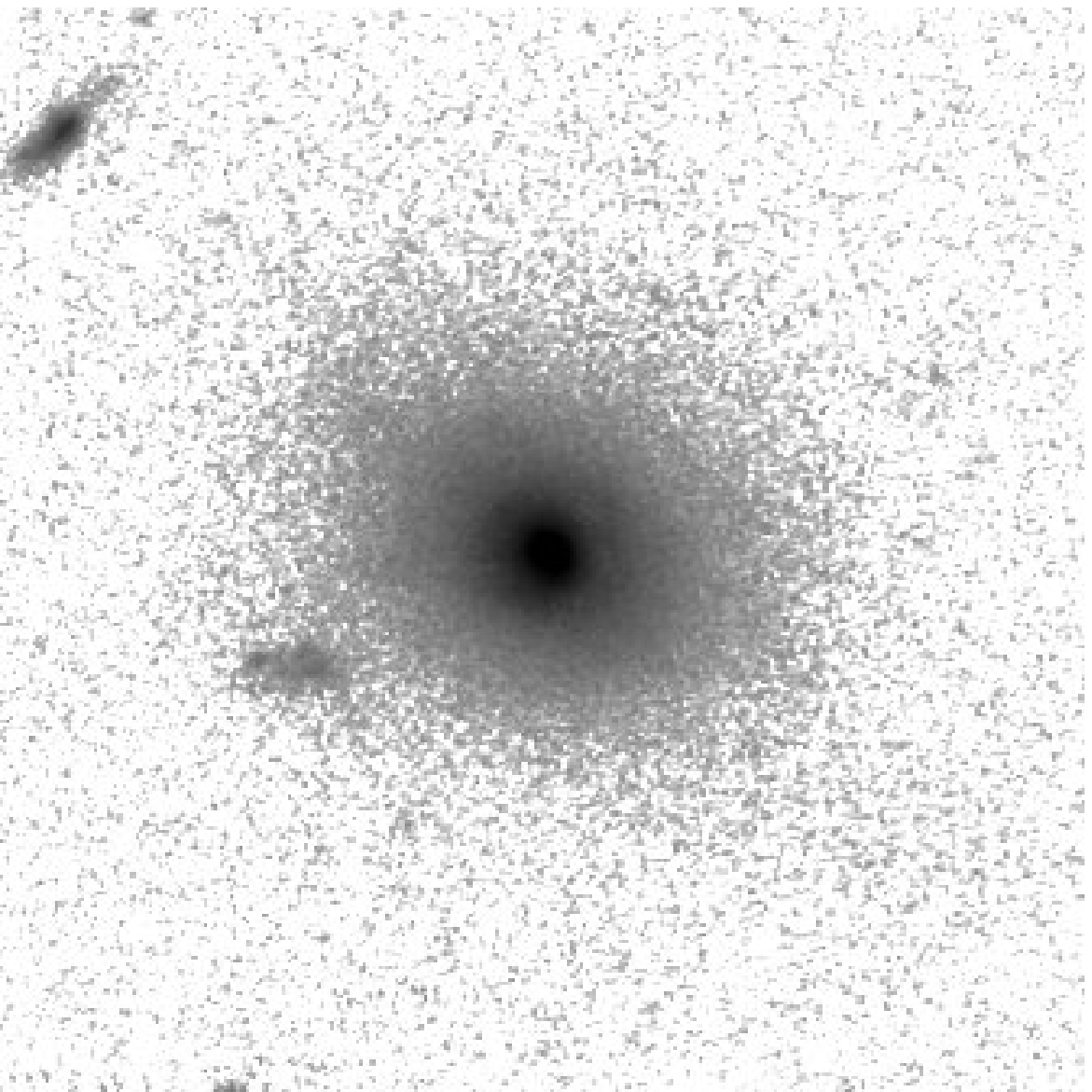}
\includegraphics[width=2cm]{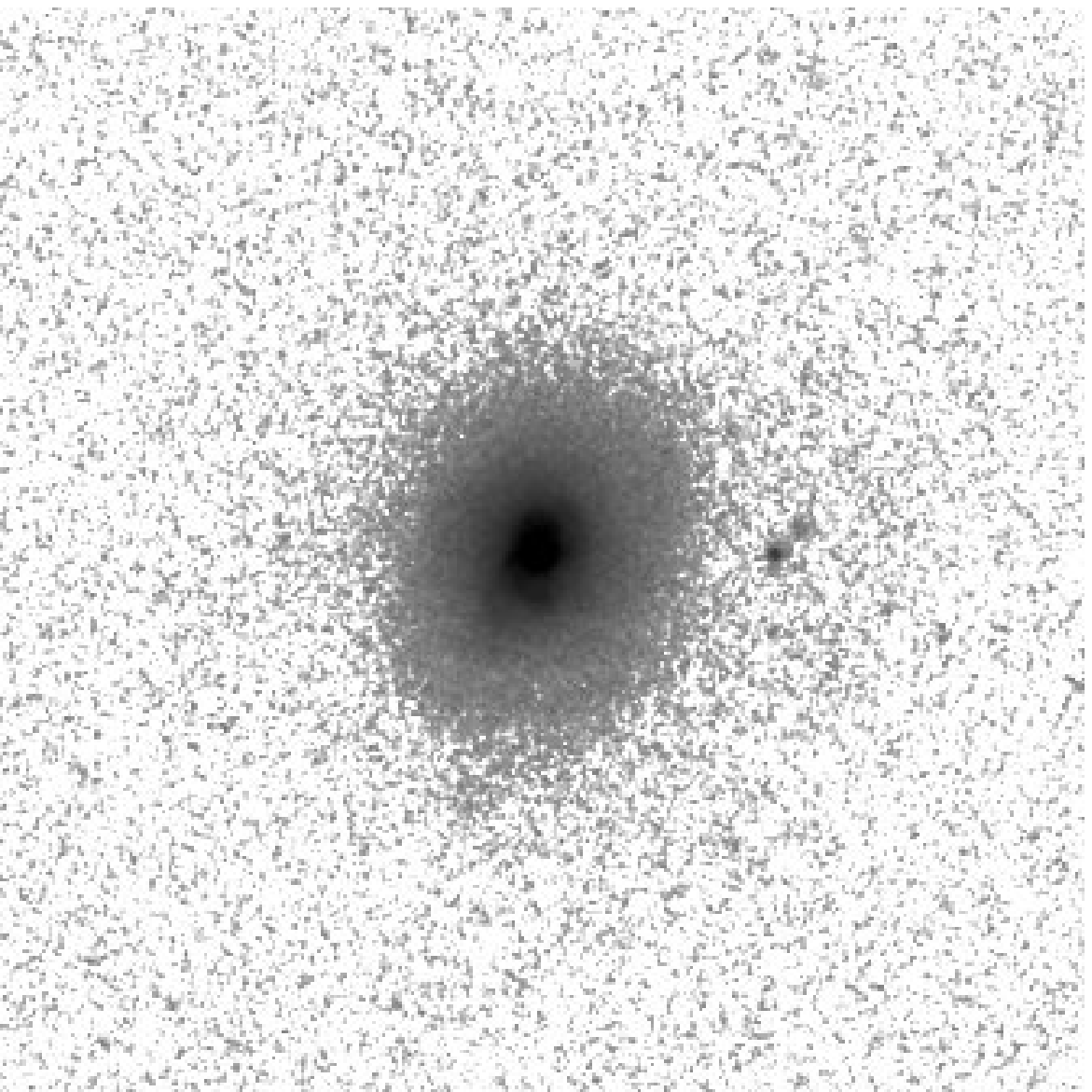}
\includegraphics[width=2cm]{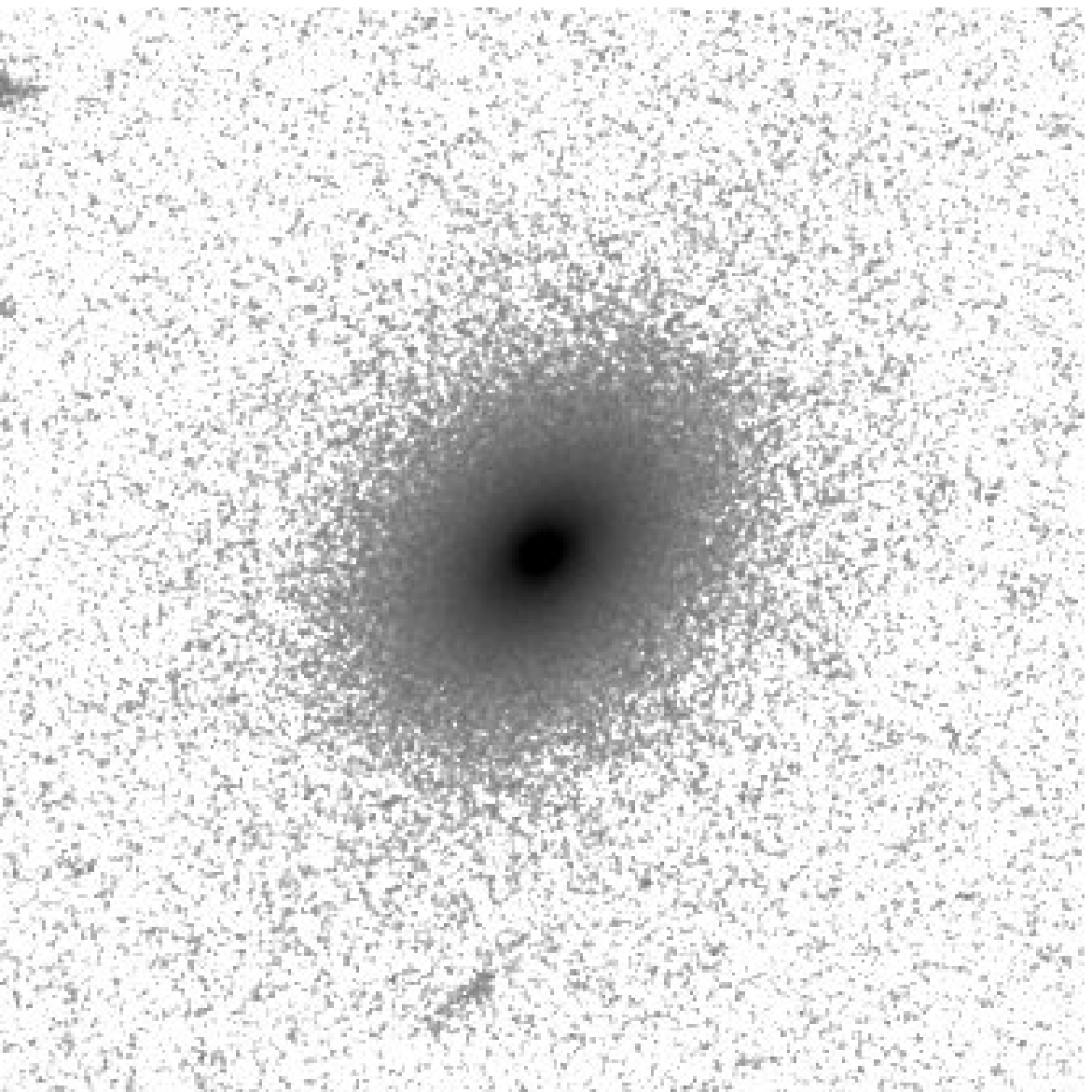}
\includegraphics[width=2cm]{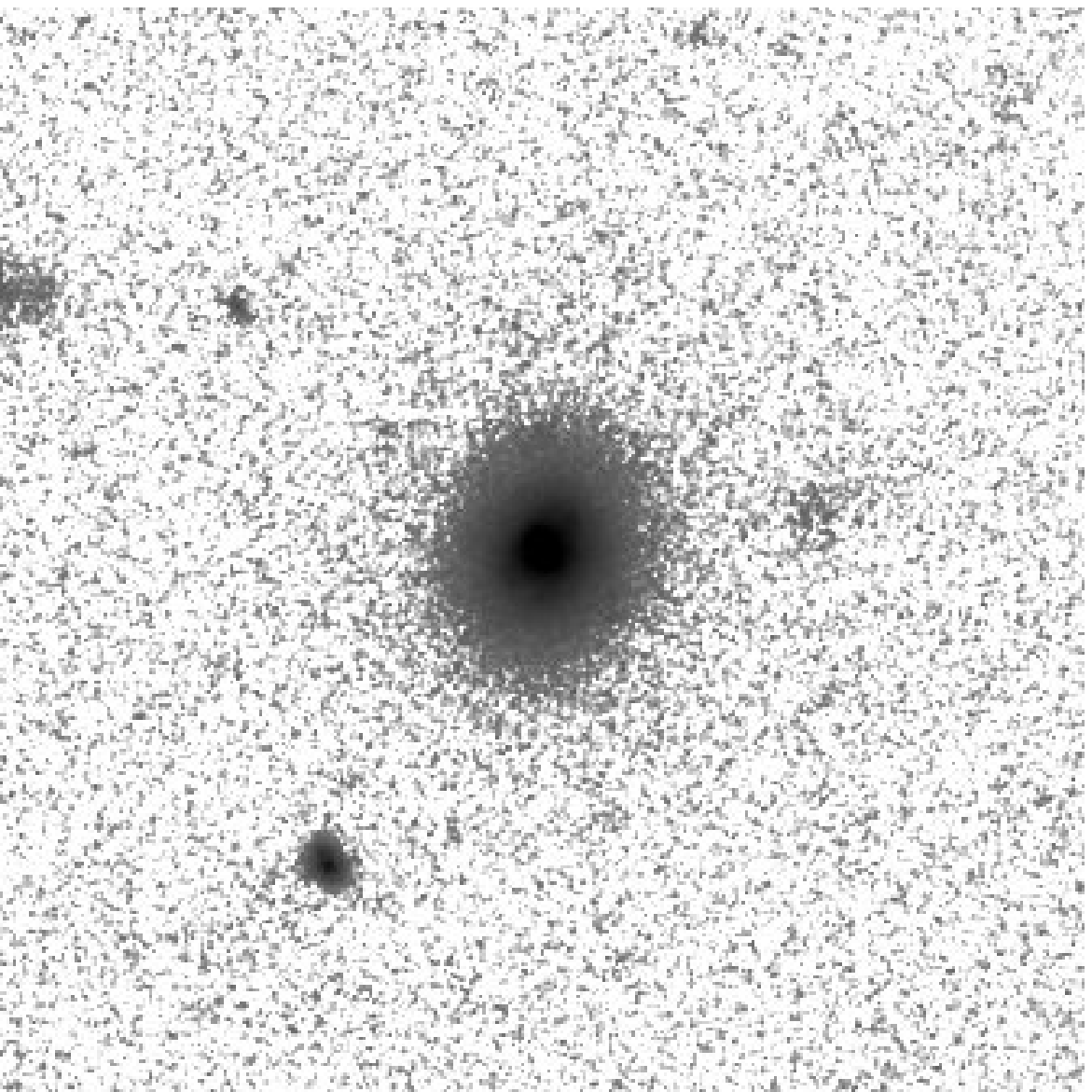}
\end{minipage}
\begin{minipage}[c]{\textwidth} 
\includegraphics[width=2cm]{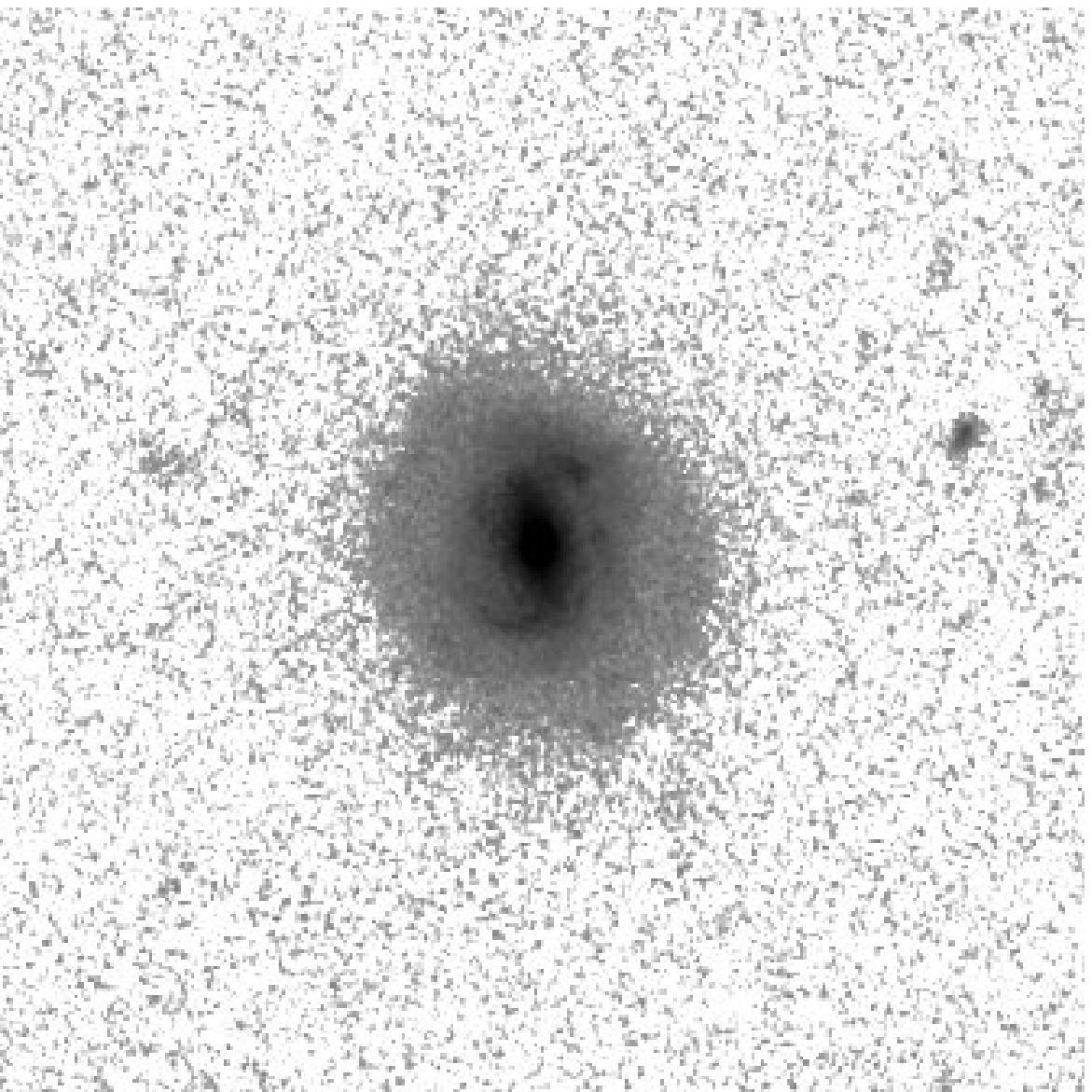}
\includegraphics[width=2cm]{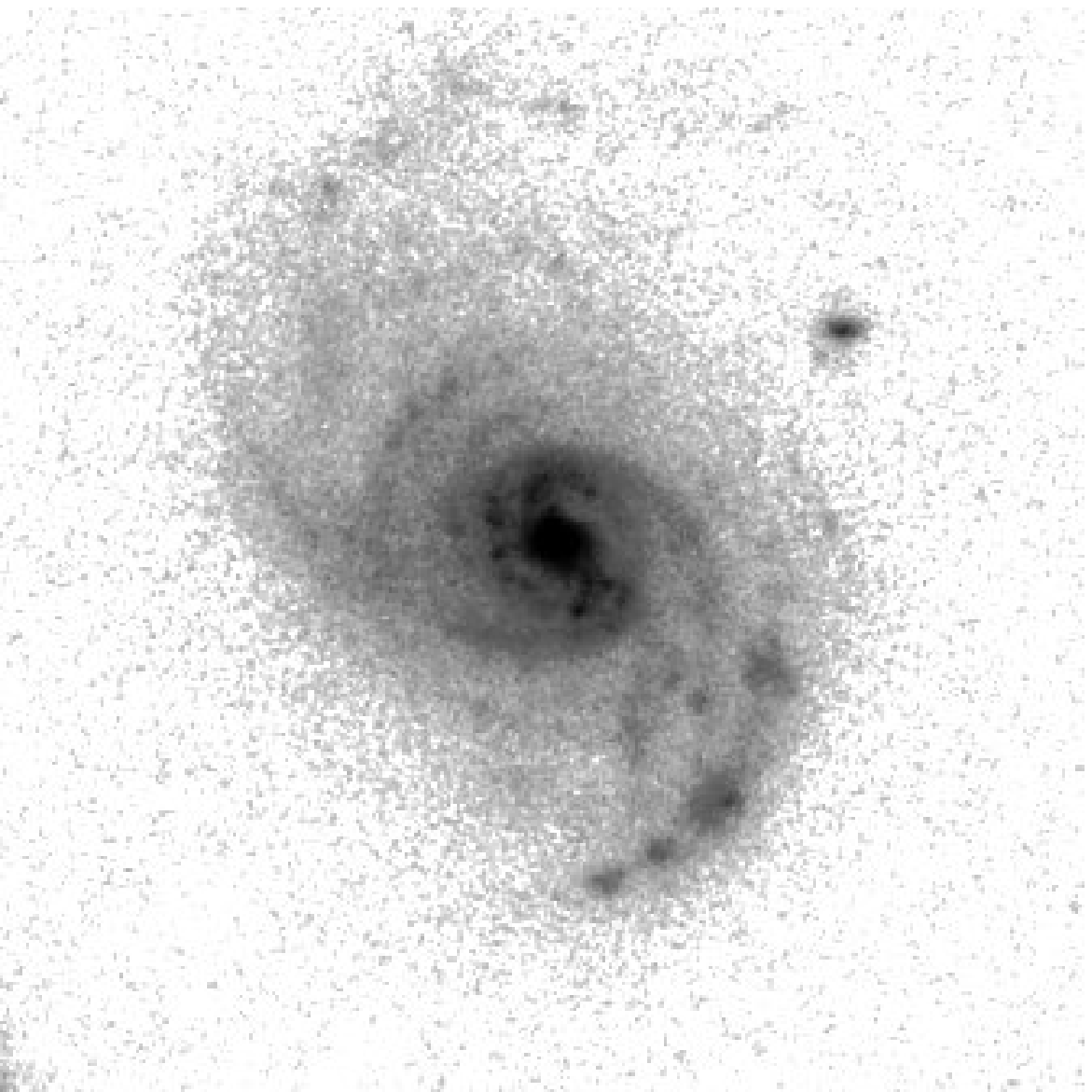}
\includegraphics[width=2cm]{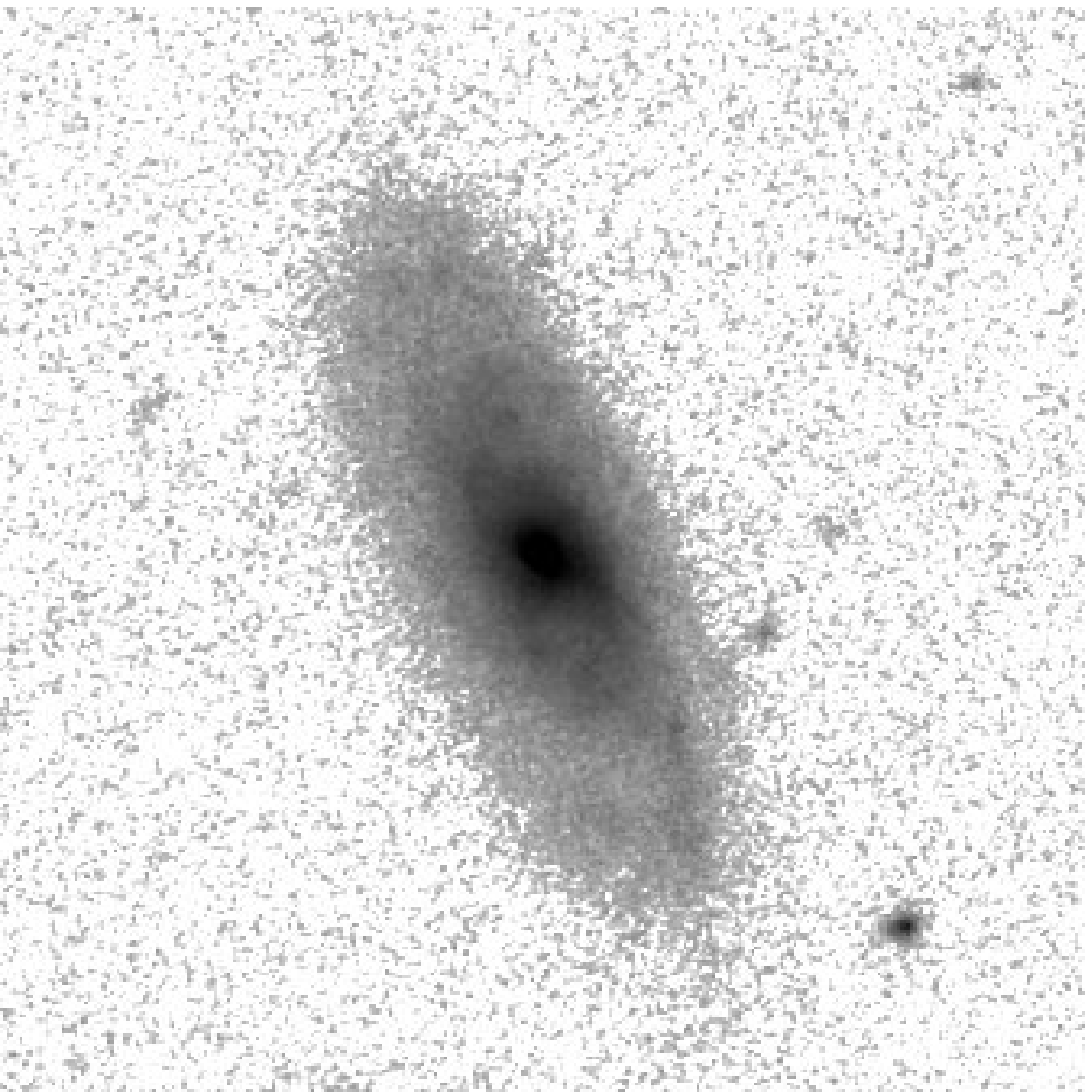}
\includegraphics[width=2cm]{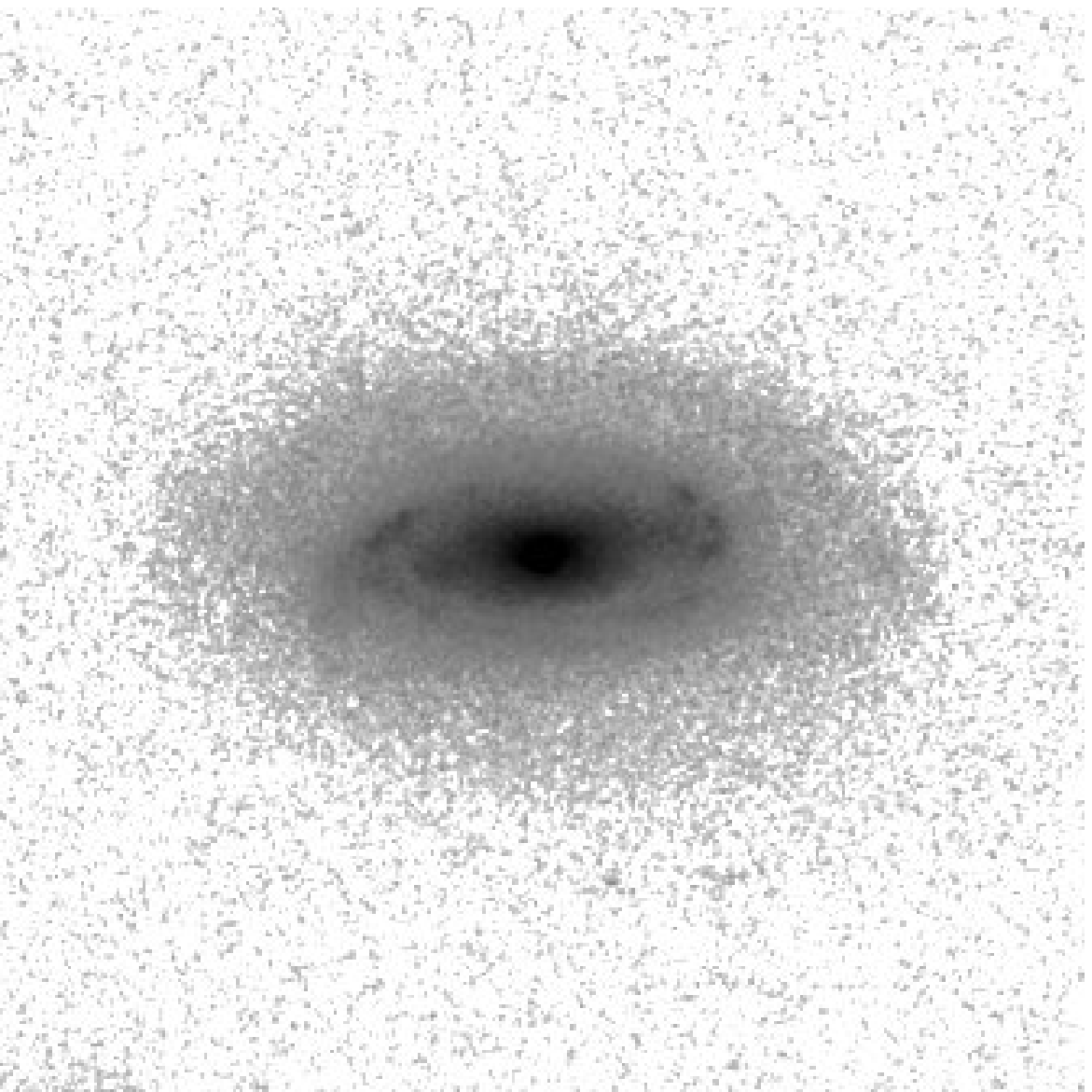}
\includegraphics[width=2cm]{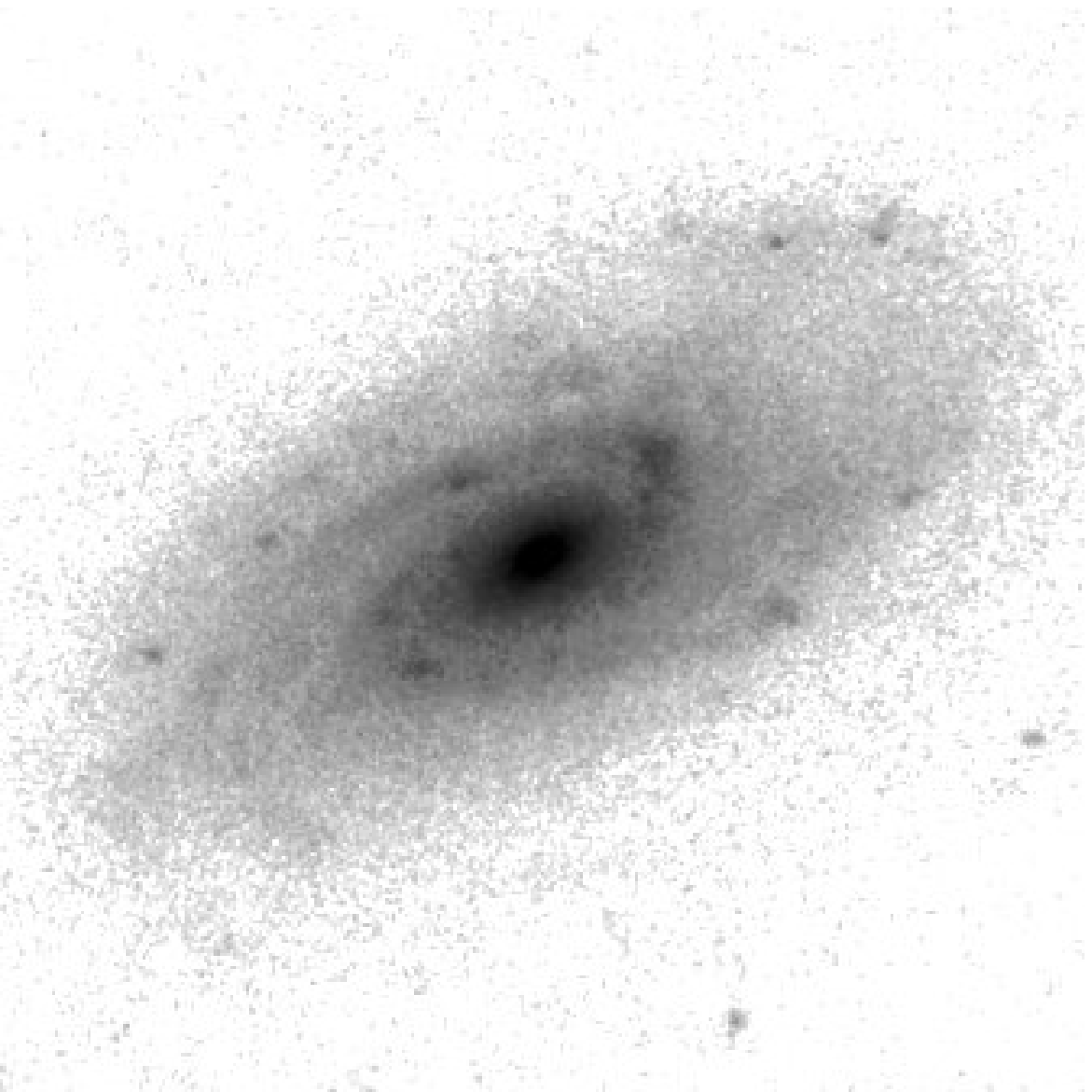}
\includegraphics[width=2cm]{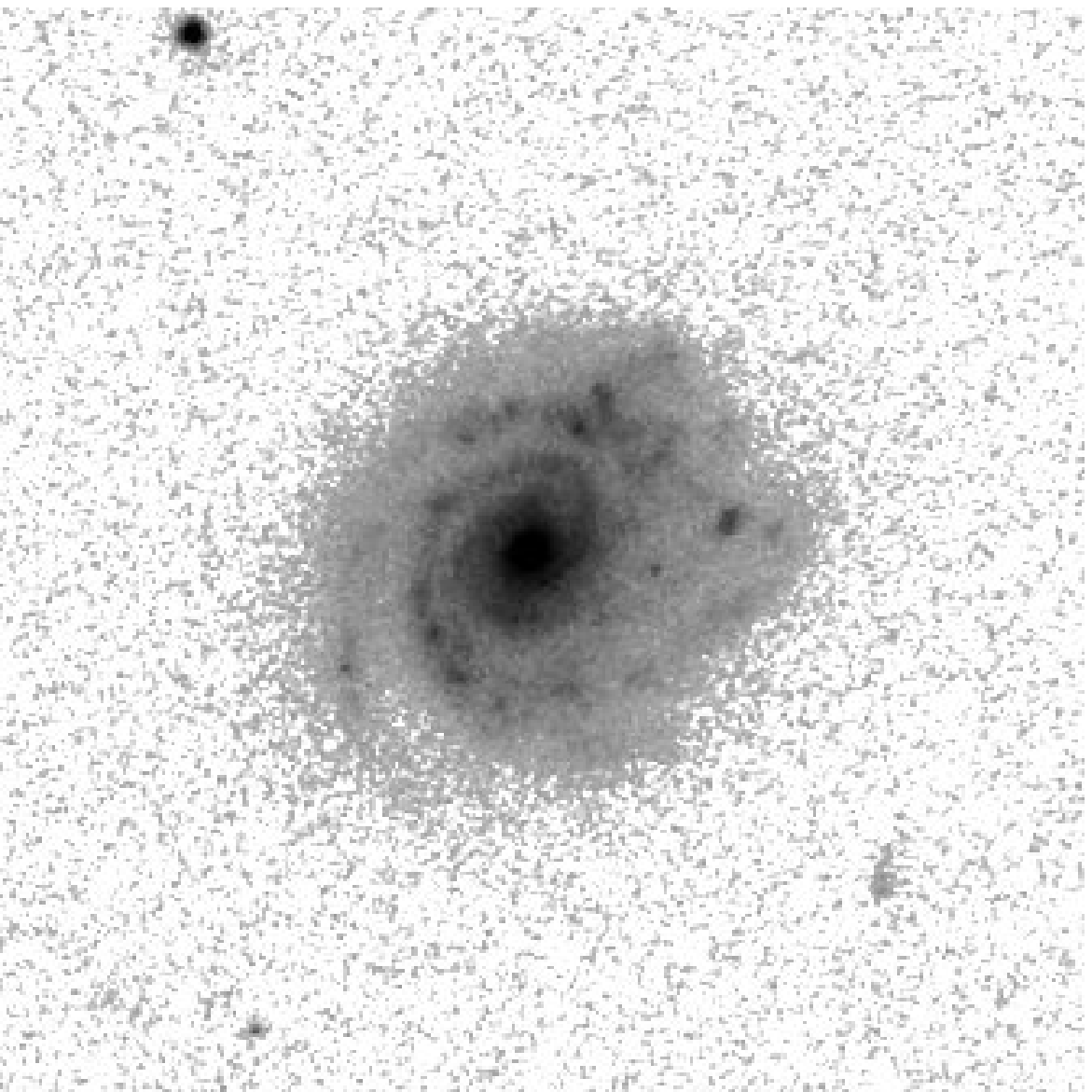}
\includegraphics[width=2cm]{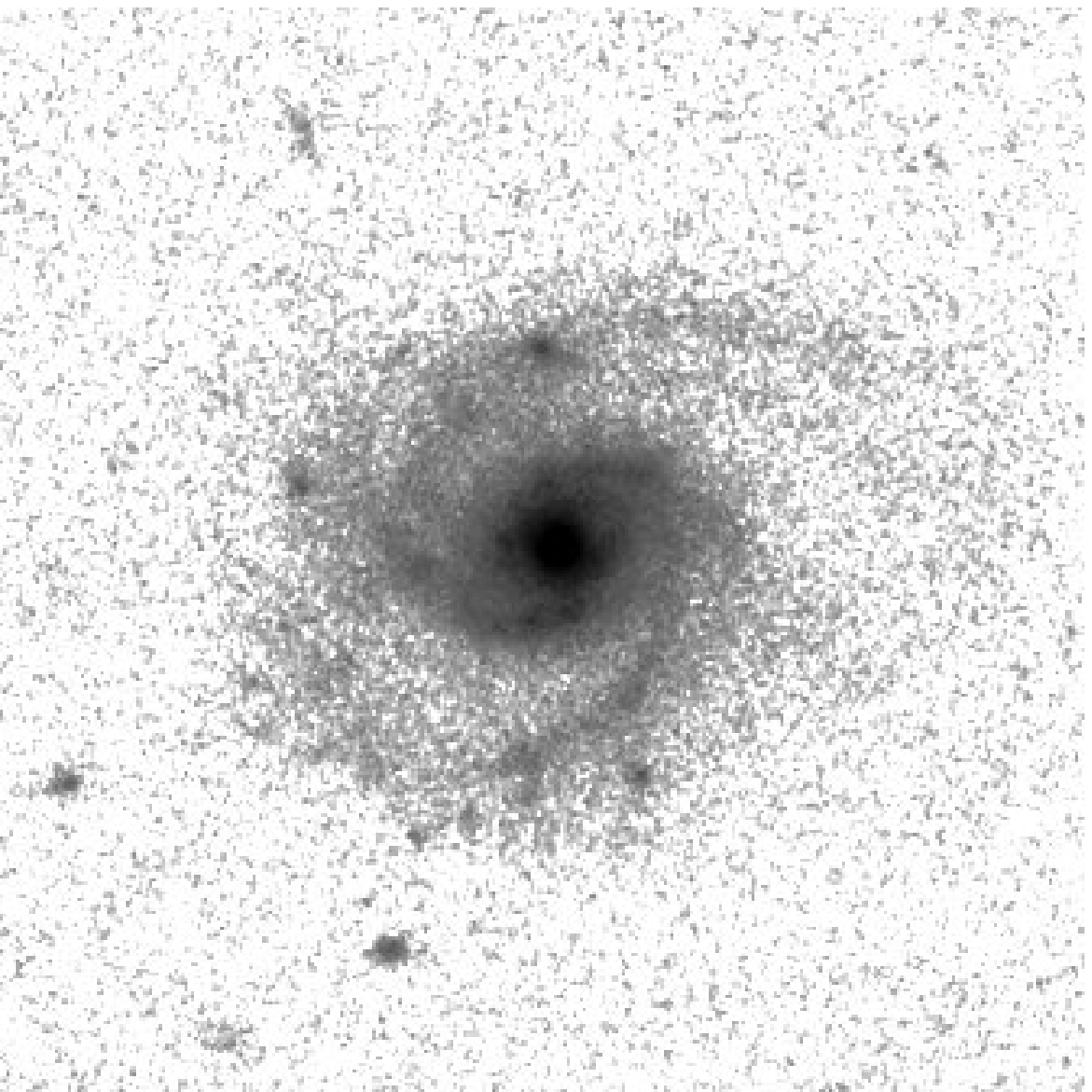}
\includegraphics[width=2cm]{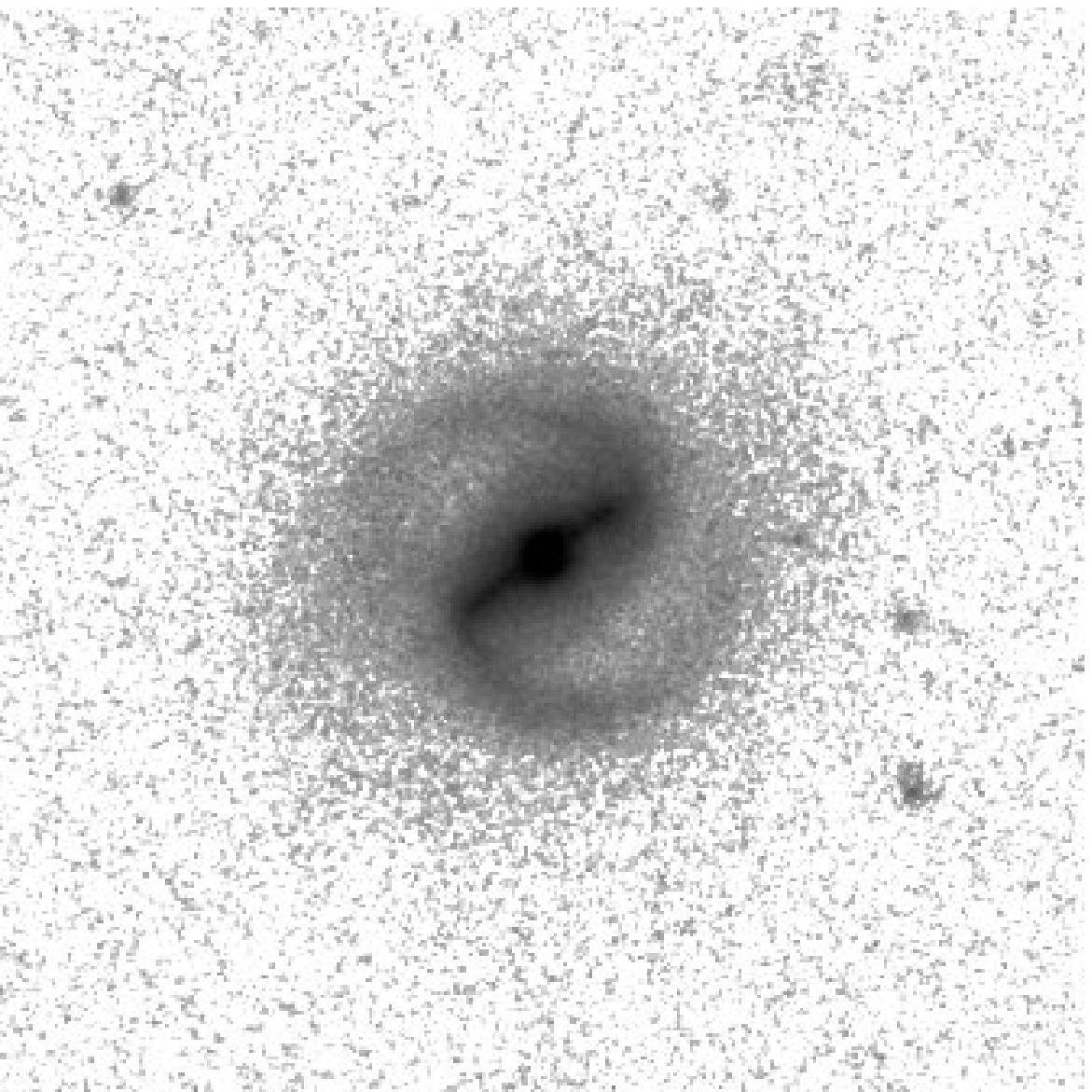}
\end{minipage}
\begin{minipage}[c]{\textwidth}
\includegraphics[width=2cm]{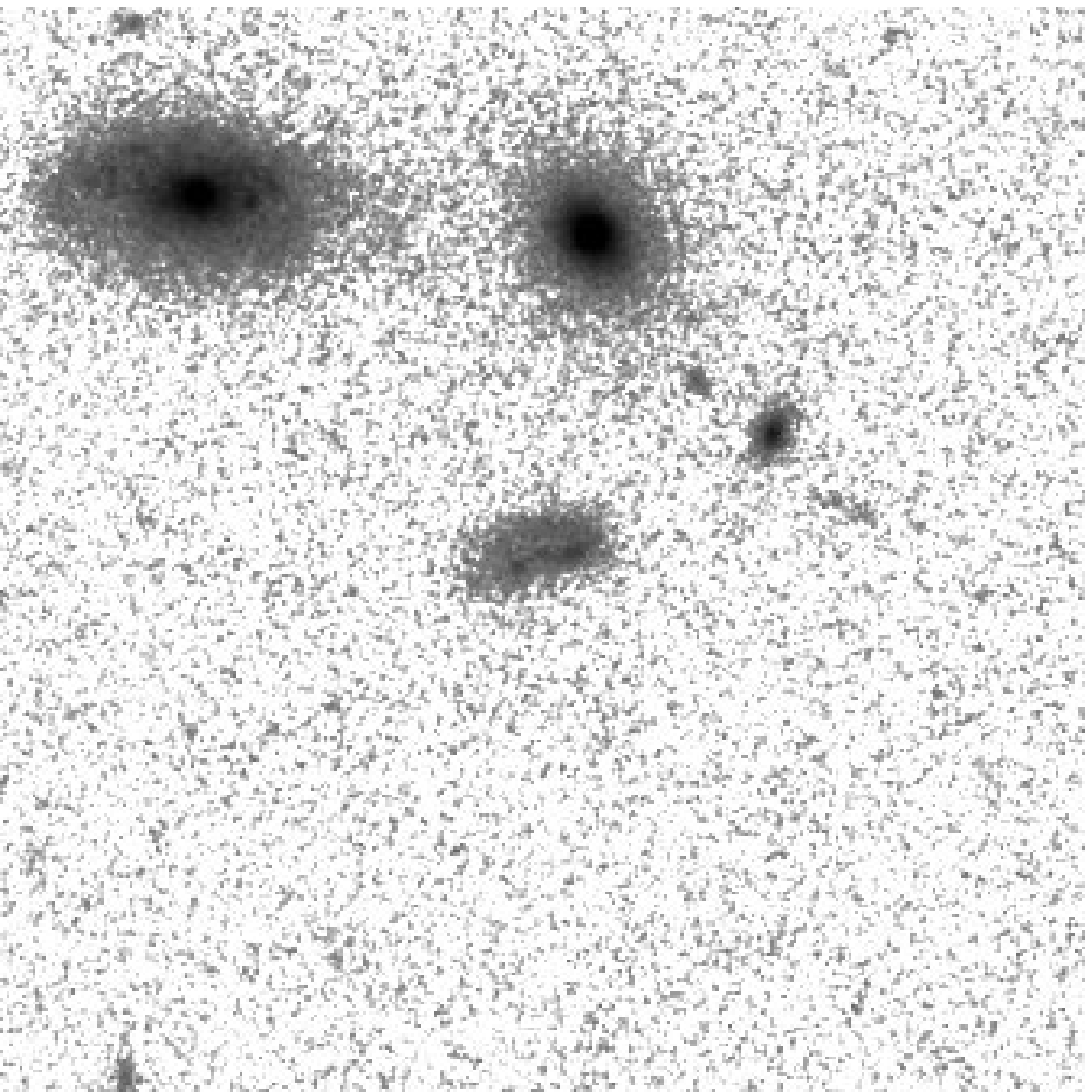}
\includegraphics[width=2cm]{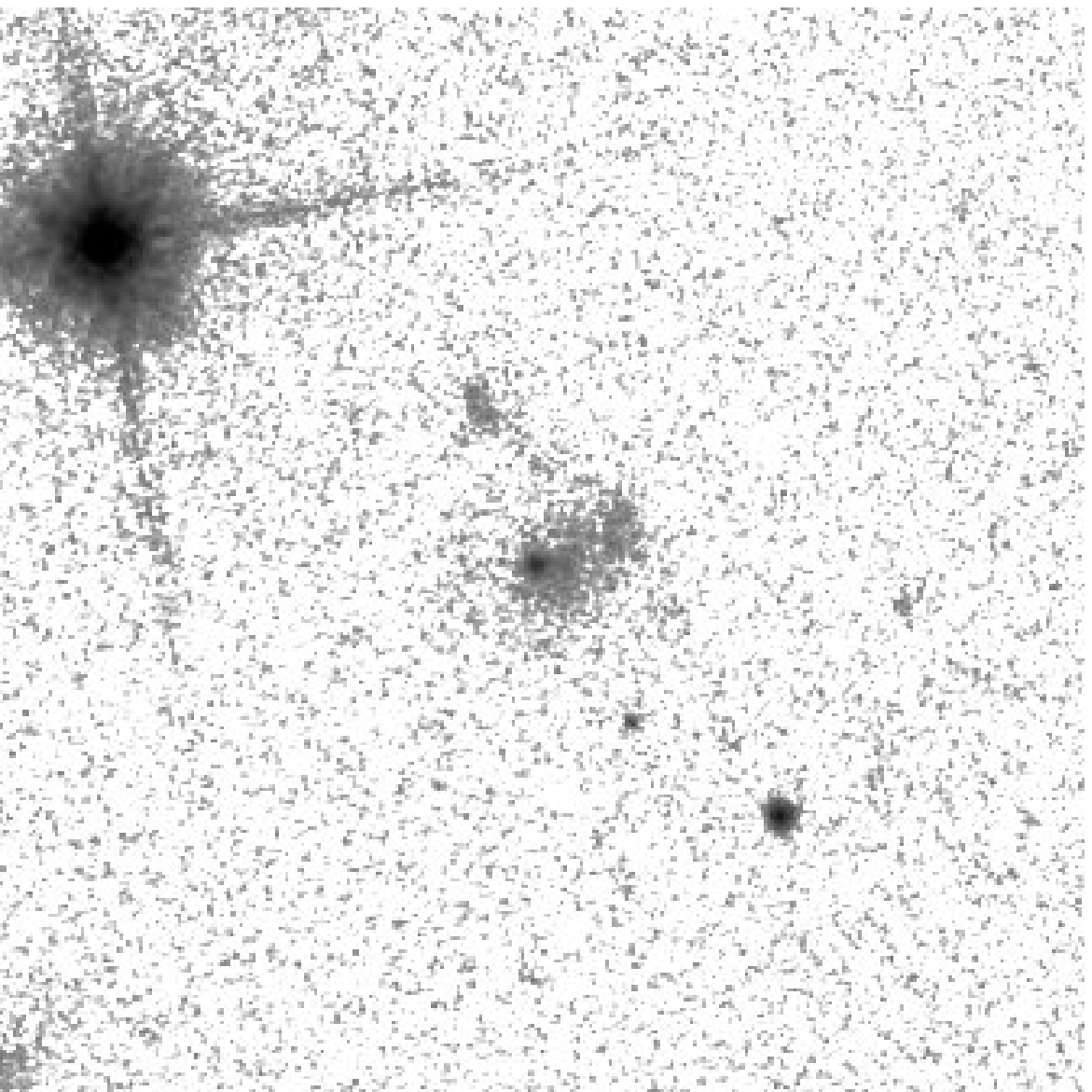}
\includegraphics[width=2cm]{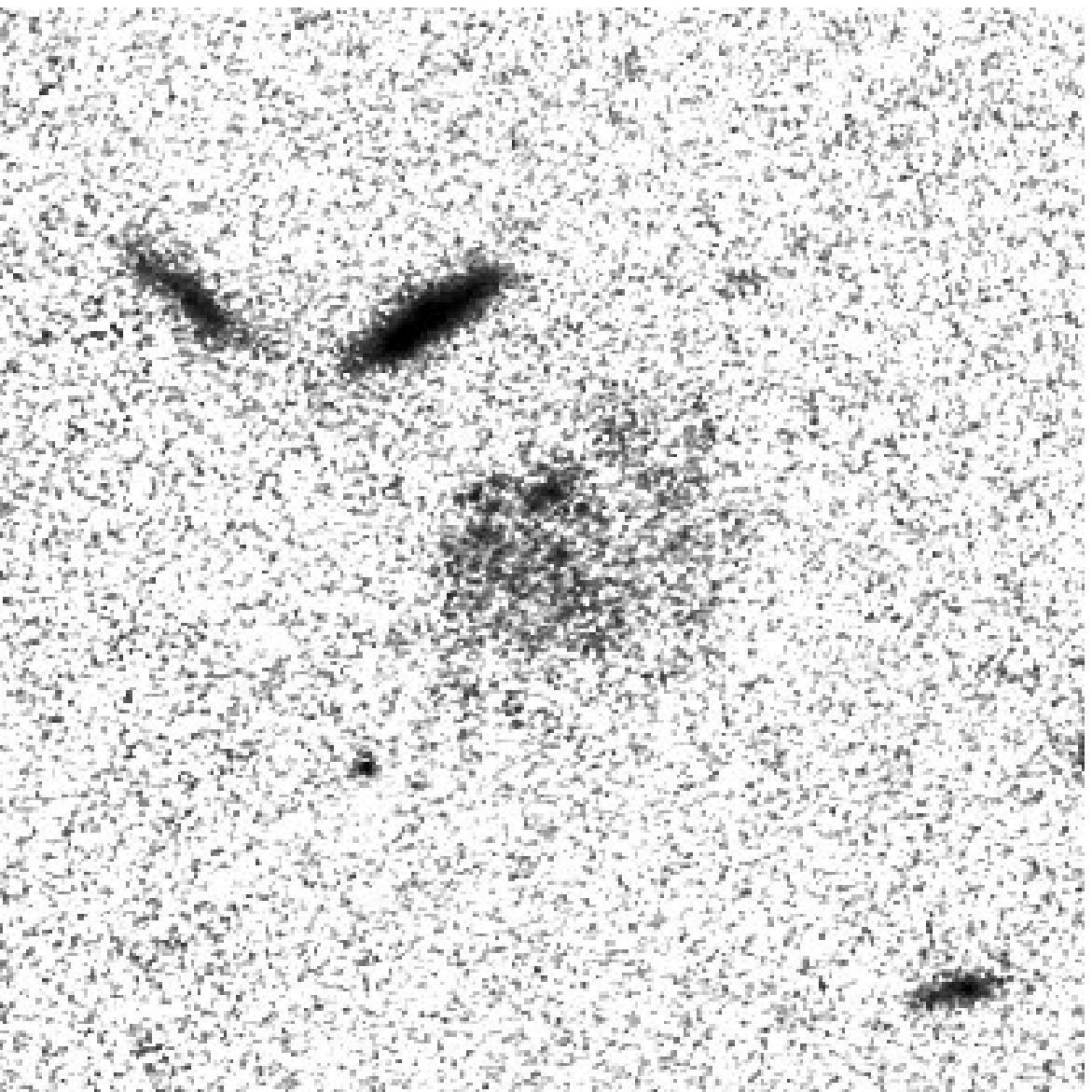}
\includegraphics[width=2cm]{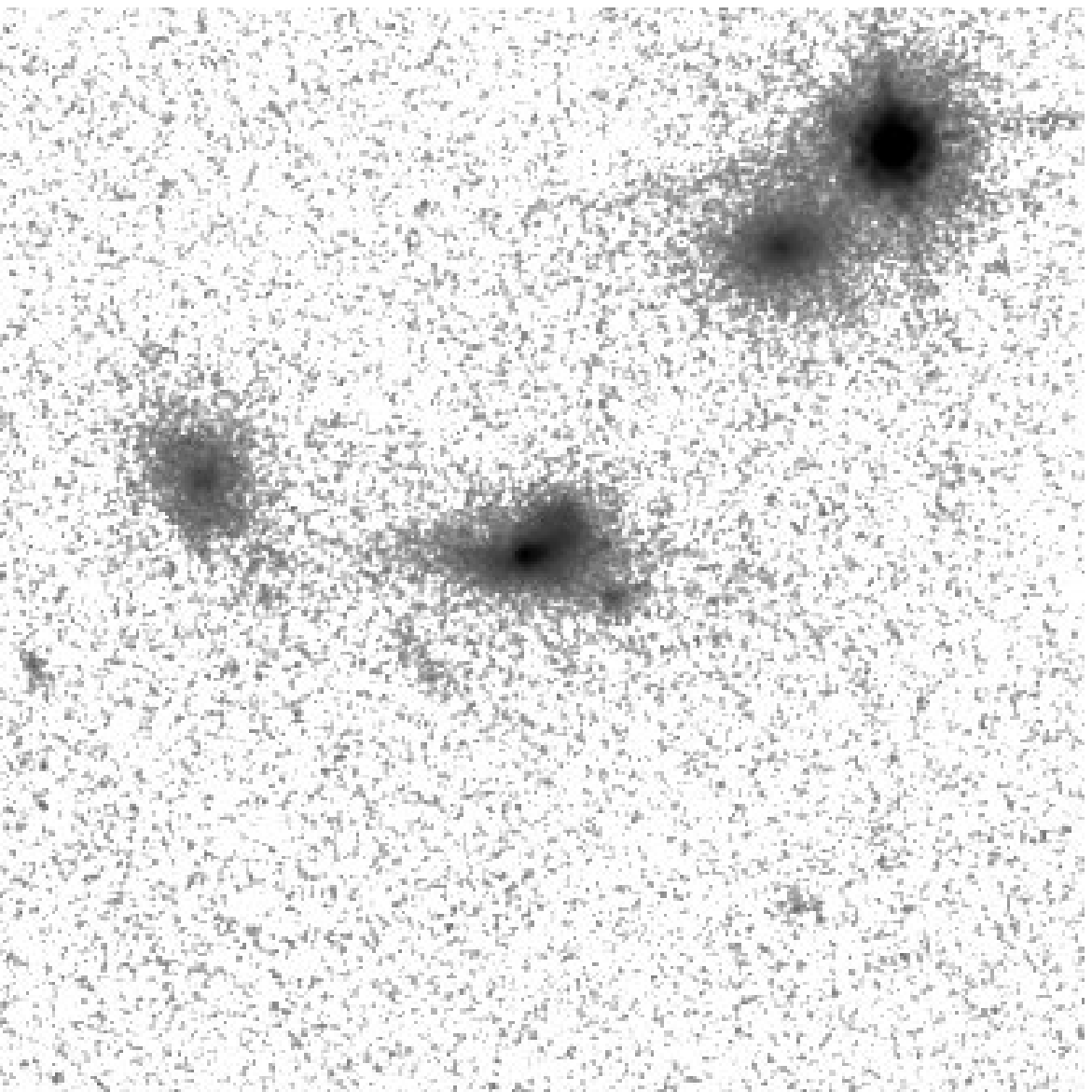}
\includegraphics[width=2cm]{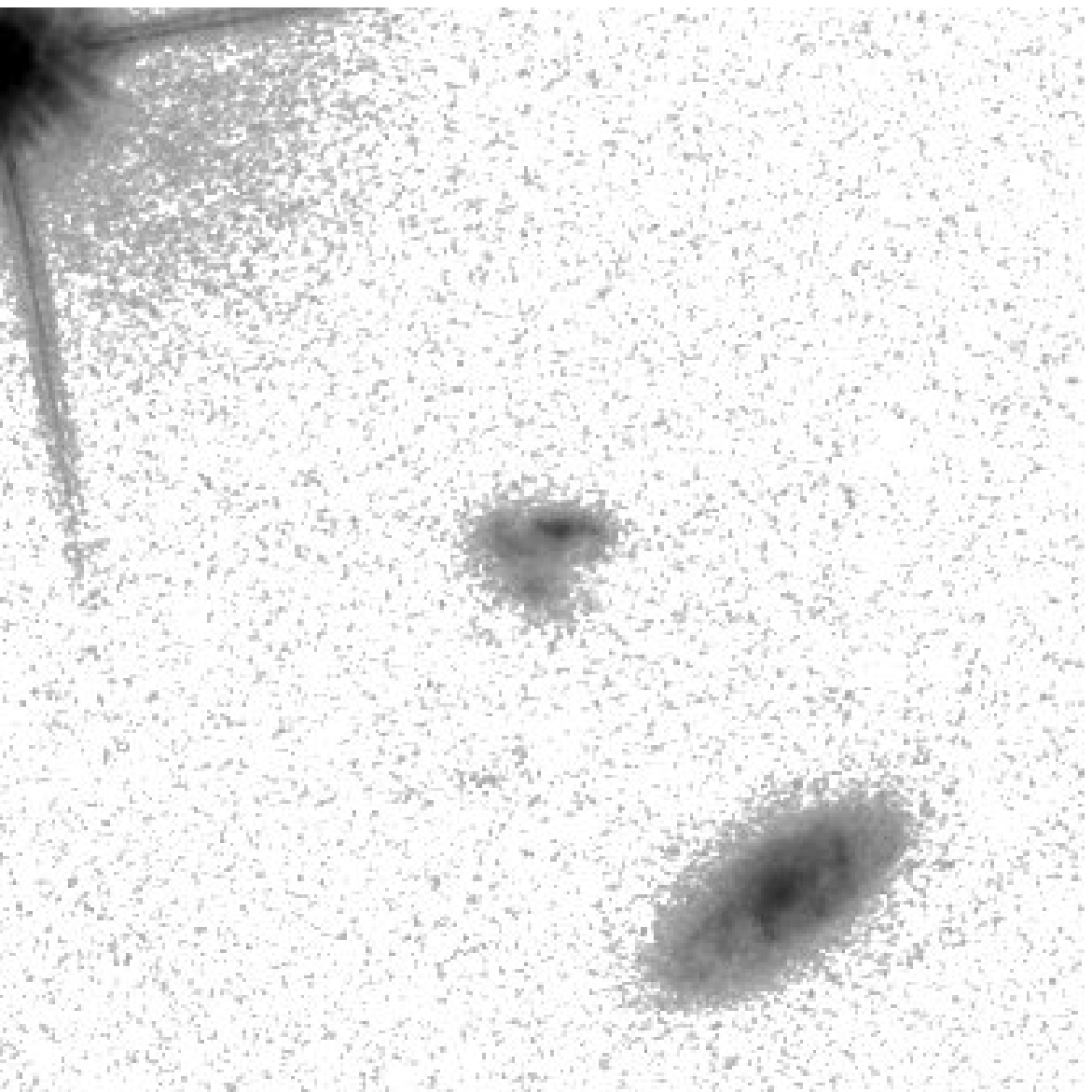}
\includegraphics[width=2cm]{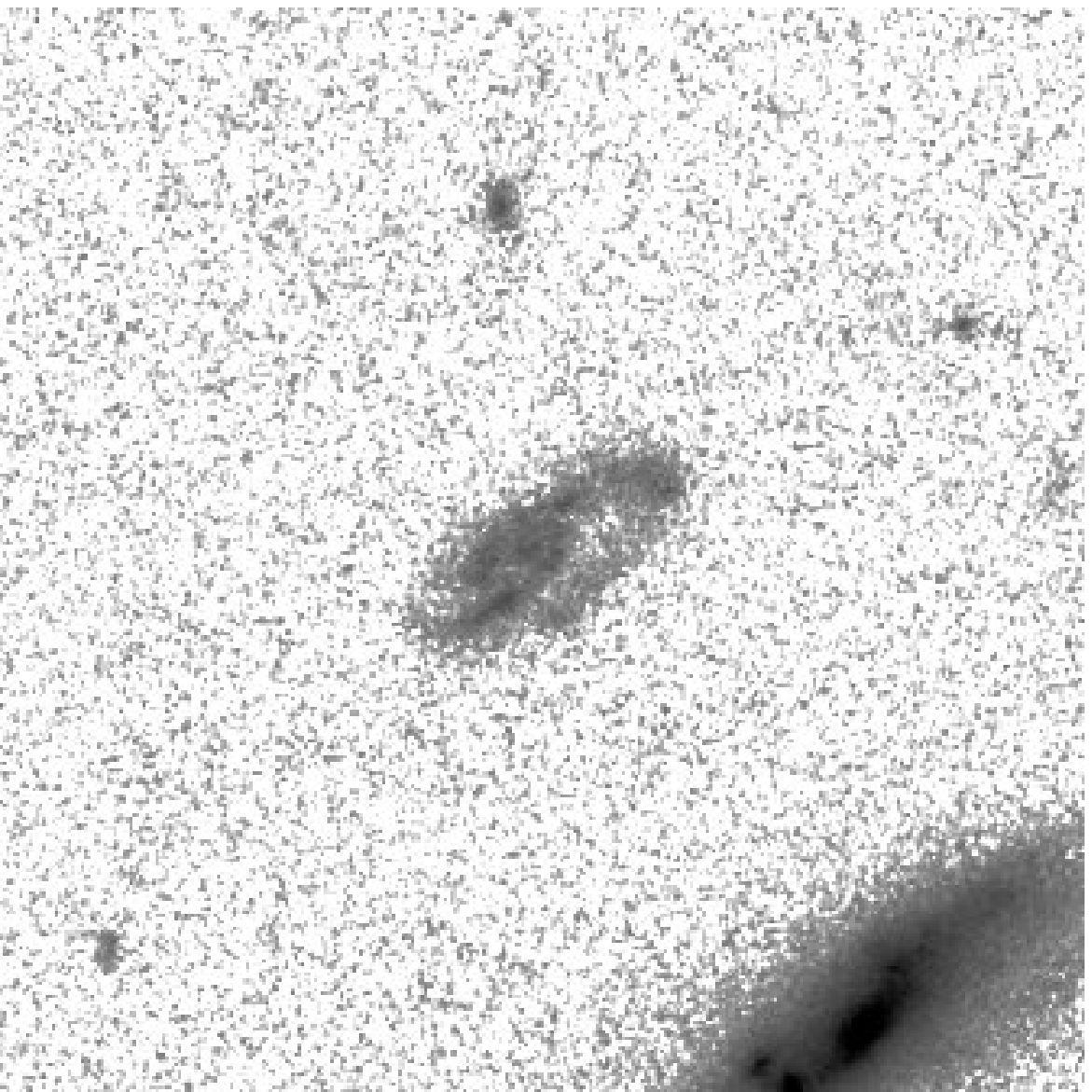}
\includegraphics[width=2cm]{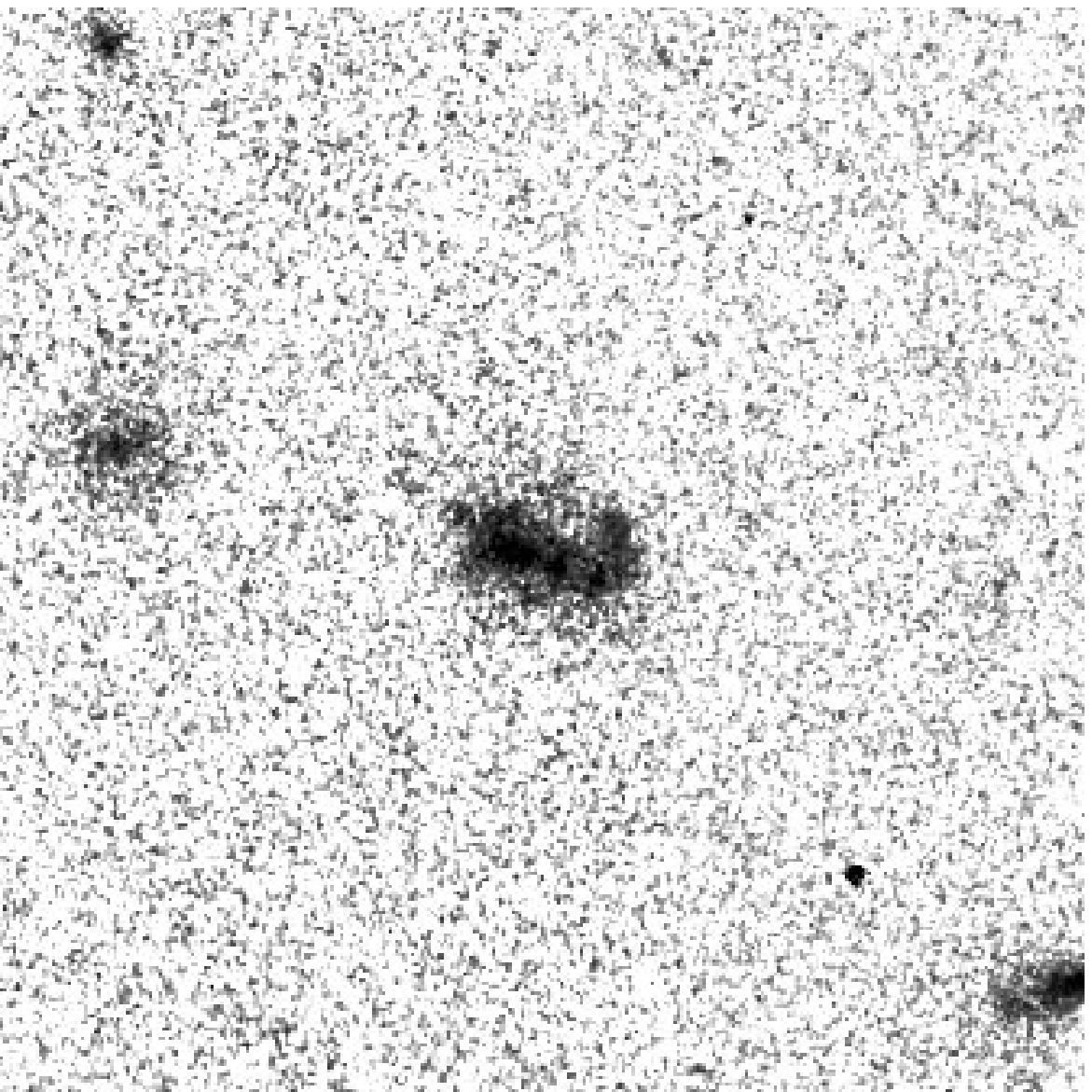}
\includegraphics[width=2cm]{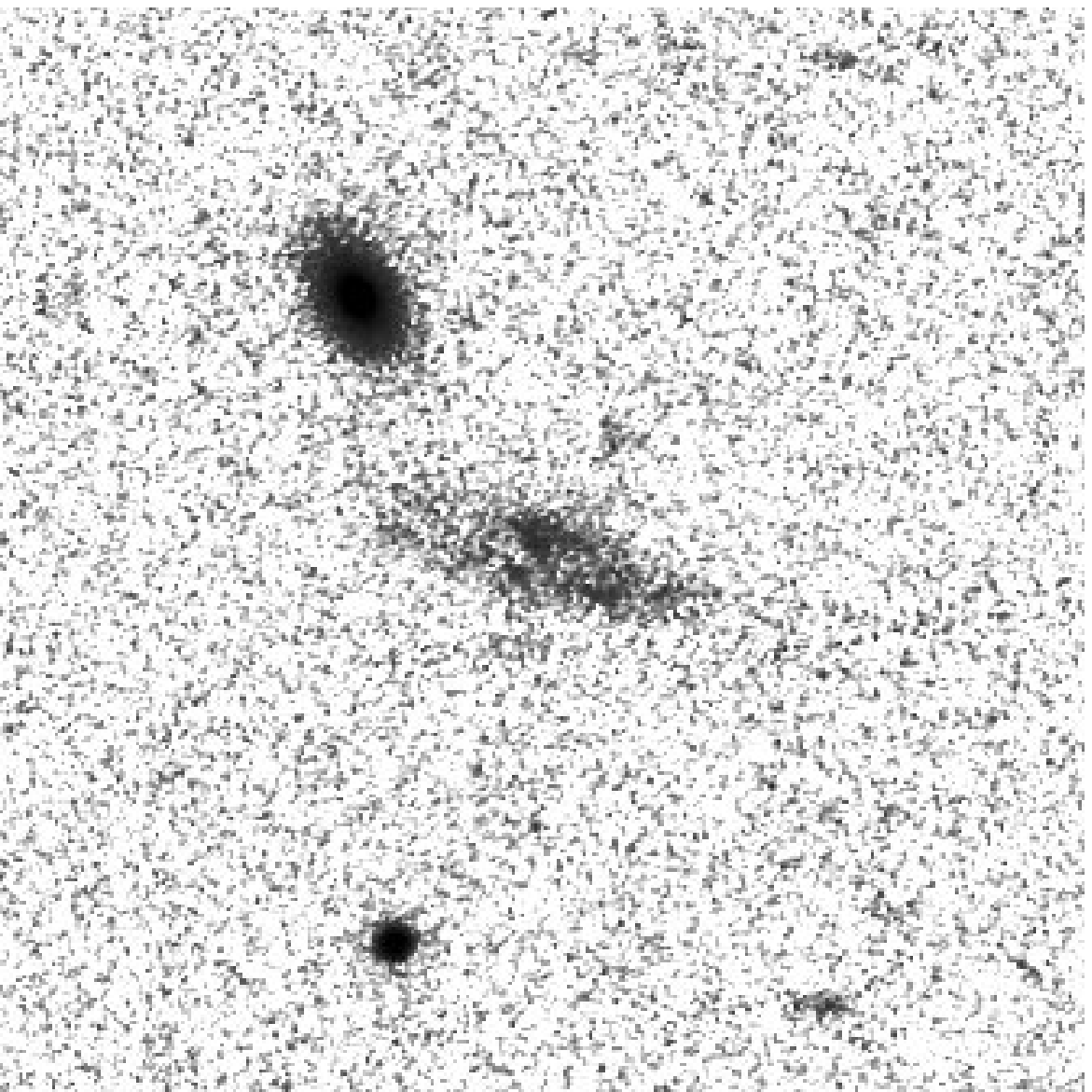}
\end{minipage}
\begin{minipage}[c]{\textwidth}
\includegraphics[width=2cm]{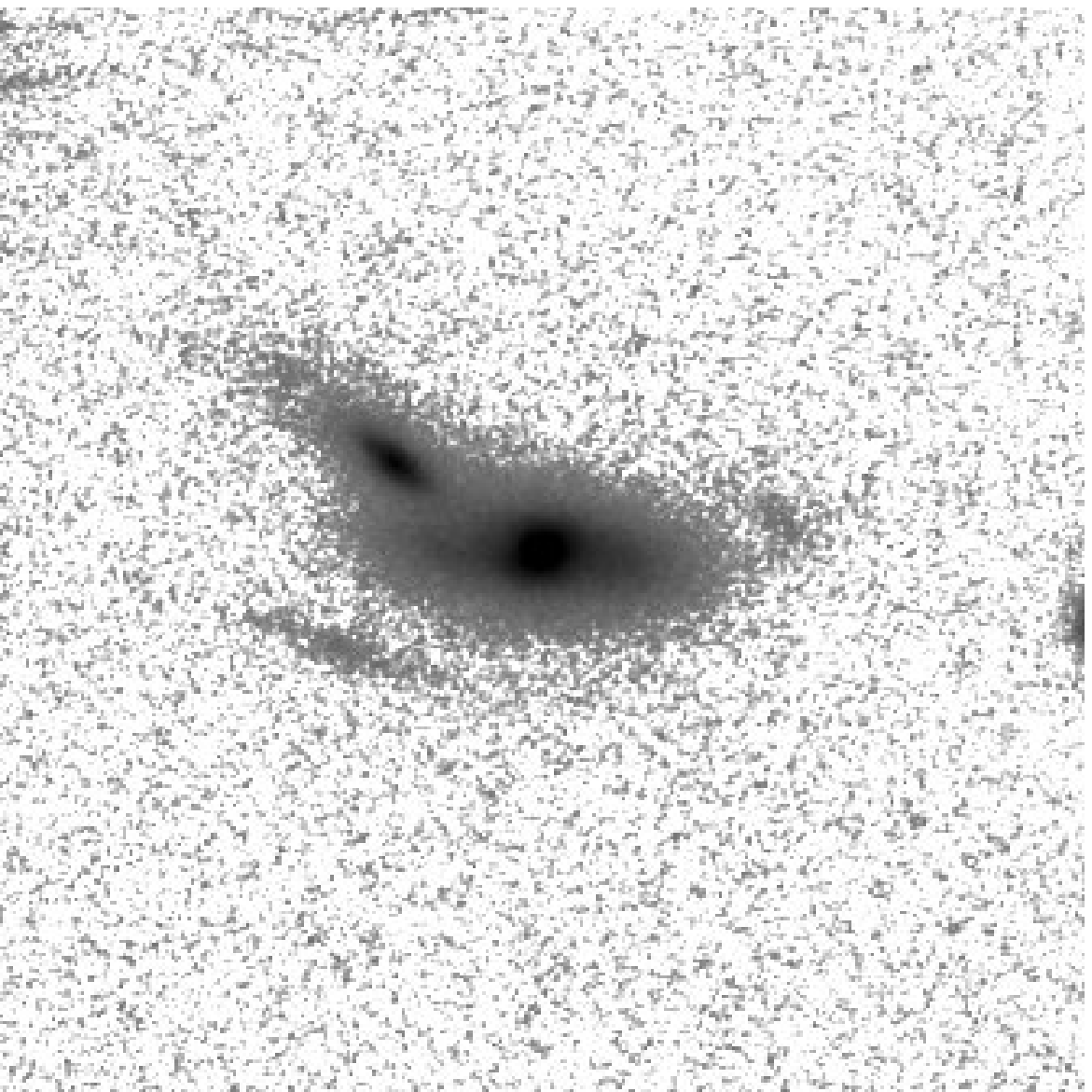}
\includegraphics[width=2cm]{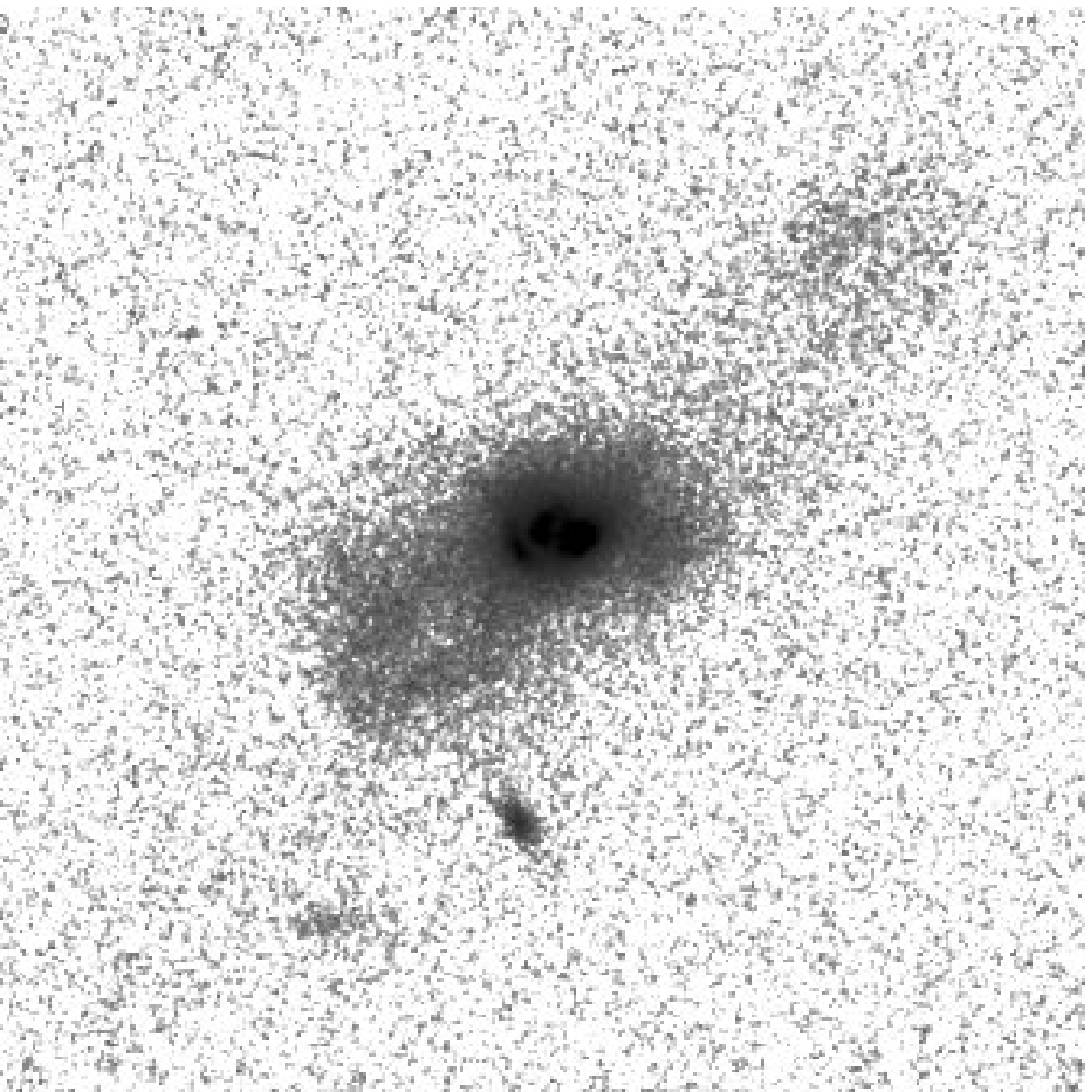}
\includegraphics[width=2cm]{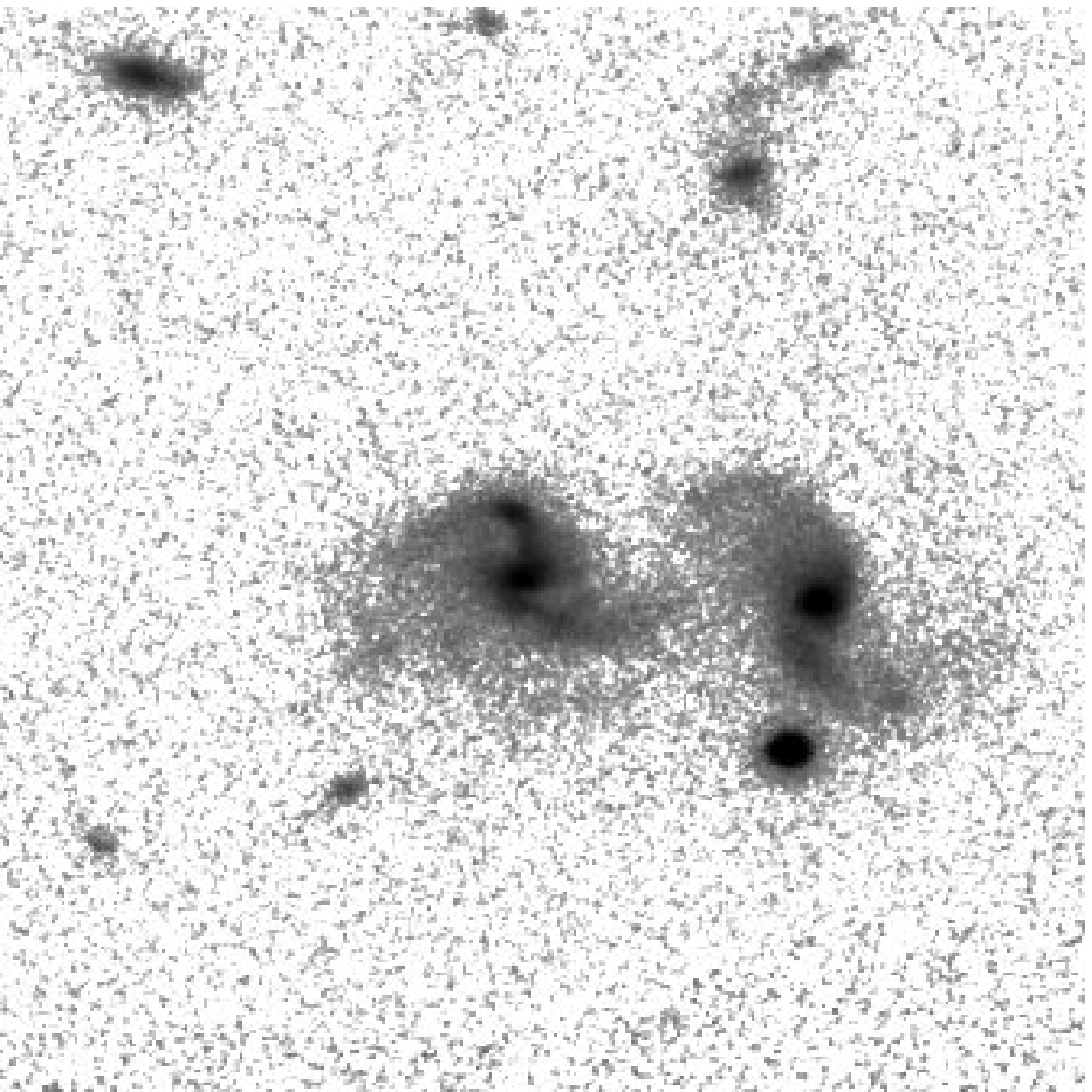}
\includegraphics[width=2cm]{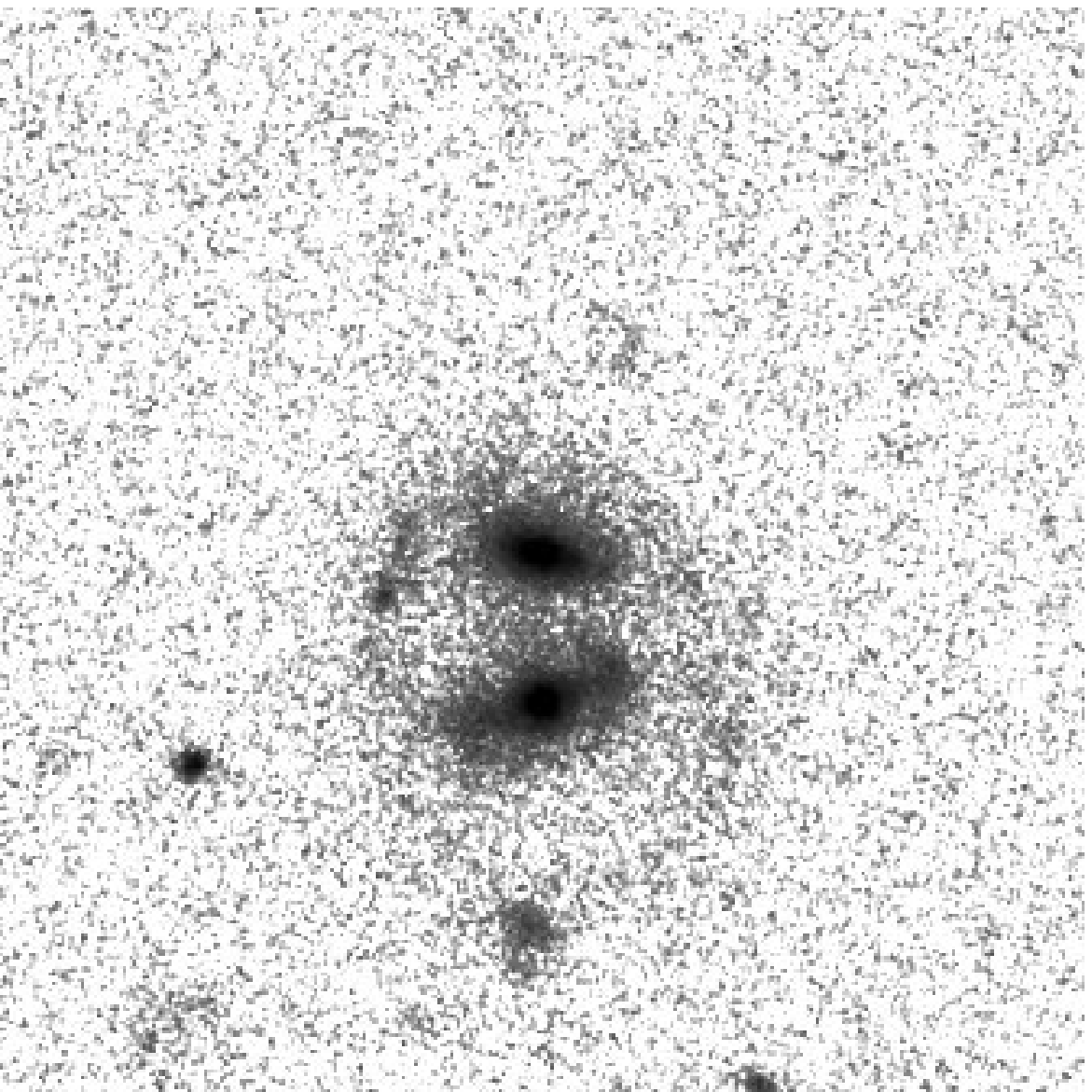}
\includegraphics[width=2cm]{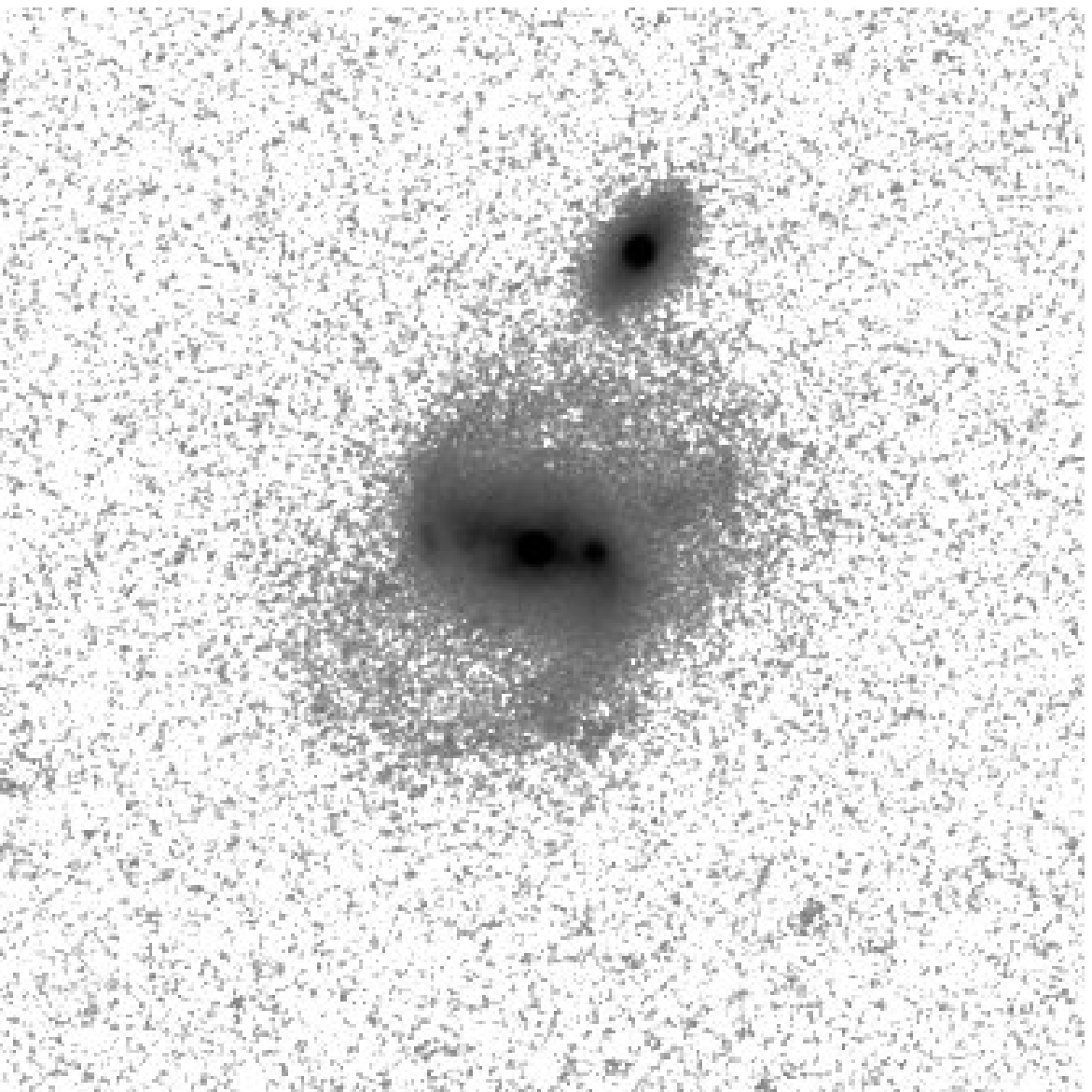}
\includegraphics[width=2cm]{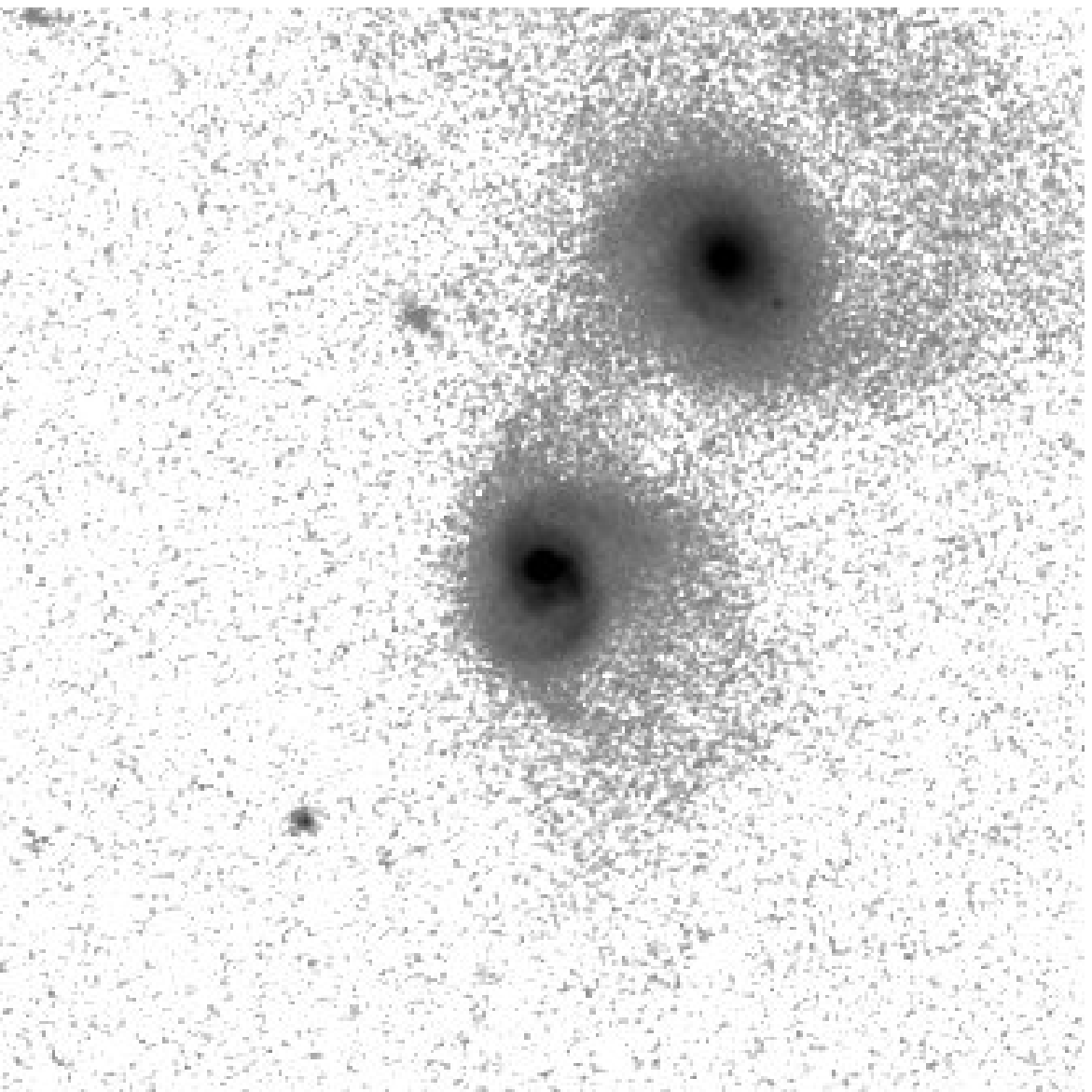}
\includegraphics[width=2cm]{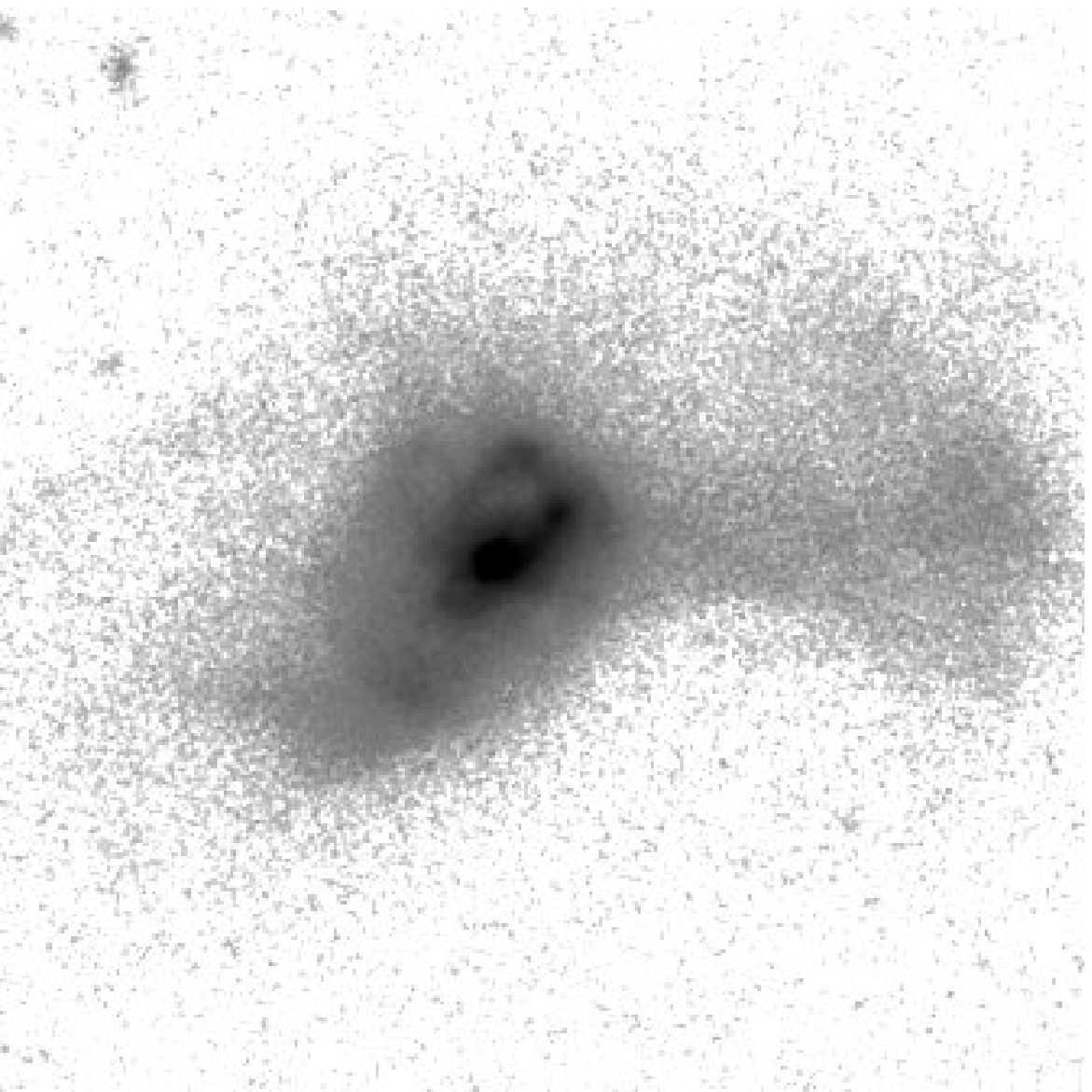}
\includegraphics[width=2cm]{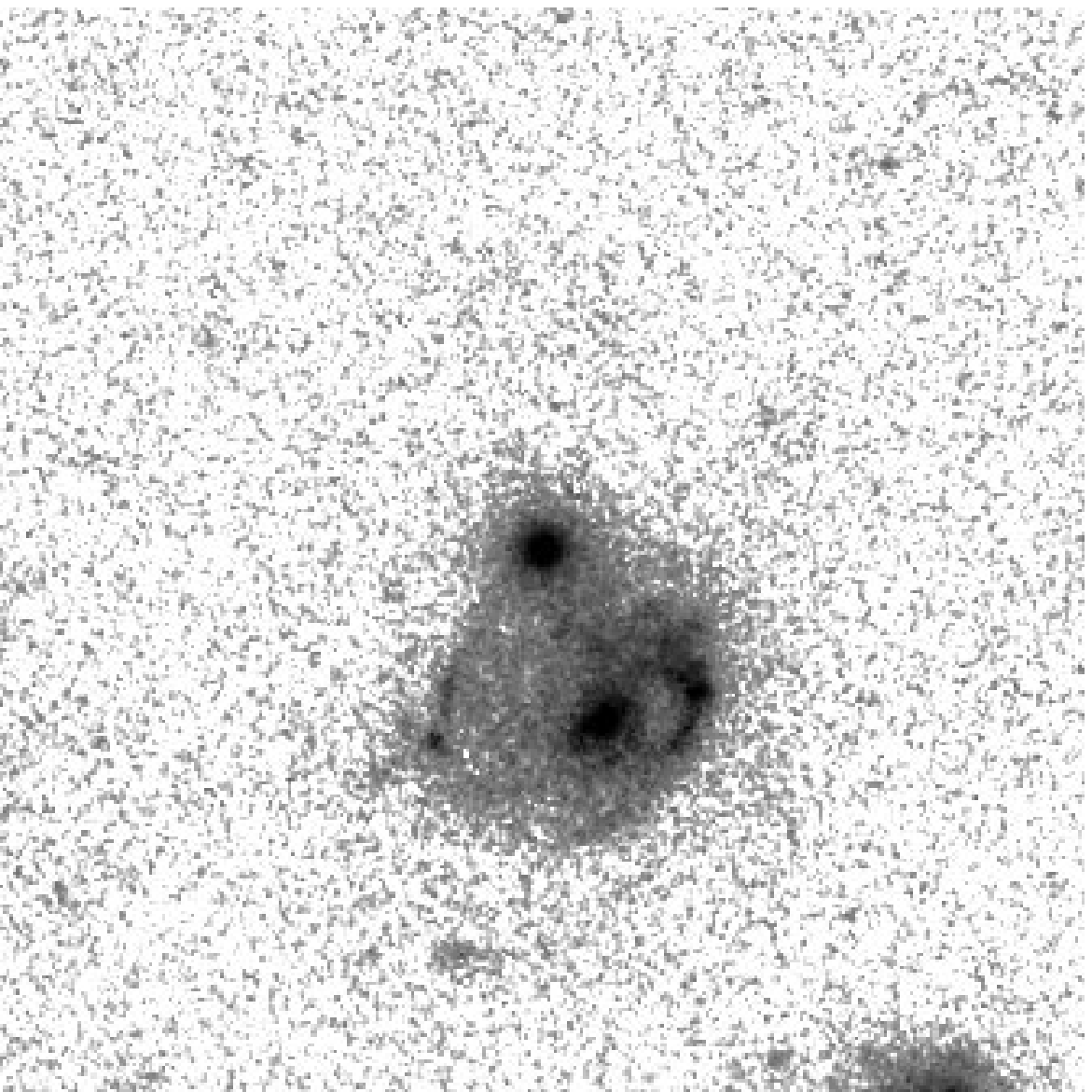}
\end{minipage}
\begin{minipage}[c]{\textwidth}
\includegraphics[width=2cm]{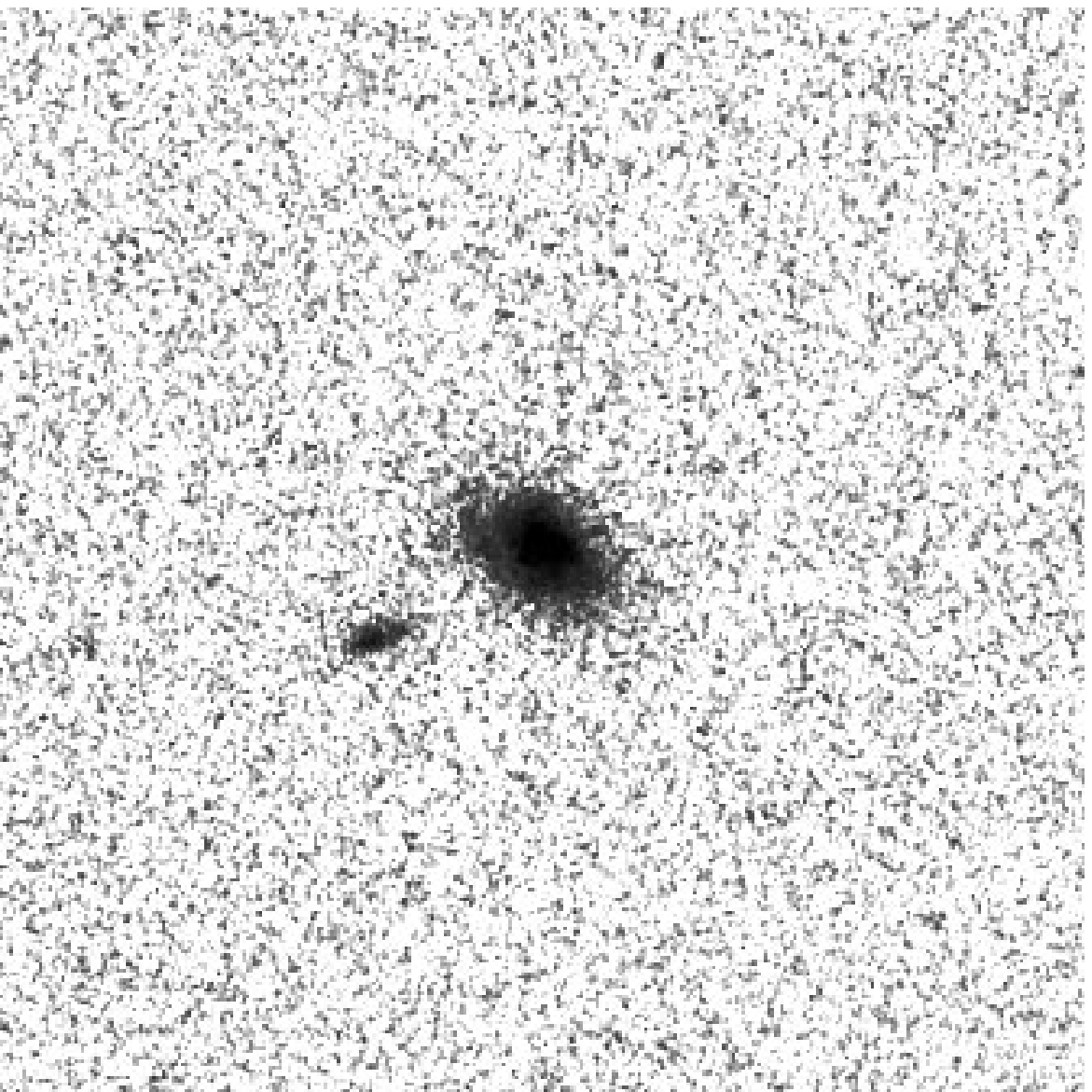}
\includegraphics[width=2cm]{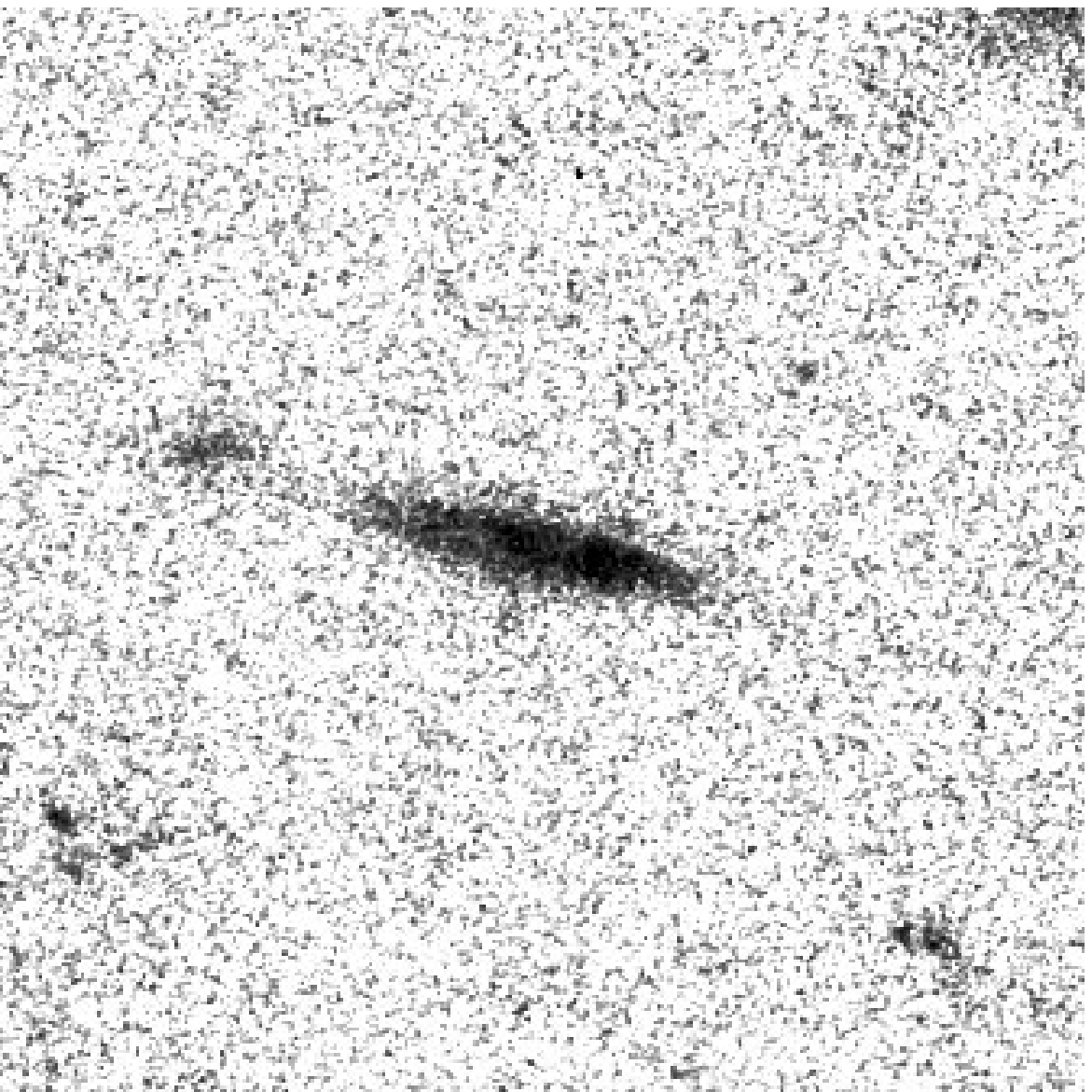}
\includegraphics[width=2cm]{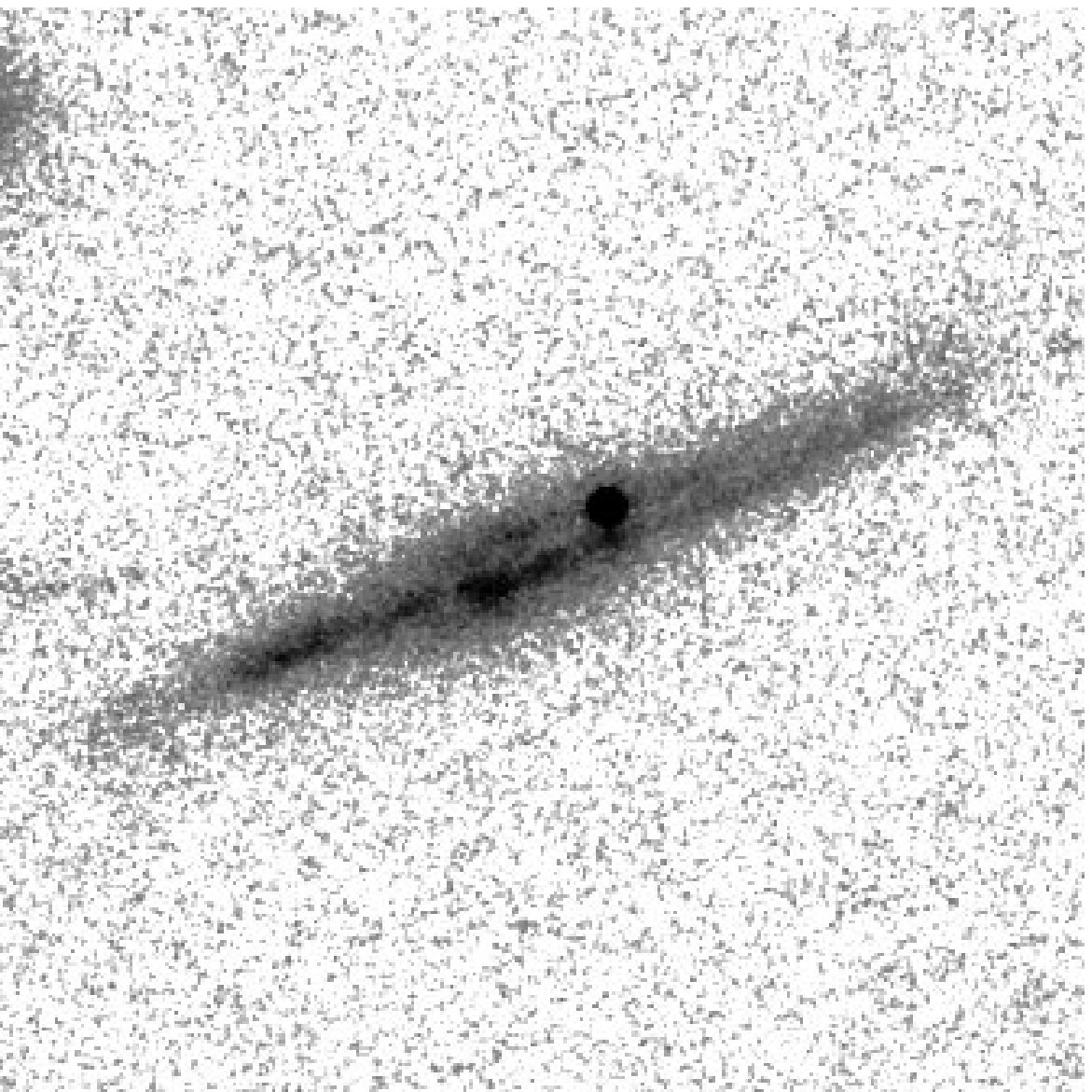}
\includegraphics[width=2cm]{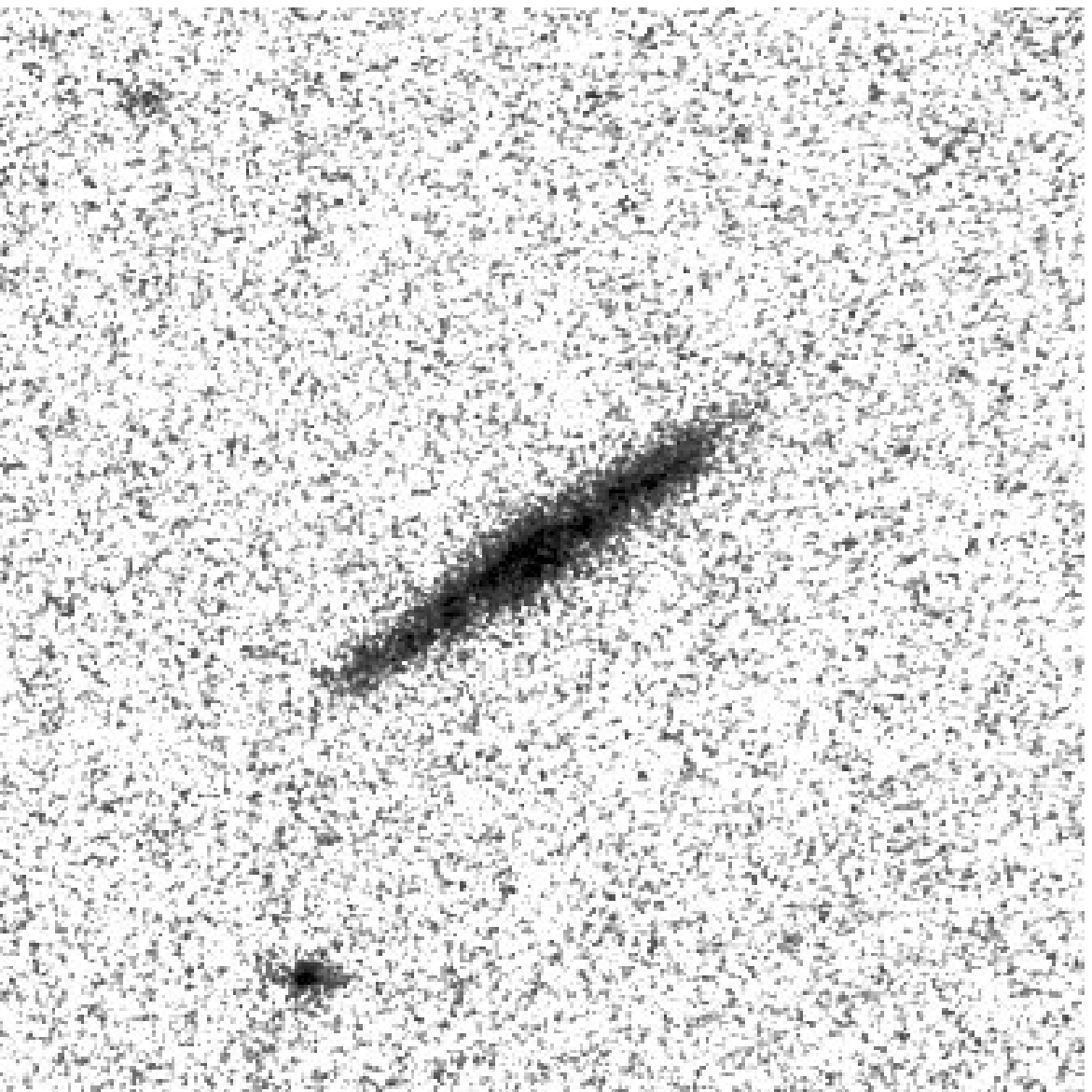}
\includegraphics[width=2cm]{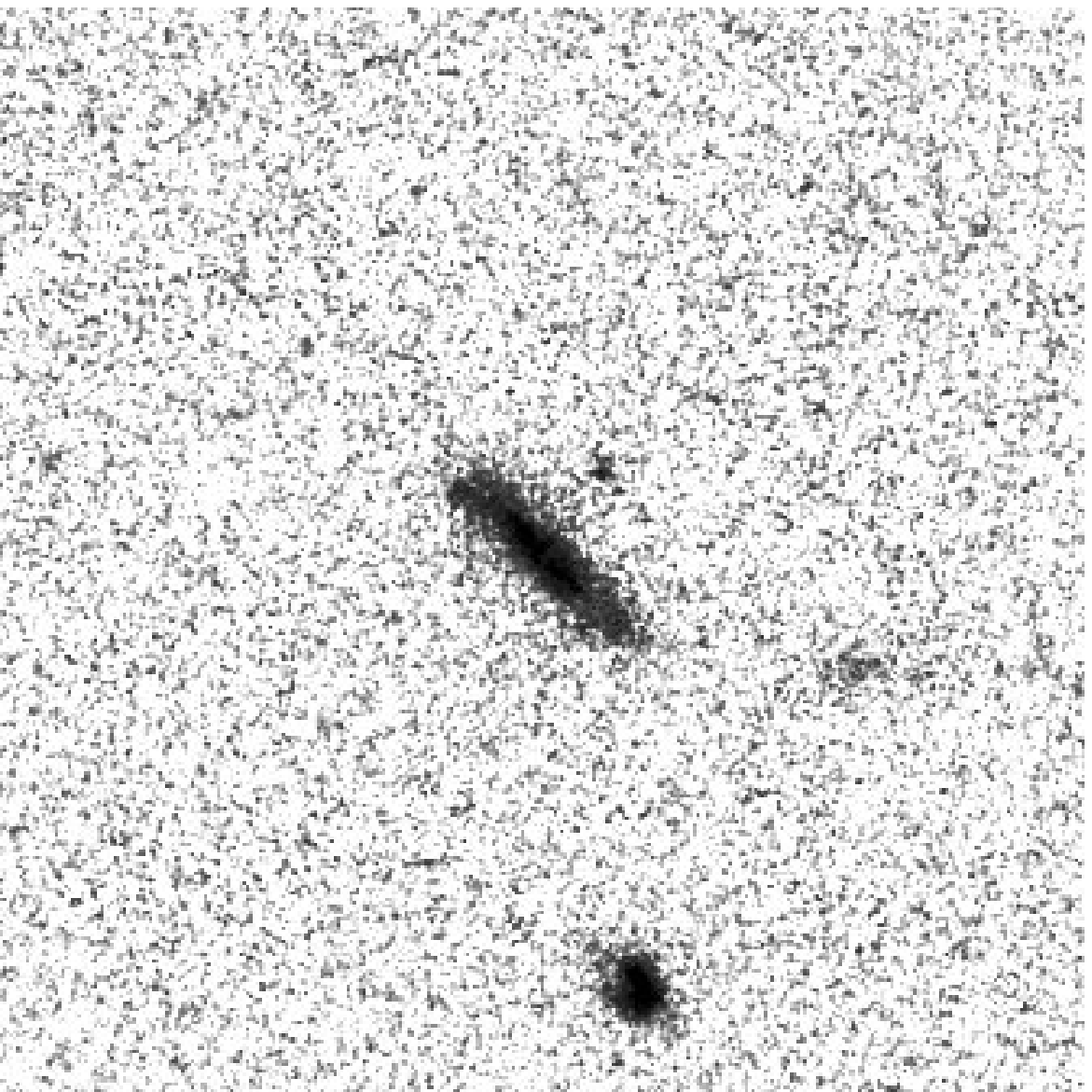}
\includegraphics[width=2cm]{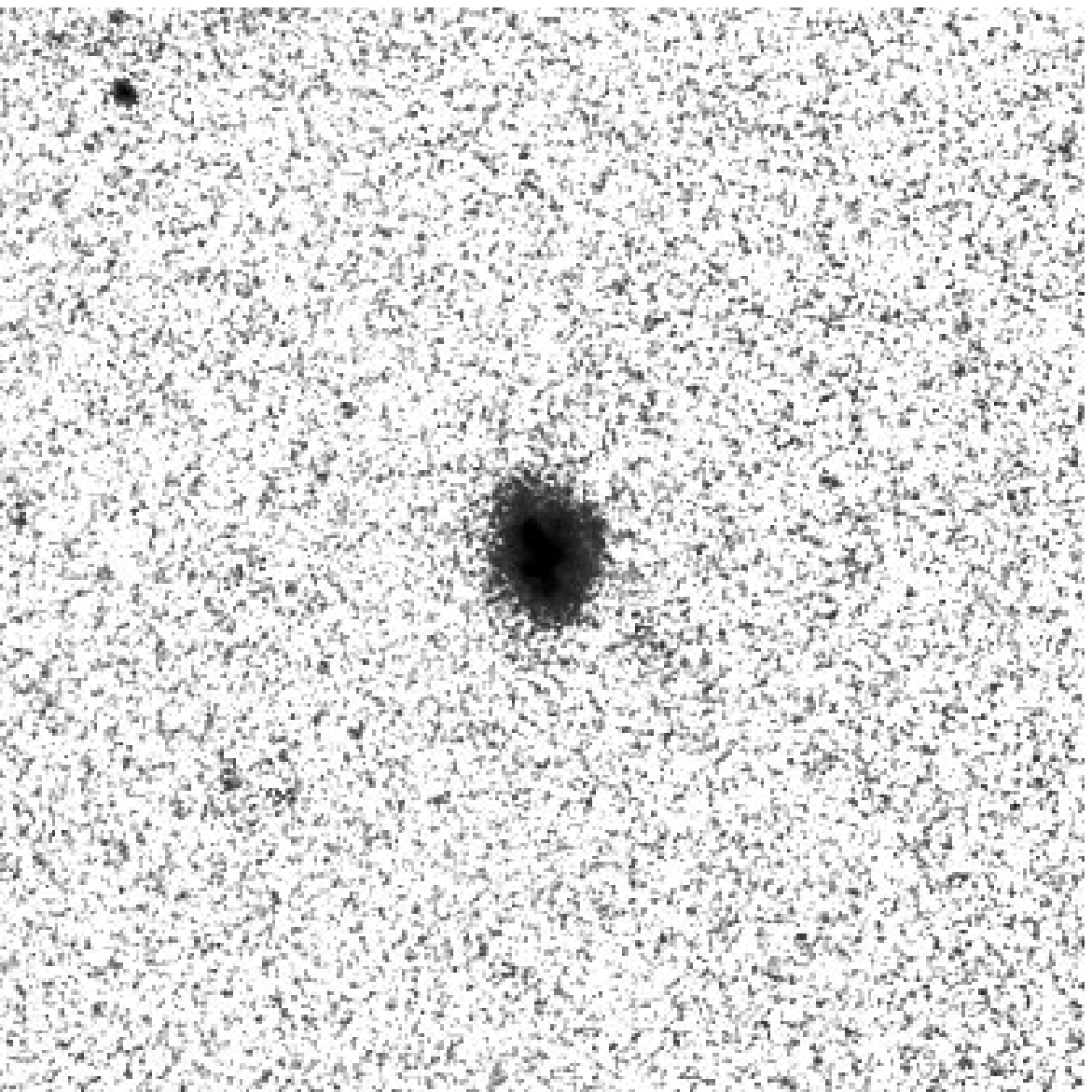}
\includegraphics[width=2cm]{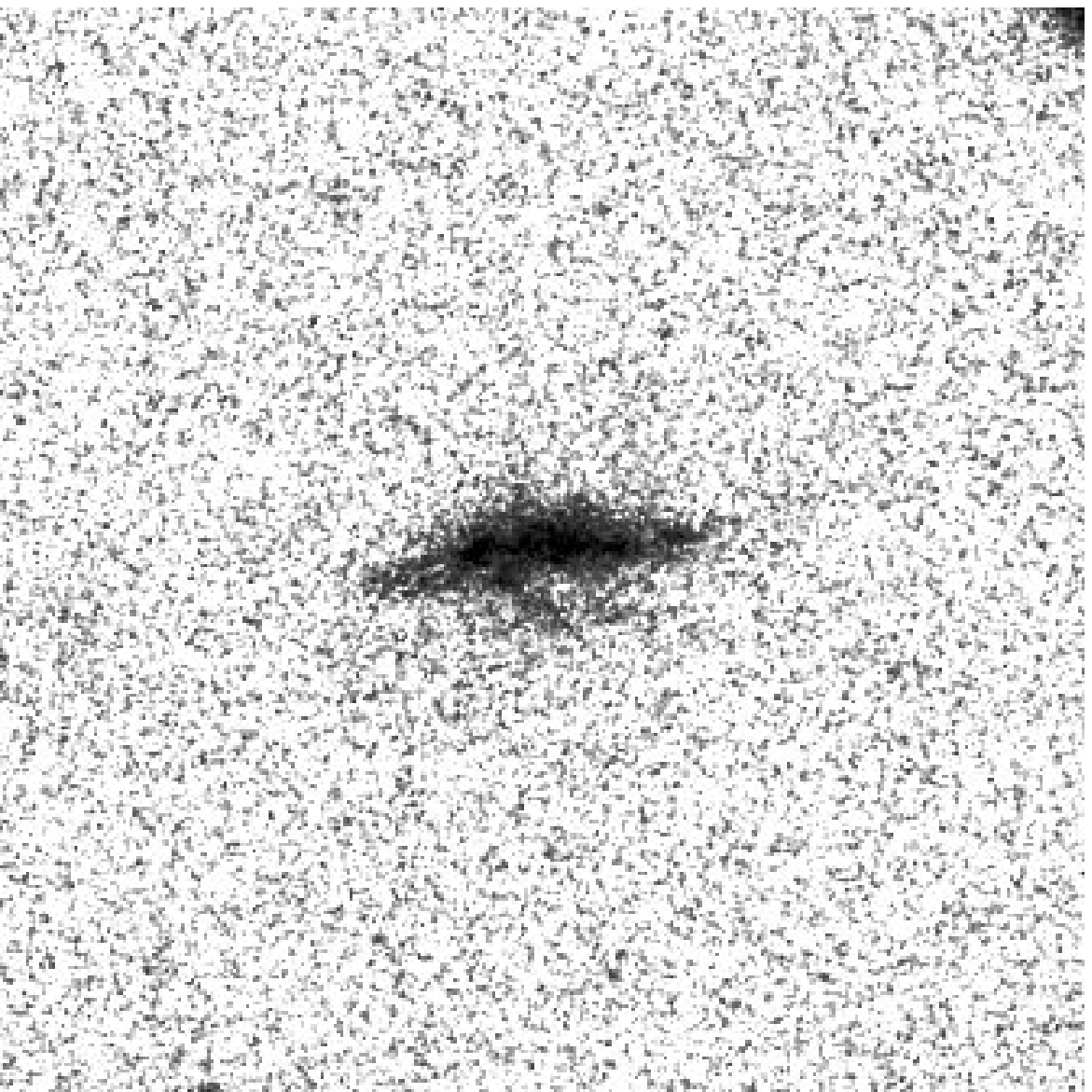}
\includegraphics[width=2cm]{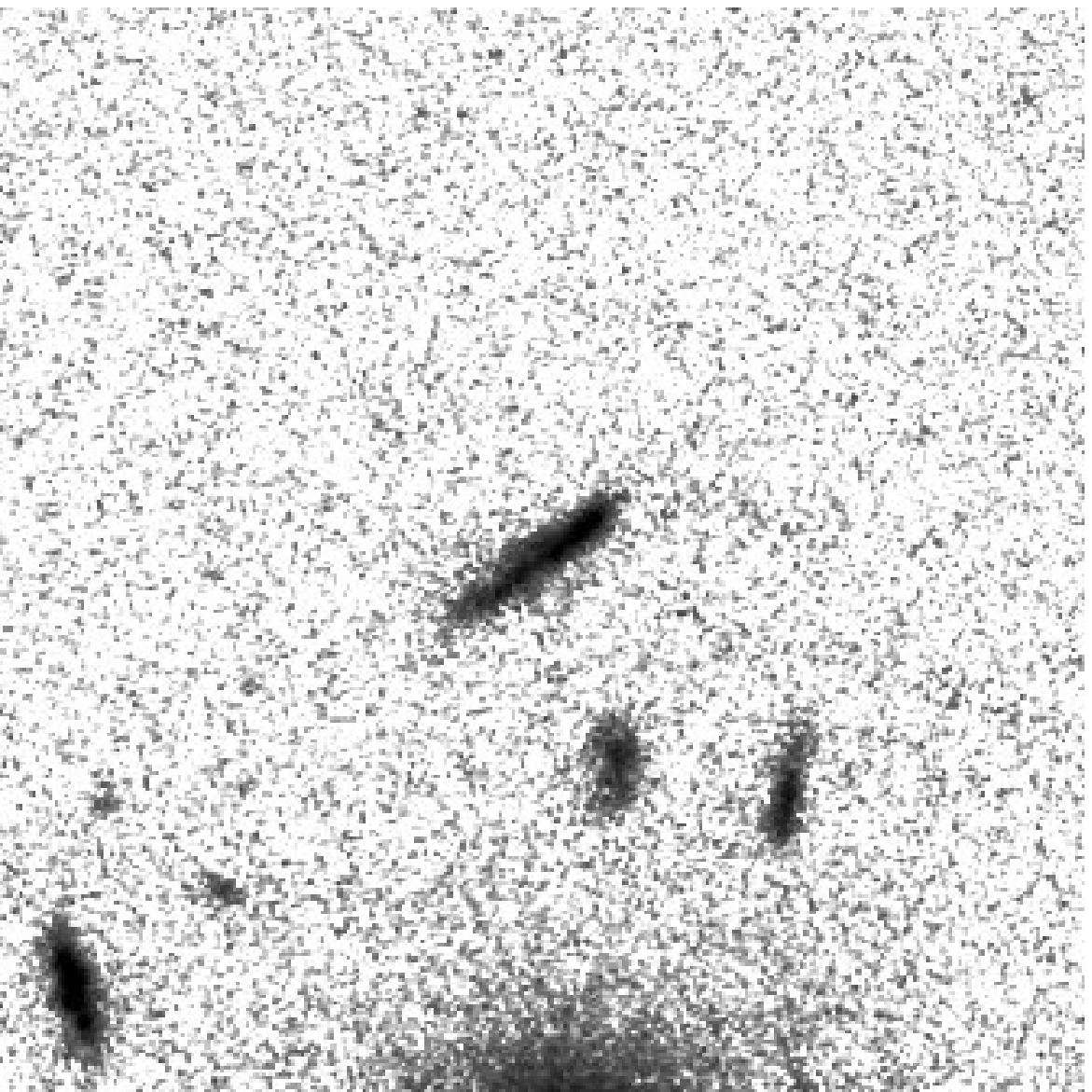}
\end{minipage}
\caption{Example of different morphological classes used in visual classification: from class 1 up to class 5 \textit{(from top to bottom)}.}
\label{Visual_class2}
\end{figure*}

\begin{figure}
 \includegraphics[width=\columnwidth]{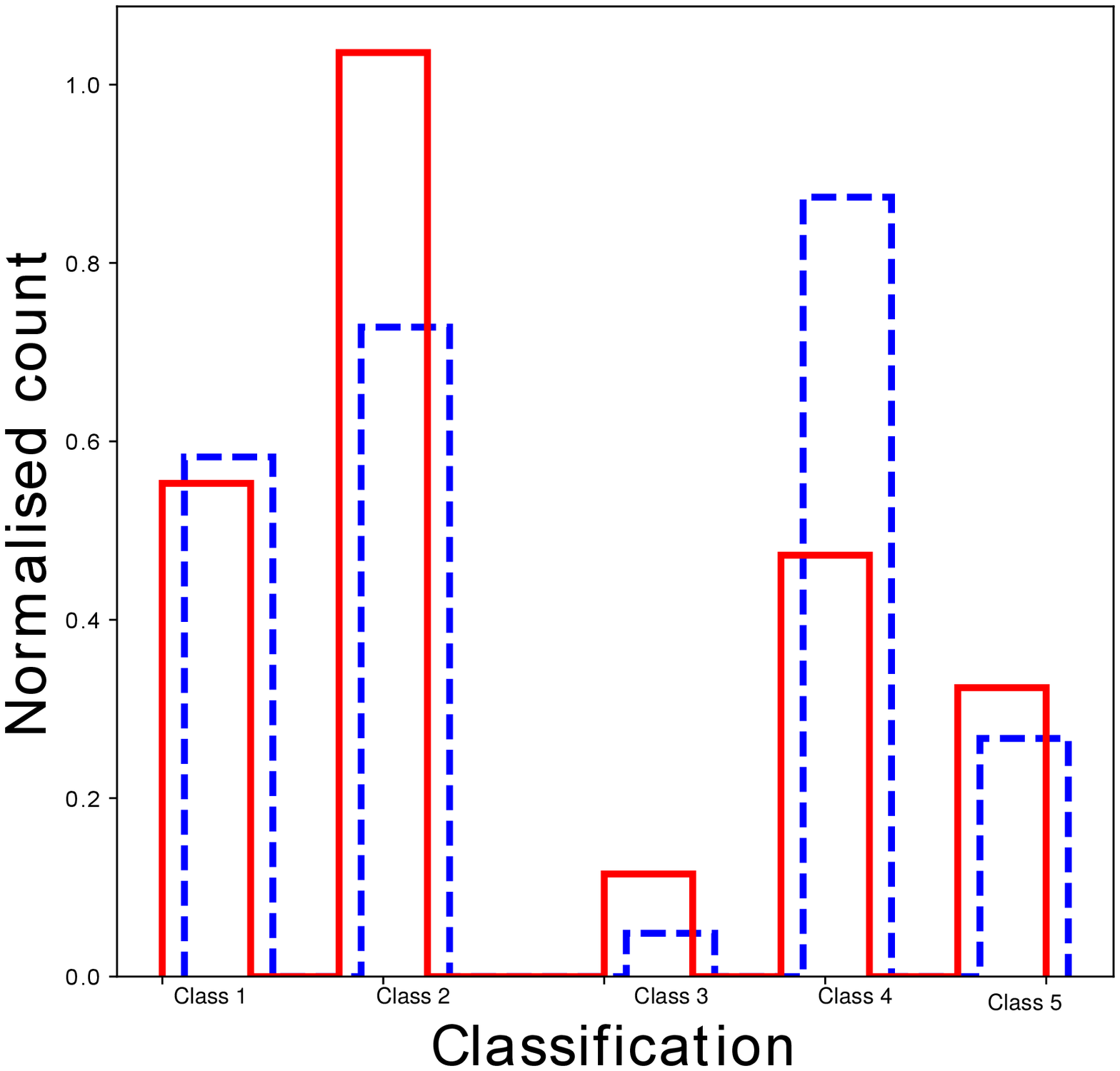}
 \caption{Normalised distributions of visually classified morphological types of FIR green valley AGN (blue dashed lines) and non-AGN (red solid lines).}
 \label{Visual_class}
\end{figure}

Using the HST/ACS F814W images we went through the visual morphological classification of all 103 and 2609 FIR AGN and non-AGN, respectively. All galaxies were classified by three classifiers independently (AM, MP, and BN) into one of the following five classes:
\begin{itemize}
\item class 1: elliptical, S0 or S0/S0a galaxies,
\item class 2: spiral galaxies,
\item class 3: irregular galaxies,
\item class 4: peculiar galaxies (with strange structures, e.g., tails, rings), signs of interactions, or clear mergers, and
\item class 5: unclassified galaxies (e.g., faint objects, low resolution images, edge-on sources, etc.).
\end{itemize}
\indent Figure \ref{Visual_class2} shows an example of galaxies belonging to some of the five classes. We considered that a galaxy belongs to certain morphological type if it was classified under the same class by two or three classifiers. Table \ref{tab_Visual_number} shows the final result of visual morphological classification. In total we classified 91\% and 86\% of FIR AGN and non-AGN green valley galaxies, respectively. Figure \ref{Visual_class} shows the distribution of both samples per morphological class. It can be seen that the most significant difference is between classes 2 and 4, where 38\% of AGN were classified as peculiar galaxies, interactions or mergers (19\% of non-AGN), while 46\% of non-AGN show spiral morphologies (26\% in case of AGN). 

\begin{table}
\caption {Number of galaxies in the final visual morphological classification.}
\label{tab_Visual_number}
\newcommand{\mc}[3]{\multicolumn{#1}{#2}{#3}}
\begin{center}
\begin{tabular}{l|c|c|}\cline{2-3}
 & FIR AGN & FIR non-AGN\\\hline
\mc{1}{|c|}{Class 1} & 26 (25\%) & 452 (17\%)\\\hline
\mc{1}{|c|}{Class 2} & 27 (26\%) & 1204 (46\%)\\\hline
\mc{1}{|c|}{Class 3} & 2 (2\%) & 87 (3\%)\\\hline
\mc{1}{|c|}{Class 4} & 39 (38\%) & 494 (19\%)\\\hline
\mc{1}{|c|}{Class 5} & 9 (9\%) & 372 (14\%)\\\hline
\mc{1}{|c|}{Total}     &103&2609 \\\hline
\end{tabular}
\end{center}
\end{table}

\subsection{Comparison with non-parametric methods}
\label{sec_nonparametric}

We are using the public COSMOS morphology catalogues as there has more information on morphology parameters to be used in the section \ref{sec_distribution_Morphology_Parameters} and section \ref{sec_same_mass_range} for analysis of AGN and non-AGN green valley galaxy morphology properties.\\
To understand better the morphological properties of our sample and high SFRs seen in \cite{Mahoro2017}, in addition to visual classifications we extracted also morphological parameters from available public catalogues in the COSMOS field: \citet[][hereafter TAS09]{Tasca2009}, \citet[][hereafter CAS07]{Cassata2007}, and \citet[][hereafter SCA07]{Scarlata2007}. In the following we describe the three used catalogues, obtained through the HST/ACS data, and compare our visual classification with non-parametric ones. All used catalogues classified galaxies into E/S0, spirals, and irregulars. None of the mentioned public catalogues classifies separately peculiar galaxies, mergers, and interaction, and irregular class may include a significant population of these. Therefore, in the following, we compared our Class 1 with E/S0, Class 2 with spirals, and Class 3\,+\,Class 4 with galaxies classified as irregulars in the three analysed catalogues. Table~\ref{tab_comparison_morph} gives the number of galaxies in different catalogues whose classification overlaps with ours, and the \% of agreement for each type. 

\begin{table*}
\begin{center}
\caption{Number of galaxies (and fraction) being in agreement between our visual and non-parametric morphological classifications. The values given in bold indicate the best match. 
\label{tab_comparison_morph}}
\begin{tabular}{|c|c|c|c|c|c|c|c|}
\hline
Catalogues                                          &                       & \multicolumn{3}{c|}{AGN}                                                       & \multicolumn{3}{c|}{Non-AGN}                                       \\ \hline
                                                    &                       & E/S0                               & spiral                  & irregular                 & E/S0                & spiral                    & irregular                  \\ \hline
\multirow{3}{*}{TAS09} & class\_int            & 23 (88\%)                         & 12 (44\%)          & 11 (27\%)            & 253 (56\%)          & 898 (75\%)           & 351 (60\%)           \\ \cline{2-8} 
                       & class\_linee          & \textbf{24 (92\%)}                         & 14 (52\%)          & 10 (24\%)            &  \textbf{303 (67\%)}         & 902 (75\%)           & 291 (50\%)           \\ \cline{2-8} 
                       & class\_galsvm 		& 19 (73\%) & \textbf{21 (78\%)} & 11 (27\%) & 197 (44\%) & \textbf{1076 (89\%)} & 379 (65\%) \\ \hline
CAS07                &                       & 16 (64\%)                         & 13 (50\%)          & \textbf{17 (41\%)}          & 199 (44\%)          & 893 (74\%)           & \textbf{381 (66\%)}          \\ \hline
SCA07               &                       & 1 (12.5\%)                           & 3 (30\%)          & 5 (39\%)          & 21 (17\%)             & 342 (86\%)           & 91 (39\%)            \\ \hline
\end{tabular}
\end{center}
\end{table*}

\subsubsection*{TAS09 morphological catalogue} 

This catalogue contains morphological information of 237912 sources in the COSMOS field, with magnitudes down to I$_{AB}$\,$\simeq$\,23.0. 
Morphological classification was carried out using three different methods:
 
\begin{enumerate}
\item Class\_int is based on three parameters: Abraham concentration index (CABR), which is a fraction of light contained in an inner 30\% and in a total flux isophote \citep{Abraham1996}; asymmetry index (ASYM), computed by rotating each galaxy 180\~deg about its centre, subtracting the rotated image from the initial one, and dividing the sum of the absolute value of pixels in the residual image by the sum of pixel values in the initial image\footnote{A correction for background noise is also applied by using the average asymmetry of the background.} \citep{Abraham1996, Cassata2007}; and Gini coefficient, defined as a cumulative distribution function of galaxy's pixel values \citep{Abraham2003}. 

\indent We cross-matched our catalogue with TAS09 catalogue. Out of 103 FIR AGN green valley galaxies, in class\_int 52 are classified as E/S0, 34 as spirals, and 17 as irregulars, while 505 of FIR non-AGN galaxies are classified as E/S0, 1535 as spirals and 569 as irregulars. The number of galaxies in each class that overlaps with our visual classification and fraction are shown in Table~\ref{tab_comparison_morph}. As can be seen the best match was found for E/S0 and spiral classes in the case of AGN and non-AGN samples, respectively.  \\

\item Class\_linee is based on five morphological parameters: \citet{Conselice2000} concentration index (CCON), measured as the logarithmic ratio of the apertures containing 80\% and 20\% of the total flux; asymmetry index (ASYM) and Gini index, measured as in class\_int; smoothness (or clumpiness), defined as the measure of the relevance of small-scale structures \citep{Conselice2000}; and moment of light at 20\% ($\rm{M_{20}}$), measured using the flux in each pixel multiplied by the squared distance to the centre of the galaxy, summed over the 20\% brightest pixels of the galaxy \citep{Lotz2004}.

\indent With this method again all FIR AGN and non-AGN from our sample were classified. 50 (602), 40 (1574), and 13 (433) of galaxies were classified as E/S0, spirals, and irregulars, respectively, in case of AGN (non-AGN). Out of these, the best match with visual classification was found for E/S0 class, as can be seen in Table~\ref{tab_comparison_morph}. In general, this classification provides better agreement with visual classification than the previous one, and the best agreement for E/S0 galaxies in comparison to all other methods.  \\

\item Class\_svm is based on learning machines called support vector machines. It uses the galSVM code and measures 7 parameters, plus the source ellipticity and surface brightness, as described in \citep{HuertasCompany2008}. With this classification in TSC09 32 (285), 54 (1777), and 17 (547) of galaxies were classified as E/S0, spirals, and irregulars, respectively, in case of AGN (non-AGN). This classification gives a better agreement with our visual classification for spiral and irregular galaxies then the previous two, and the best agreement for spiral galaxies when compared with all other methods. 
\end{enumerate}
We are using the public COSMOS morphology catalogues as there has more information on morphology parameters to be used in the section \ref{sec_distribution_Morphology_Parameters} and section \ref{sec_same_mass_range} for analysis of AGN and non-AGN green valley galaxy morphology properties.\\

\subsubsection*{CAS07 morphological catalogue}

This catalogue contains morphological classification of 232022 COSMOS galaxies with $I_{AB}\,\leq$ 25. Classification is based on four morphological parameters: ASYM, CCON, Gini, and M$_{20}$ moment of light, defined as mentioned previously. The authors also measured axial ratio, as a ratio between minor and major axis b/a using SExtractor \citep{Bertin1996}. \\
\indent We found in total 101 and 2595 AGN and non-AGN counterparts, respectively, when cross-matching our catalogue of visual classification and CAS07 catalogue. In case of AGN, CAS07 catalogue contains 32 galaxies classified as E/S0, 45 as spirals, and 24 as irregulars (in comparison to our visual classification where we have 25, 26, 2, 39, and 9 galaxies classified as Class 1, 2, 3, 4, and 5 respectively). In case of non-AGN, 359 were classified as E/S0, 1538 as spirals, and 698 as irregular galaxies (in comparison to our visual classification where we have 451, 1201, 86, 489, and 368 galaxies classified as Class 1, 2, 3, 4, and 5, respectively). When compared with our visual classification, this catalogue gives the best match for irregular galaxies (including interactions and mergers), as can be seen in Table~\ref{tab_comparison_morph}. 

\subsubsection*{SCA07 morphological catalogue}

This catalogue is based on the HST/ACS photometric catalogue of \cite{Leauthaud2007}, and uses the Zurich Estimator of Structural Types (ZEST) classification code \citep{Scarlata2007}. Morphological classification is available for 56000 galaxies with $I_{AB}\,\leq$ 24 from the COSMOS field. The authors used the combination of four morphological parameters to classify galaxies into E/S0, disc (spiral), and irregular: ASYM, CCON, Gini, and M$_{20}$ moment of light, defined as mentioned above. \\
\indent SCA07 catalogue contains 33 and 906 of our FIR AGN and non-AGN galaxies, respectively, much less than the previous two catalogues. Of these, 5 AGN were classified as ellipticals, 10 as spirals, 8 as irregulars, and 10 stayed unclassified (in comparison to 8, 10, 2, 11, and 2 galaxies classified visually in our work as Class 1, 2, 3, 4, and 5 respectively). On the other hand 37, 692, and 163 of non-AGN FIR galaxies were classified as ellipticals, spirals, and irregulars, respectively (compared to 125, 397, 33, 201, and 150 galaxies classified in our work as Class 1, 2, 3, 4, and 5 respectively). Matching with our visual classification is again shown in Table~\ref{tab_comparison_morph}, and it can be observed that in most cases this catalogue gives poorer comparison than the previous two.\\
\indent  We used this public data to compare our visual classification with automated classification and we get good matching Class\_svm \citep{HuertasCompany2008} classification in general and more results are summarised in Table \ref{tab_comparison_morph}.

\subsection{FIR green valley AGN and non-AGN in standard morphological diagrams}
\label{sec_distribution_Morphology_Parameters}

In this section we analyse the location of our FIR AGN and non-AGN galaxies in some of the standard morphological diagrams. We use the parameters measured in TSC09 (class\_int) and CAS07. SCA07 was not used in this section due to the small number of counterparts, especially in case of AGN. 

\subsubsection{Gini vs. concentration indices}
\label{Gini_coefficient versus_Concentration_index}

\begin{figure*}
\centering
\begin{minipage}[c]{.49\textwidth}
\includegraphics[width=0.85\textwidth,angle=0]{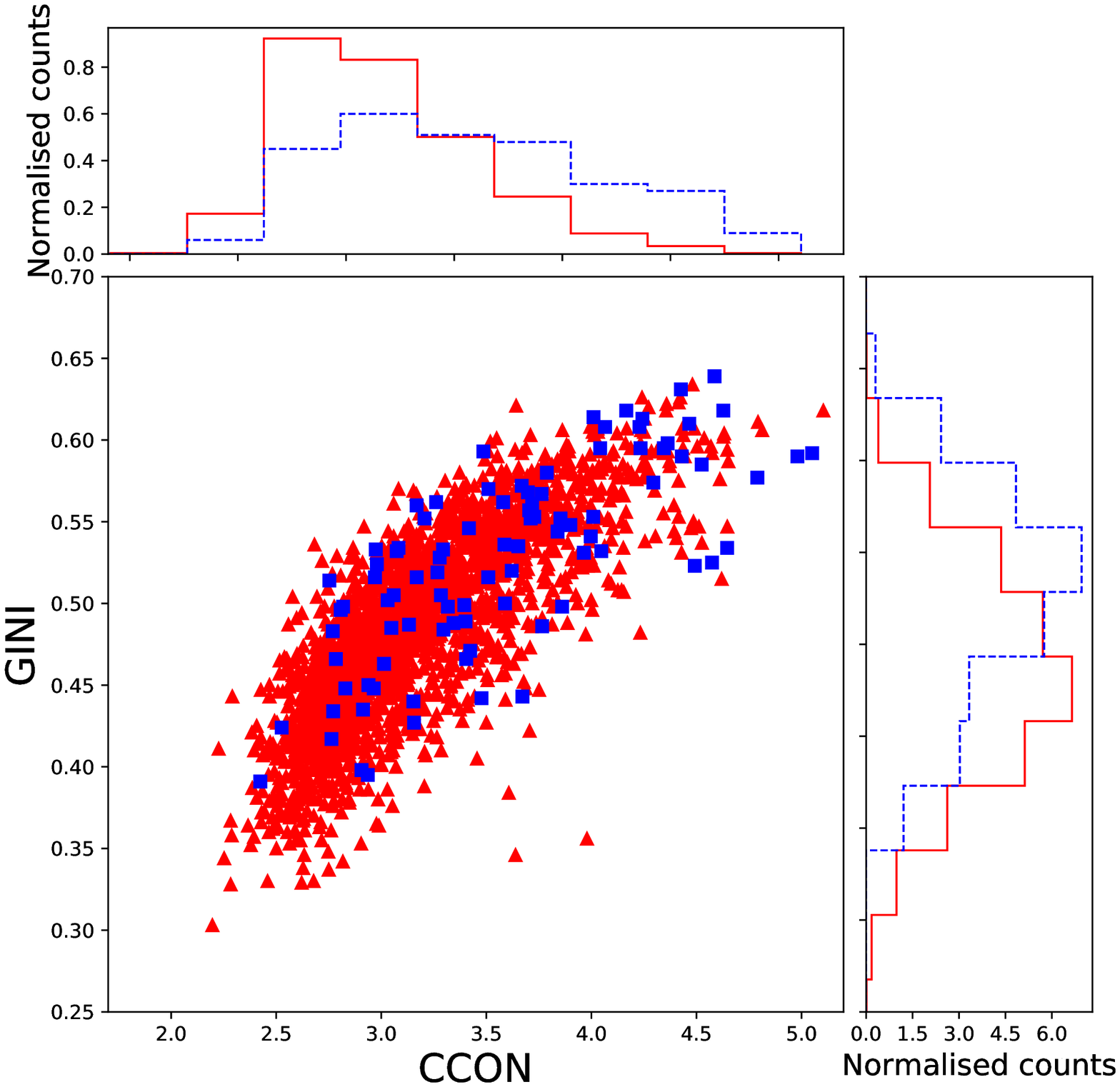}
\end{minipage}
\begin{minipage}[c]{.49\textwidth}
\includegraphics[width=0.85\textwidth,angle=0]{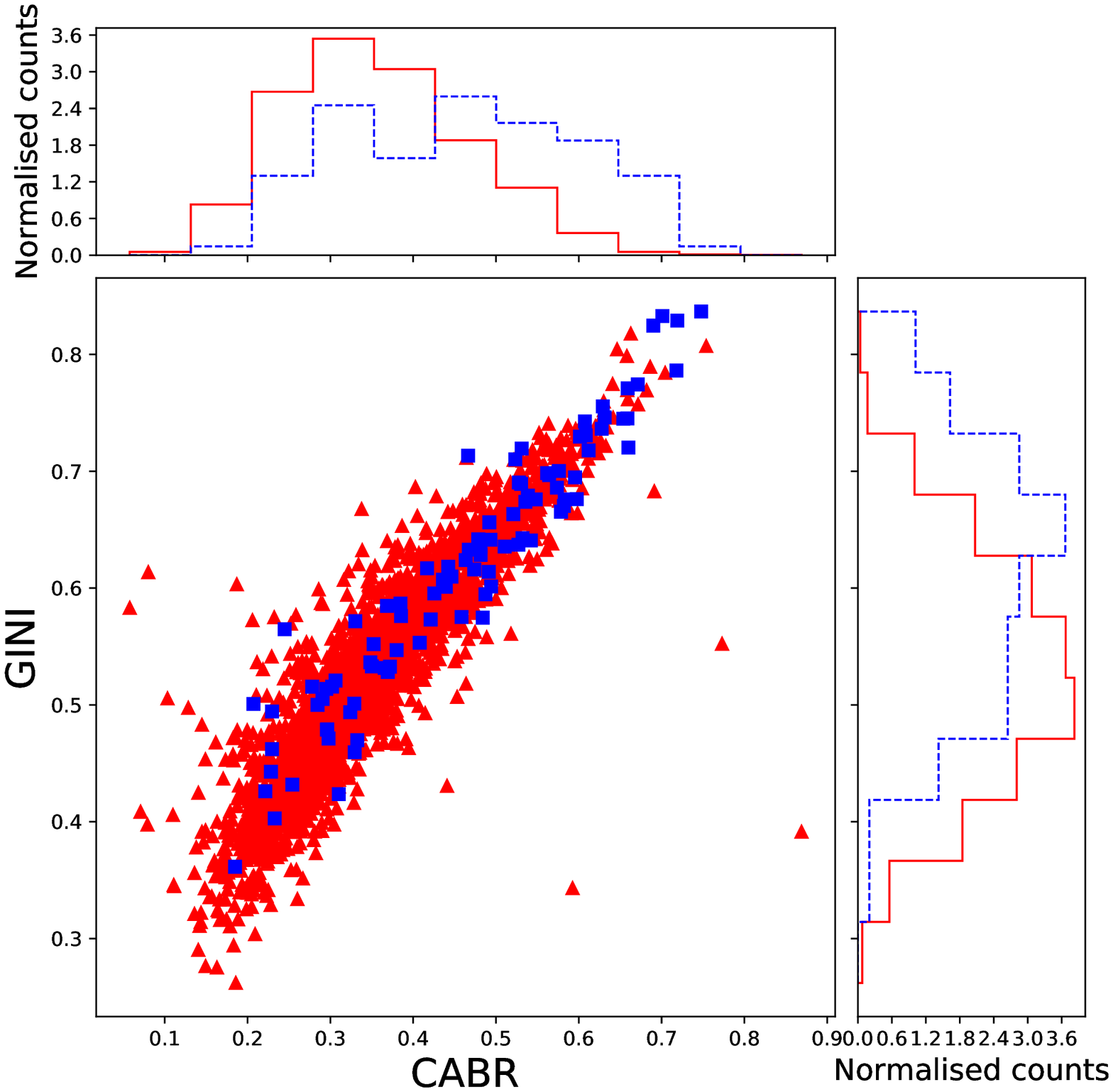}
\end{minipage}
\caption[ ]{Morphological diagrams representing the relation between Gini and CCON \textit{(left)} and Gini and CABR \textit{(right)} indices. In both plots, FIR AGN and non-AGN green valley galaxies are marked with blue squares and red triangles, respectively. \textit{Top and right histograms:} normalised distributions of corresponding parameters represented on the central plots of FIR AGN (blue dashed lines) and non-AGN (red solid lines) galaxies.} 
\label{fig_morph_diagrams_CONC}
\end{figure*}

In Figure \ref{fig_morph_diagrams_CONC} we show the location of our FIR AGN and non-AGN samples on Gini vs. CCON (left plot) and vs. CABR (right plot) diagrams, plotted using CAS07 and TAS09 measurements, respectively. It can be seen that both samples follow the same trend and show linear correlation between the three parameters, as shown previously for field \citep[e.g.,][]{Conselice2000, Abraham2003, Lotz2004, Cassata2007, Povic2013a} and cluster \citep[e.g.,][]{Cibinel2013, Pintos2016, BeyoroAmado2018} galaxies. Top and right-hand panels of each diagram represent normalised distributions of corresponding parameters for AGN (blue dashed lines) and non-AGN (red solid lines) galaxies. It can be seen that the two samples do not show the same distribution in any of the three parameters, with AGN showing slightly higher light concentrations. \\
\indent In case of CABR index, 50\% of AGN (non-AGN) samples have values between 0.32\,-\,0.56 (0.26\,-\,0.41), with median concentrations of 0.46 (0.33). In case of CCON index, similar results have been obtained, with 50\% of AGN (non-AGN) galaxies being located in the range 3.08\,-\,4.00 (2.82\,-\,3.39), with median concentrations of 3.50 (3.06). \\
\indent Gini coefficient measured in CAS07 and TAS09 gives similar results, although Gini measured in TAS09 has slightly higher values. In case of CAS07 measurements, 50\% of AGN (non-AGN) samples have values between 0.49\,-\,0.57 (0.44\,-\,0.52), with median Gini of 0.53 (0.49), while in TAS09 50\% of AGN (non-AGN) are found between 0.52 and 0.69 (0.45 and 0.53), with median Gini of 0.61 (0.52).\\
\indent We carried out a two-sided Kolmogorov-Smirnov (KS) test for all comparisons and found in all cases that the two samples do not belong to the same parent distribution (probability parameter being $\ll$\,0).  
 
\subsubsection{ASYM vs. concentration indices}

\begin{figure*}
\centering
\begin{minipage}[c]{.49\textwidth}
\includegraphics[width=0.85\textwidth,angle=0]{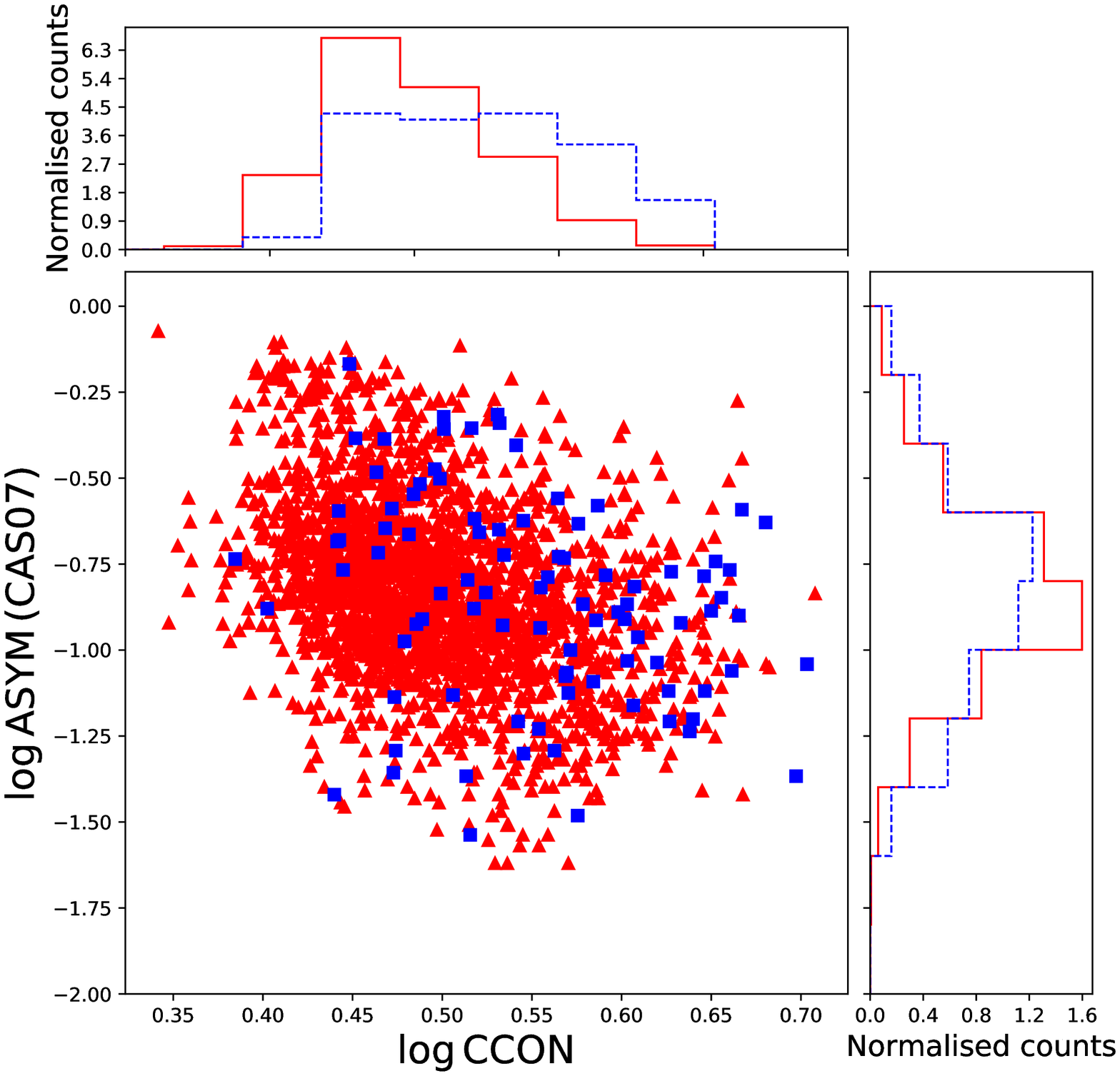}
\end{minipage}
\begin{minipage}[c]{.49\textwidth}
\includegraphics[width=0.85\textwidth,angle=0]{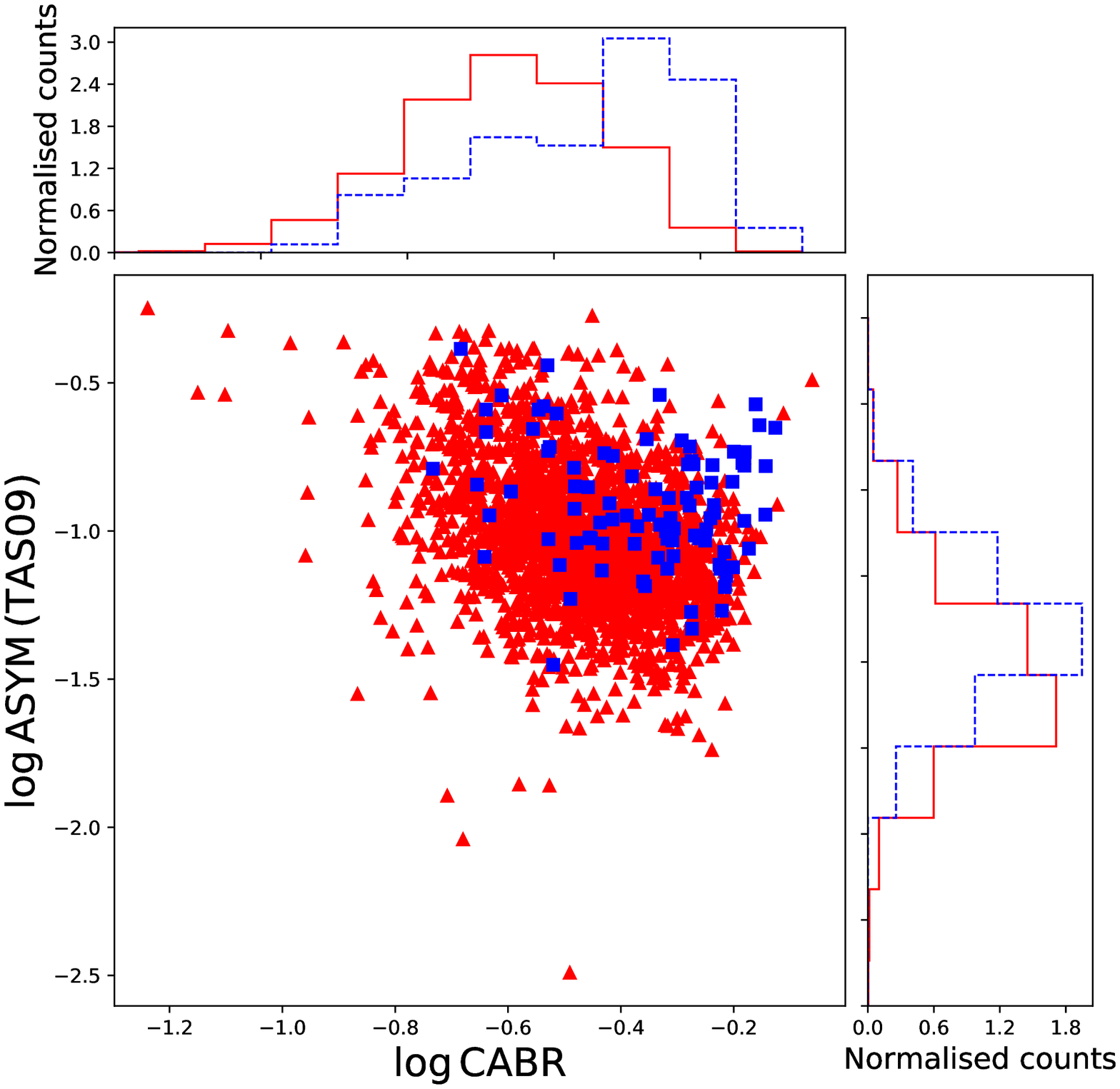}
\end{minipage}
\caption[ ]{Morphological diagrams representing the relation between ASYM and CCON \textit{(left)} and ASYM and CABR \textit{(right)} indexe. Top and right histograms show the normalised distributions of corresponding parameters represented on the central plots. For symbols and lines description see Figure~\ref{fig_morph_diagrams_CONC}.}
\label{fig_morph_diagrams_ASYM_conc}
\end{figure*}

In this section we compare ASYM index with concentration parameters such as CCON and CABR, measured in CAS07 and TAS09, respectively, as shown in Figure~\ref{fig_morph_diagrams_ASYM_conc} (central plots). We find an anticorrelation in both cases, with higher concentrated galaxies having lower asymmetries, as expected \citep{Abraham1996, Abraham2003, Conselice2000, Lotz2004}. When comparing the ASYM normalised distributions (right histograms) between AGN and non-AGN samples, very small or no differences have been found. In case of TAS09, AGN show slightly larger asymmetries, with 50\% of sample being in the range 0.09\,-\,0.16 (0.06\,-\,0.13), and having median concentrations of 0.12 (0.09). In this case a KS test rejected the hypothesis that the two samples are coming from the same parent distribution, showing the probability parameter to be $\ll$\,0. On the other hand, in CAS07 measurements there are no significant differences between the two samples. 50\% of AGN (non-AGN) have ASYM between 0.09\,-\,0.23 (0.1\,-\,0.22), with median concentrations of 0.15 (0.15), and KS test does not reject the notion that distributions are coming from the same sample (with probability parameter being 0.09). This parameter is very sensitive to noise, as shown in \cite{Povic2015}, and should not be used separately in morphological classification of galaxies. 

\subsubsection{M$_{20}$ vs. Gini and CCON}

\begin{figure*}
\centering
\begin{minipage}[c]{.49\textwidth}
\includegraphics[width=0.85\textwidth,angle=0]{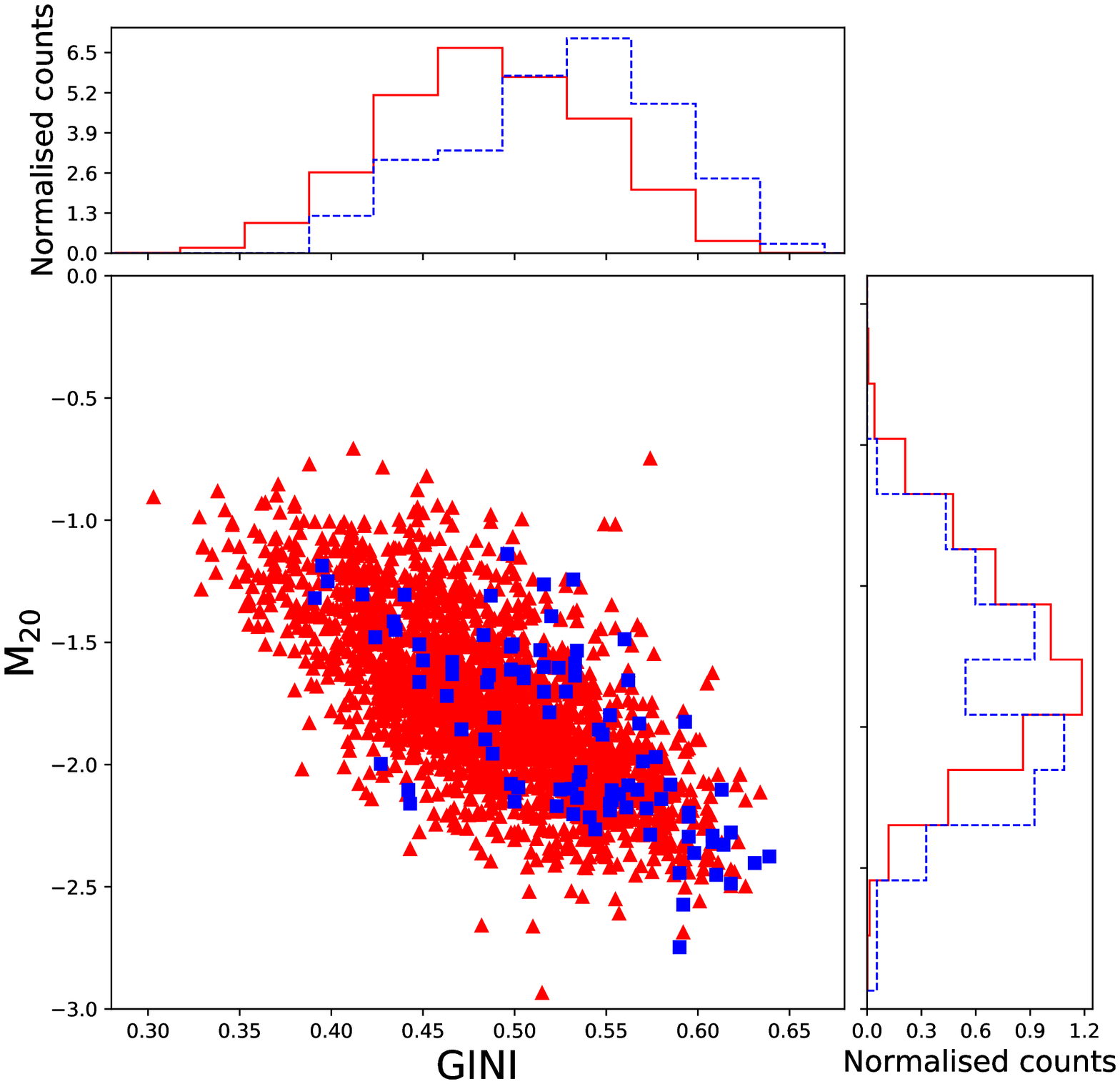}
\end{minipage}
\begin{minipage}[c]{.49\textwidth}
\includegraphics[width=0.85\textwidth,angle=0]{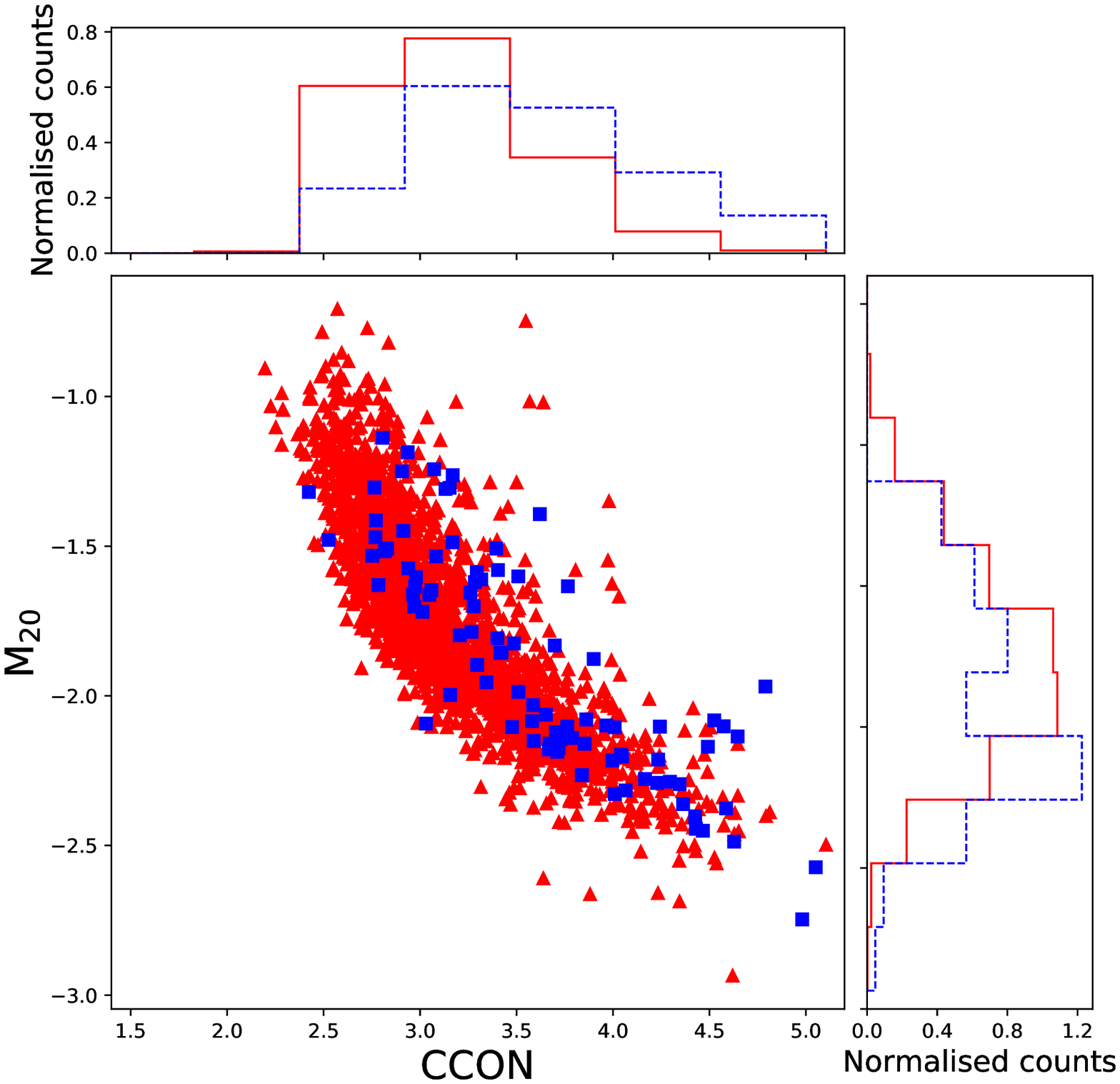}
\end{minipage}
\caption[ ]{Morphological diagrams representing the relation between M$_{20}$ moment of light and Gini \textit{(left)} and M$_{20}$ and CCON \textit{(right)}. Top and right histograms show the normalised distributions of corresponding parameters represented on the central plots. For symbols and lines description see Figure~\ref{fig_morph_diagrams_CONC}.} 
\label{Cassata_fig_morph_diagrams_20}
\end{figure*}

Figure \ref{Cassata_fig_morph_diagrams_20} shows the relationship between M$_ {20}$ moment of light and Gini (left central plot) and CCON (right central plot) indices, respectively, measured by CAS07. This parameter was used previously for detecting interactions and mergers \citep{Lotz2004, Lotz2010}. Once again we found that both samples follow the standard trends, showing the linear relation in the two cases, as shown previously for other galaxy samples \citep{Lotz2004, Lotz2010, Povic2013a, Povic2015, Pintos2016, BeyoroAmado2018}. AGN shows lower values of M$_ {20}$, with 50\% of sample being in a range of -2.16 and -1.58, and median M$_ {20}$\,=\,-1.89, while 50\% of non-AGN green valley galaxies have M$_ {20}$ in range -1.99 and -1.52, and median M$_ {20}$\,=\,-1.78. KS test also suggests that the two distributions are different (with probability factor being $\ll$\,0.)

\subsection{Morphological parameters for the same range of stellar mass}
\label{sec_same_mass_range}

\begin{figure*}
\centering
\begin{minipage}[c]{.49\textwidth}
\includegraphics[width=0.85\textwidth,angle=0]{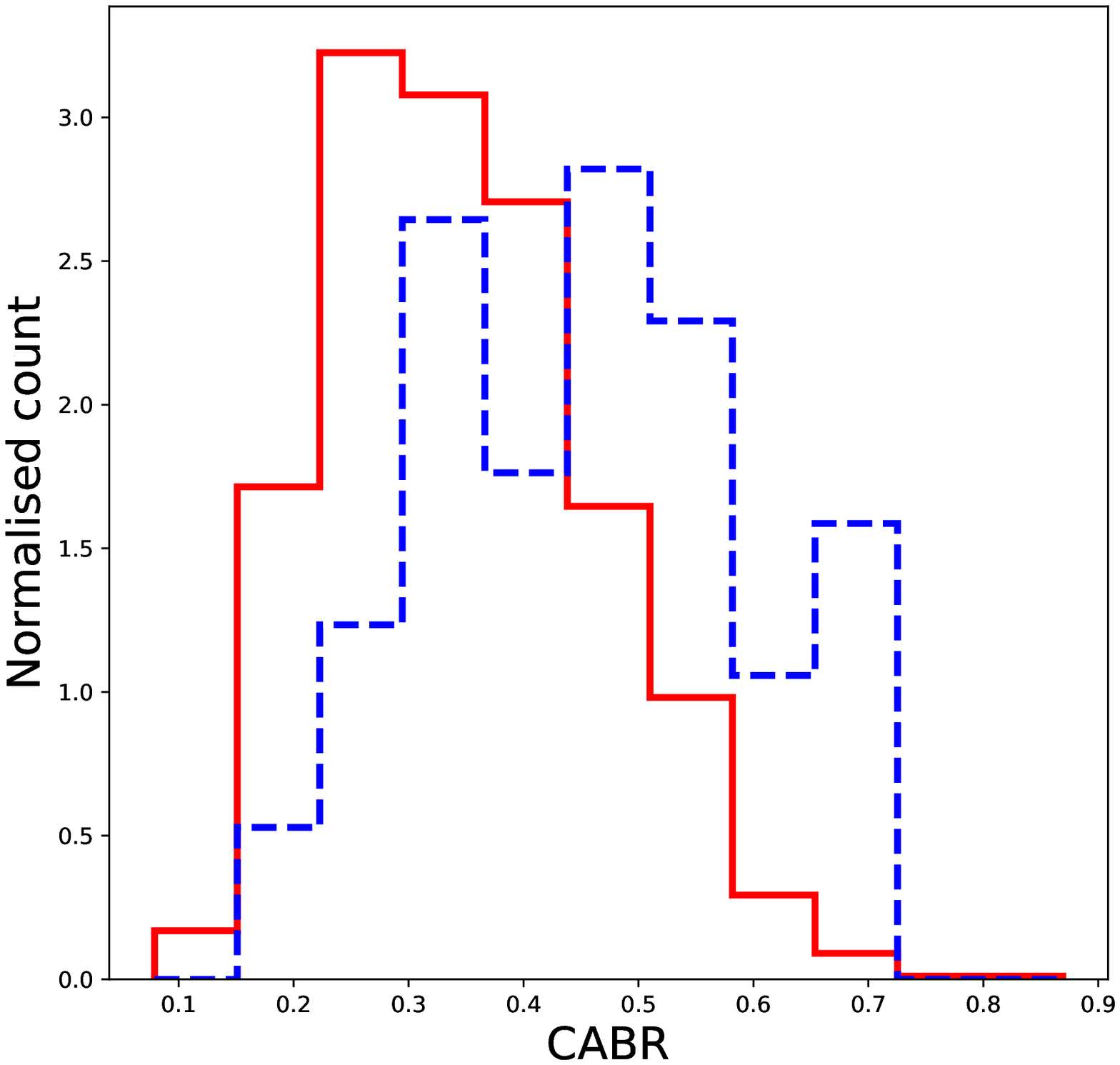}
\end{minipage}
\begin{minipage}[c]{.49\textwidth}
\includegraphics[width=0.85\textwidth,angle=0]{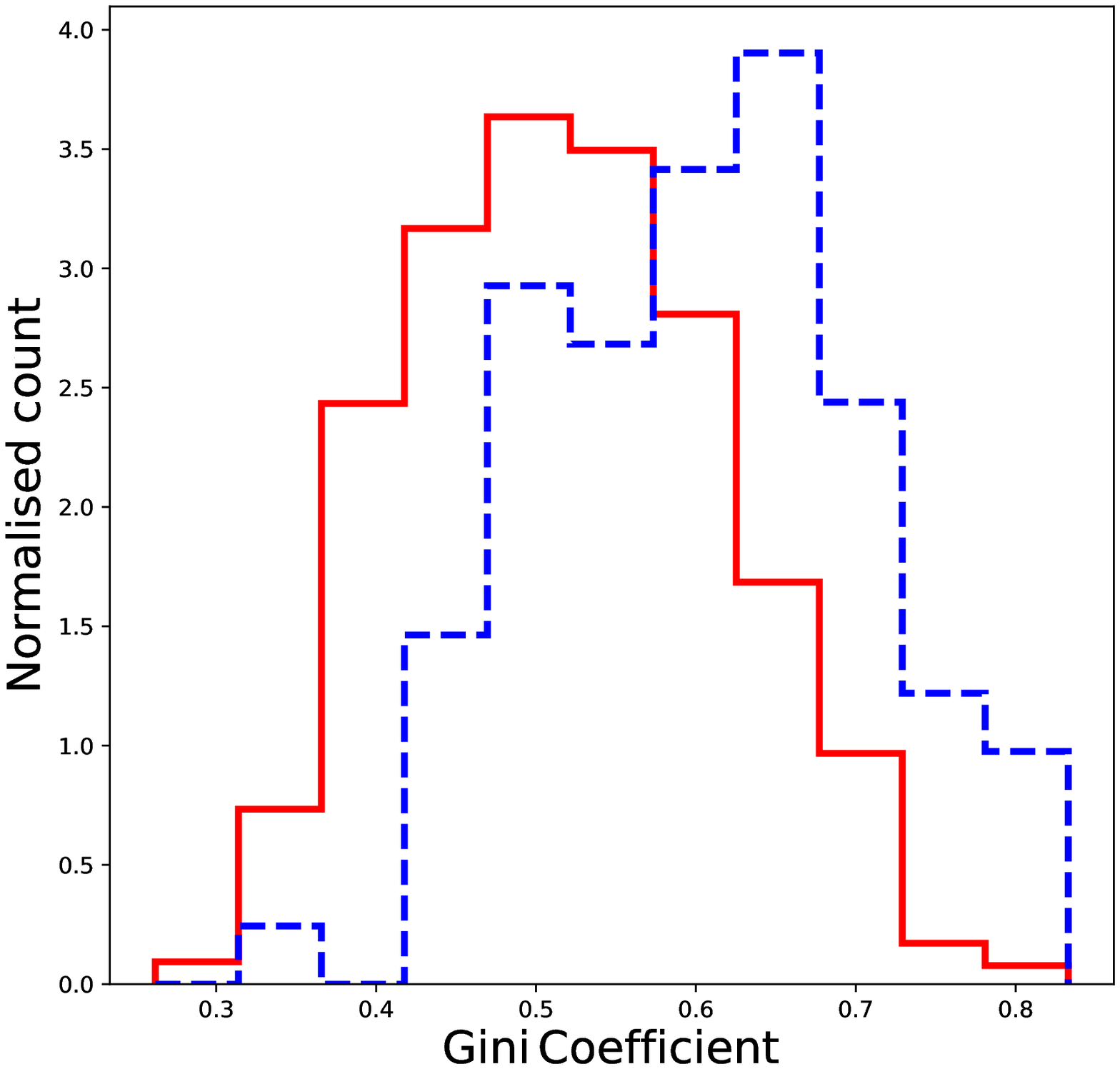}
\end{minipage}
\begin{minipage}[c]{.49\textwidth}
\includegraphics[width=0.85\textwidth,angle=0]{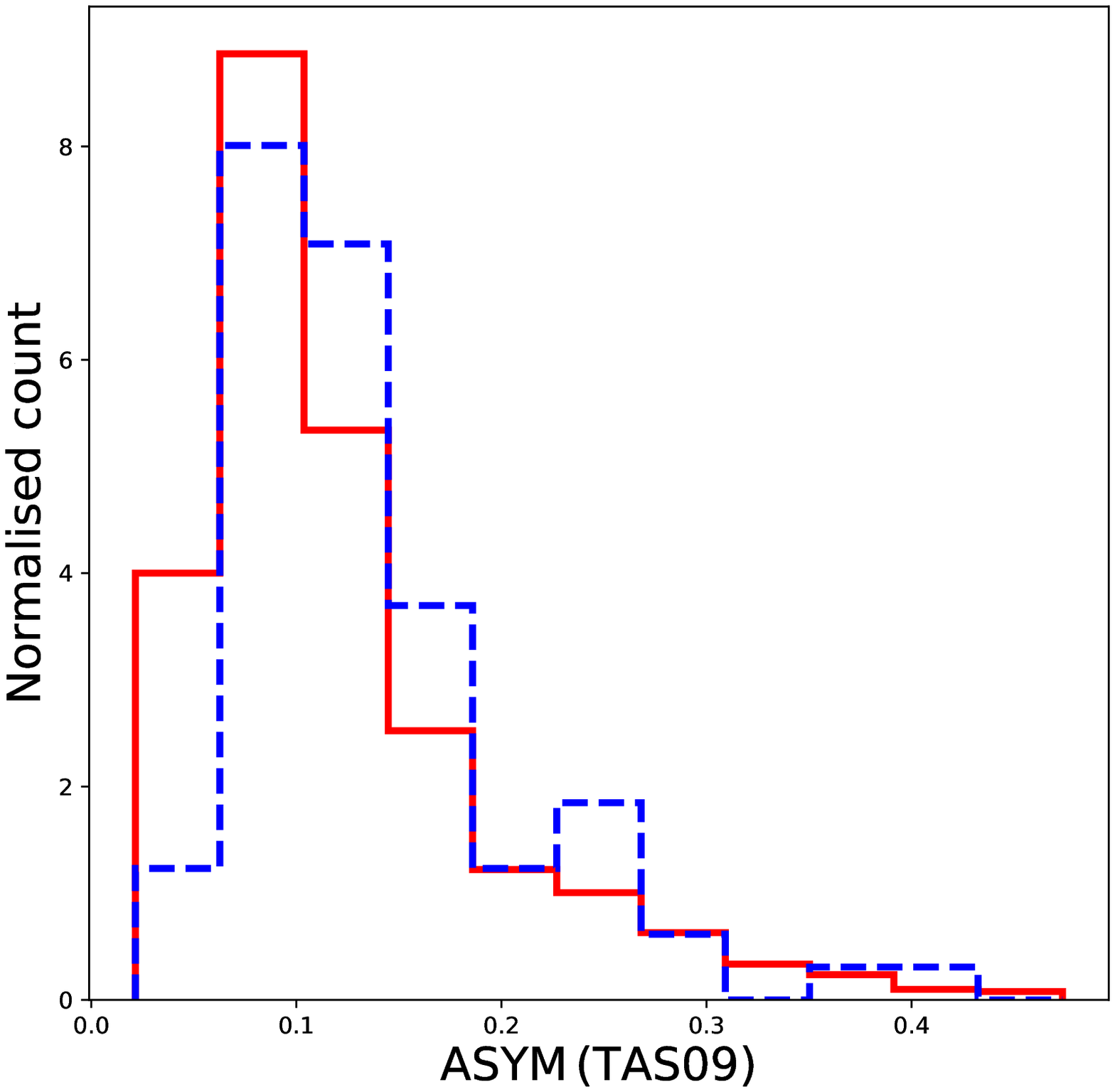}
\end{minipage}
\begin{minipage}[c]{.49\textwidth}
\includegraphics[width=0.85\textwidth,angle=0]{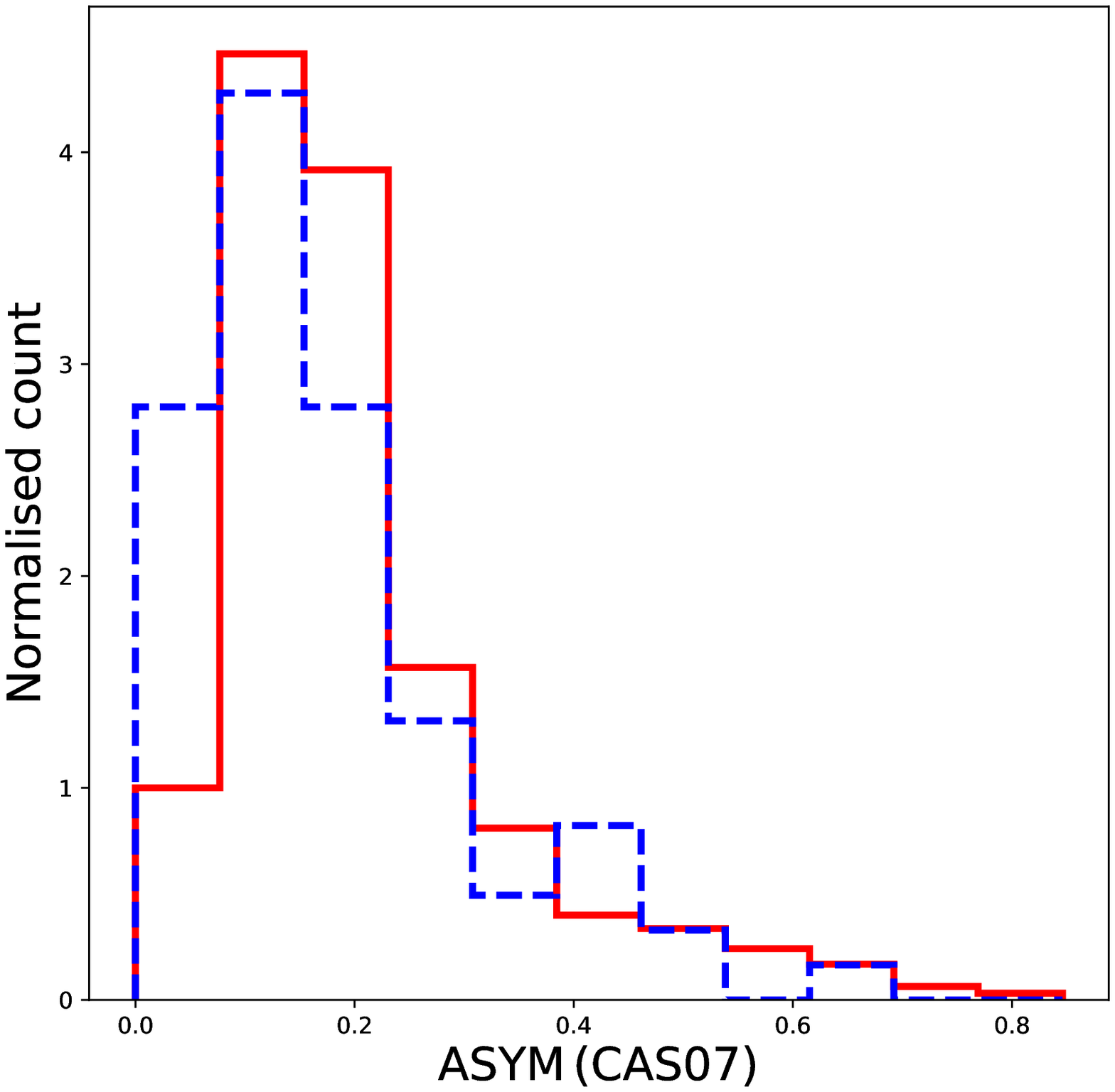}
\end{minipage}
\begin{minipage}[c]{.49\textwidth}
\includegraphics[width=0.85\textwidth,angle=0]{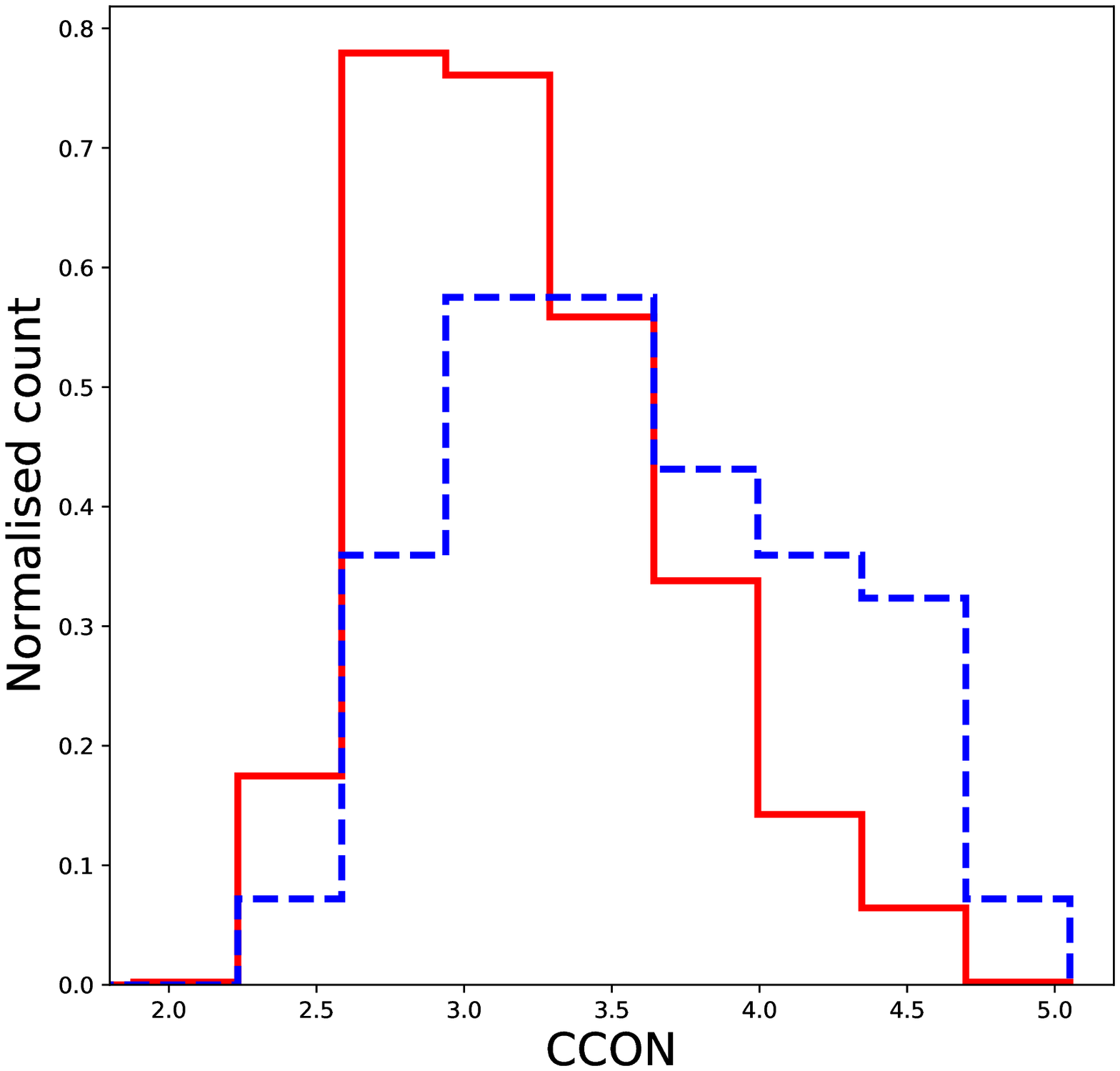}
\end{minipage}
\begin{minipage}[c]{.49\textwidth}
\includegraphics[width=0.85\textwidth,angle=0]{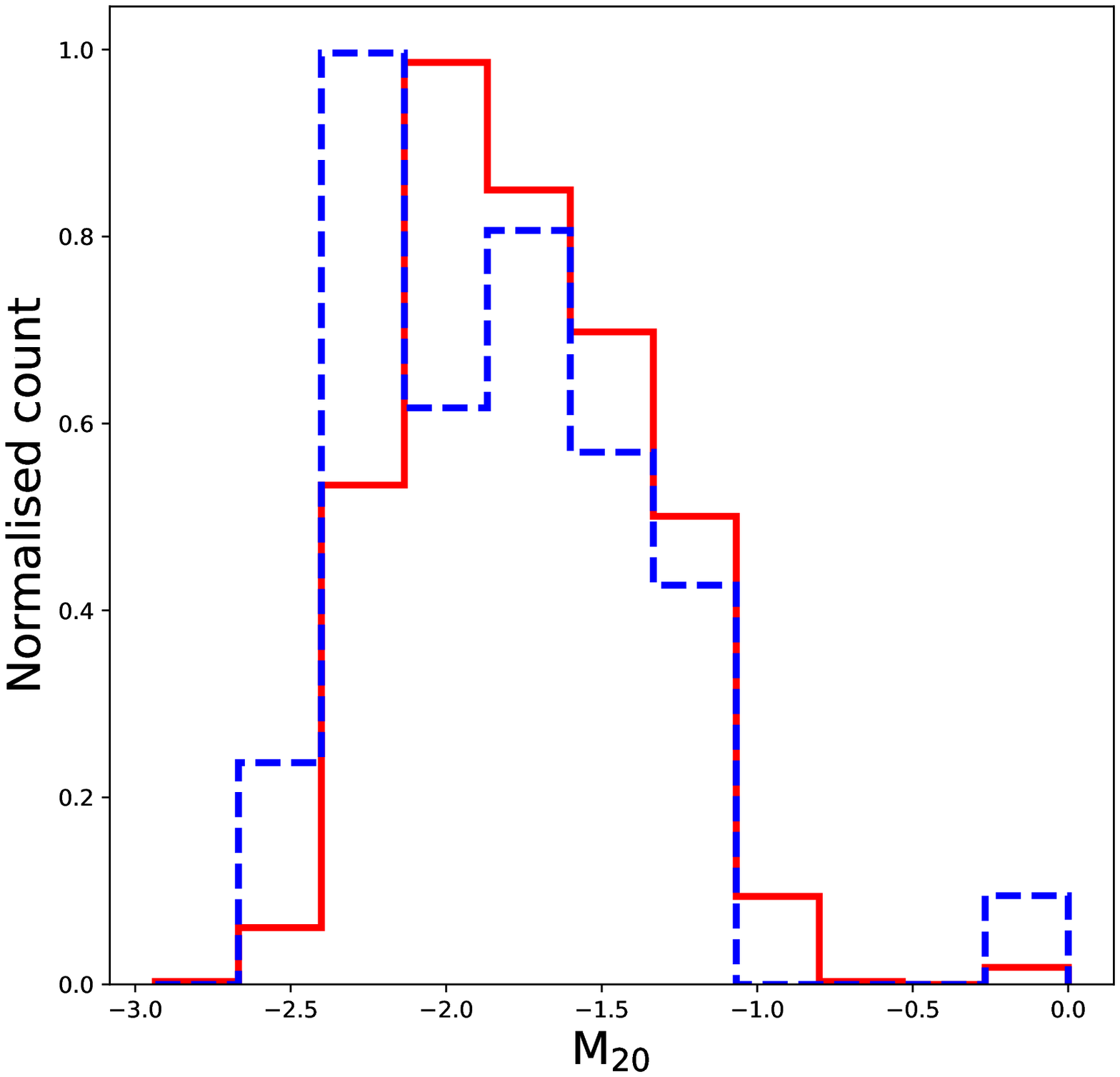}
\end{minipage}
\caption[ ]{Distributions of different morphological parameters of FIR AGN (blue-dashed lines) and non-AGN (red solid lines) within the same stellar mass range of logM*\,=\,10.6M$_\odot$\,-\,11.6M$_\odot$. From top to bottom and from left to right the following parameters are represented CABR, Gini, and ASYM measured in TAS09, and ASYM, CCON, and M$_ {20}$ measured in CAS07.} 
\label{Mass_A_C_G_fig_morph_diagrams}
\end{figure*}

Previous studies showed that AGN are hosted by more massive galaxies \citep[e.g.][]{Kauffmann2003, Leslie2016, Ellison2016, Nkundabakura2016} and, therefore, it is crucial to make the comparison of morphological parameters between AGN and non-AGN considering fixed stellar mass range. In this section, we analysed different parameters in the same mass range of $\rm{\log M_{*}\,=\,10.6M{_\odot}\,-\,11.6M{_\odot}}$, where we count 84\% and 55\% of our FIR AGN and non-AGN green valley galaxies, respectively \citep{Mahoro2017}.\\
\indent In Figure \ref{Mass_A_C_G_fig_morph_diagrams} we compare the distributions of five morphological parameters, CABR, Gini, ASYM (measured in both TAS09 and CAS07), CCON, and M$_ {20}$ moment of light, of those AGN and non-AGN within the selected stellar mass range. It can be seen again, as in previous section, that even when considering the same stellar mass range, indices such as CABR, GINI, and CCON (M$_ {20}$) have higher values (lower in case of M$_ {20}$), showing higher concentrations in case of active galaxies. In all of these cases KS test showed that AGN and non-AGN distributions are not coming from a parent distribution. The difference between concentration index measurements of AGN and non-AGN can also be seen in Table~\ref{tab_morph_param_stats_samemassrange}, where we measured basic statistics. In all cases it can be seen that both median values and the range that covers 50\% of sample show that active galaxies have higher light concentrations. In addition to this, we observed that these trends are maintained independently on morphological type, as can be seen again in Table~\ref{tab_morph_param_stats_samemassrange}, where we also provide comparisons of visual Class 1, 2, and 4 types. We discuss this more in Sec.~\ref{sec_discussion}. \\
\indent Regarding ASYM parameter similar findings were obtained as in previous section, with slight differences between TAS09 and CAS07 measurements and showing slightly lower asymmetries in case of active galaxies. However, these differences are within the uncertainties, and as mentioned previously this parameter is very sensitive to noise and is not a reliable indicator of galaxy morphology when considered independently \citep{Povic2015}.     

\begin{table*}
\begin{center}
\caption{Morphological parameters of AGN and non-AGN galaxies within the same stellar mass range}
\label{tab_morph_param_stats_samemassrange}
\begin{tabular}{|c|c|c|c|c|c|c|c|c|}
\hline
& & & \textbf{CABR} & \textbf{GINI} & \textbf{ASYM (TAS09)} & \textbf{ASYM (CAS07)} & \textbf{CCON} & \textbf{M20}\\
\hline
\textbf{AGN} & \textbf{all} & \textbf{median} & 0.46 & 0.61 & 0.12 & 0.15 & 3.49 & -1.96\\
\hline
& & \textbf{Q1 - Q3} & 0.32 - 0.54 & 0.52 - 0.68 & 0.09 - 0.16 & 0.09 - 0.23 & 3.07 - 4.02 & -2.19 - (-1.59)\\
\hline
 & \textbf{Class 1} & \textbf{median} & 0.55 & 0.68 & 0.10 & 0.08 & 3.70 & -2.13\\
\hline
& & \textbf{Q1 - Q3} & 0.48 - 0.59 & 0.64 - 0.70 & 0.09 - 0.13 & 0.06 - 0.12 & 3.46 - 4.21 & -2.29 - (-1.83)\\
\hline
 & \textbf{Class 2} & \textbf{median} & 0.38 & 0.54 & 0.09 & 0.14 & 3.34 & -1.99\\
\hline
& & \textbf{Q1 - Q3} & 0.29 - 0.49 & 0.50 - 0.63 & 0.07 - 0.16 & 0.10 - 0.19 & 2.93 - 3.96 & -2.17 - (-1.47)\\
\hline
 & \textbf{Class 4} & \textbf{median} & 0.38 & 0.57 & 0.14 & 0.20 & 3.31 & -1.63\\
\hline
& & \textbf{Q1 - Q3} & 0.32 - 0.54 & 0.51 - 0.66 & 0.10 - 0.17 & 0.13 - 0.30 & 2.97 - 3.86 & -2.07 - (-1.48)\\
\hline
\textbf{non-AGN} & \textbf{all} & \textbf{median} & 0.32 & 0.51 & 0.10 & 0.18 & 3.13 & -1.75\\
\hline
& & \textbf{Q1 - Q3} & 0.25 - 0.41 & 0.43 - 0.59 & 0.07 - 0.14 & 0.13 - 0.25 & 2.83 - 3.50 & -2.02 - (-1.45)\\
\hline
 & \textbf{Class 1} & \textbf{median} & 0.48 & 0.63 & 0.07 & 0.09 & 3.60 & -2.03\\
\hline
& & \textbf{Q1 - Q3} & 0.40 - 0.55 & 0.58 - 0.68 & 0.05 - 0.09 & 0.07 - 0.13 & 3.27 - 3.88 & -2.18 - (-1.90)\\
\hline
 & \textbf{Class 2} & \textbf{median} & 0.35 & 0.51 & 0.08 & 0.15 & 3.18 & -1.84\\
\hline
& & \textbf{Q1 - Q3} & 0.28 - 0.41 & 0.45 - 0.57 & 0.06 - 0.11 & 0.11 - 0.18 & 2.86 - 3.48 & -2.06 - (-1.56)\\
\hline
 & \textbf{Class 4} & \textbf{median} & 0.27 & 0.47 & 0.15 & 0.26 & 2.94 & -1.51\\
\hline
& & \textbf{Q1 - Q3} & 0.22 - 0.35 & 0.42 - 0.54 & 0.11 - 0.22 & 0.19 - 0.37 & 2.71 - 3.31 & -1.78 - (-1.28)\\
\hline
\end{tabular}
\end{center}
\end{table*}

\section{DISCUSSION}
\label{sec_discussion}

\subsection{FIR green valley galaxies on morphological diagrams}
\label{sec_discussion_morph_diag}

Figures~\ref{fig_morph_diagrams_CONC}, \ref{fig_morph_diagrams_ASYM_conc}, and \ref{Cassata_fig_morph_diagrams_20} show the standard behavior  of morphological parameters, where different concentration indices are correlated with each other and anticorrelated with asymmetry index, independently of the presence or absence of AGN. These trends are very well known from previous studies and have been observed at different redshifts and in different environments \citep[e.g.,][]{Abraham1996,Abraham2003,Conselice2000,Lotz2004,Lotz2010,Cassata2007,Tasca2009,Povic2013a, Povic2015, Pintos2016}. In this work we confirm that known relationships are maintained also in case of FIR green valley galaxies, both active and inactive. In addition, in Sec.~\ref{sec_distribution_Morphology_Parameters} when analysing the whole samples of AGN and non-AGN, it was observed in both TAS09 and CAS07 classifications that active galaxies show higher light concentrations. Since AGN are known to be hosted by more massive galaxies \citep[e.g.,][]{Kauffmann2003, Leslie2016, Ellison2016, Nkundabakura2016}, in Sec.~\ref{sec_same_mass_range} we made comparisons of morphological parameters taking into account the same mass range of AGN and non-AGN galaxies, and we found the same trends independently on morphology. We are currently analysing in details how different AGN contribution can affect various morphological parameters such as CABR, CCON, Gini, M$_ {20}$, ASYM, and smoothness at both low and high reshifts, taking into account both survey depth and resolution (Getachew et al., in preparation).     

\subsection{Morphologies of FIR AGN and non-AGN and main-sequence of star formation}
\label{sec_discussion_ms}

As shown in Sec.~\ref{sec_visual_class}, we found a high percentage of 38\% of FIR AGN green valley galaxies to have disturbed morphologies, with peculiar structures and clear signs of interactions or mergers. This could explain higher values of SFR in these AGN as observed in comparison with non-AGN and discussed in \cite{Mahoro2017}. Testing this was one of our initial goals, as mentioned in Sec.\ref{sec_intro}. However, the majority of FIR AGN hosts still show non-disturbed morphologies, including 25\% of them being E/S0. This suggests that in addition to interactions and mergers, other phenomena are also responsible for higher SFRs in these galaxies. On the other hand, the majority of non-AGN (46\%) were classified as spirals. 

\begin{figure*}
\centering
\begin{minipage}[c]{0.49\textwidth}
\includegraphics[width=7.5cm,angle=0]{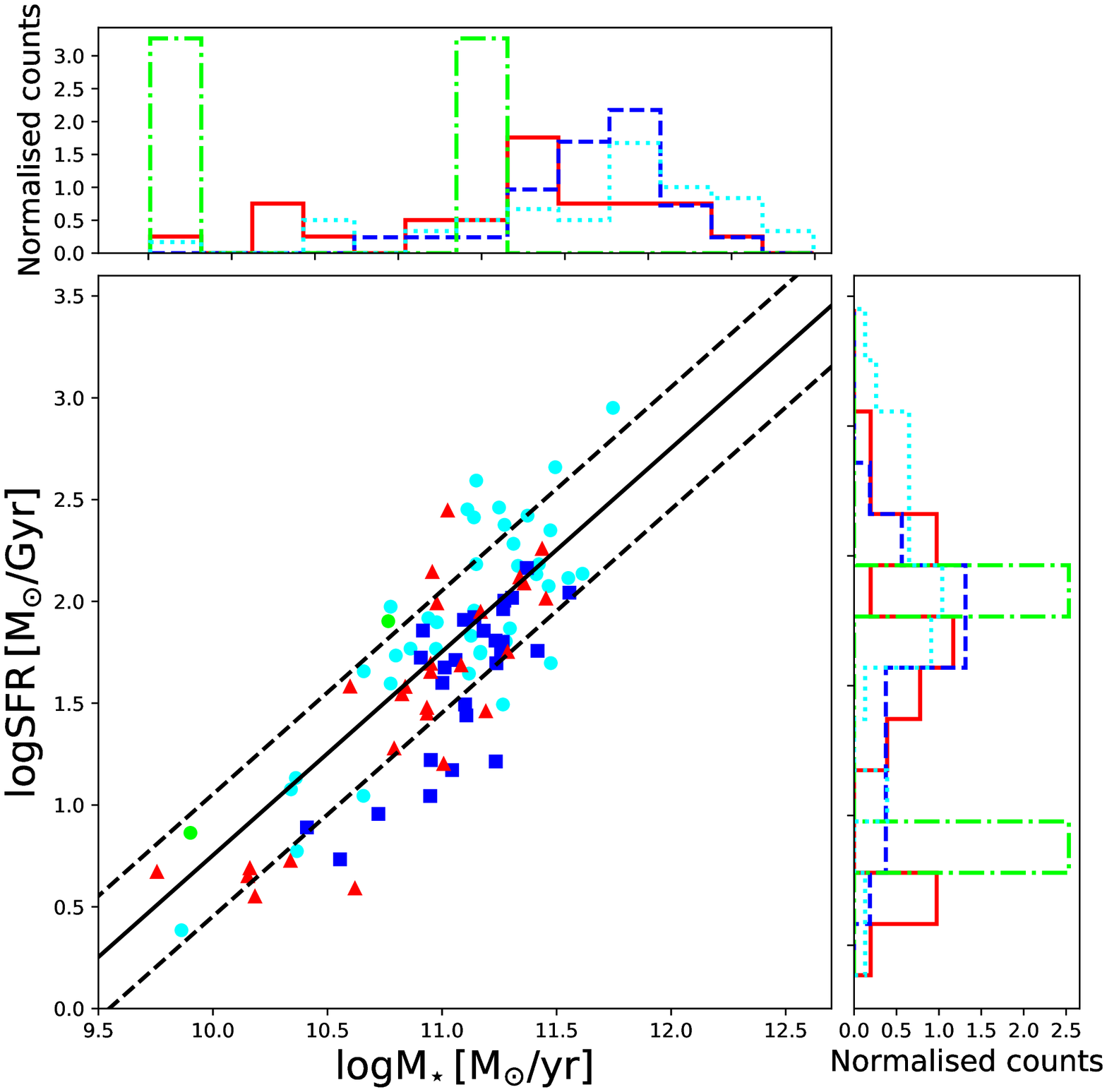}
\end{minipage}
\begin{minipage}[c]{0.49\textwidth}
\includegraphics[width=7.5cm,angle=0]{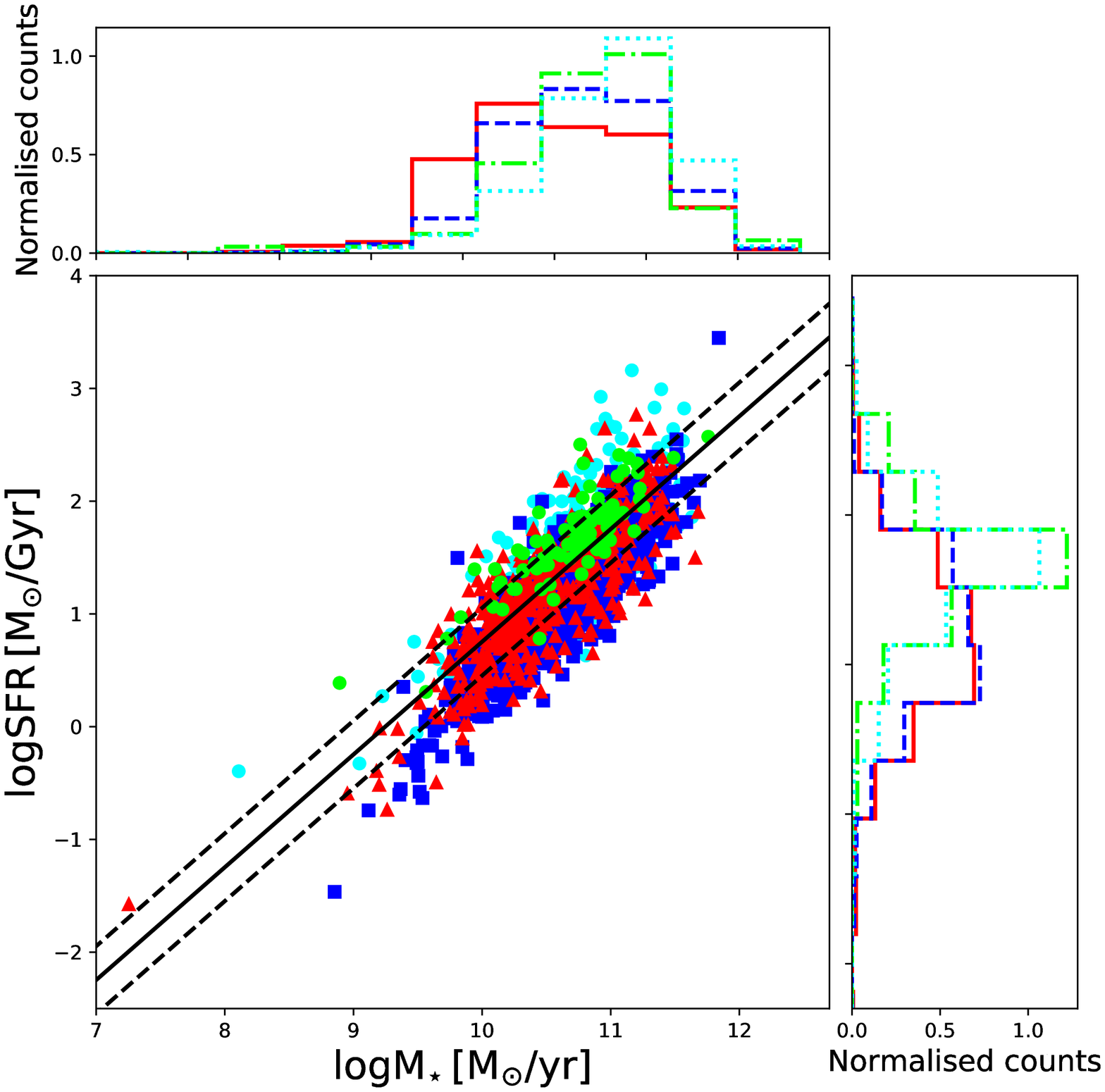}
\end{minipage}
\caption[]{Relation between the SFR and stellar mass for FIR AGN (left plot) and non-AGN (right plot) green valley galaxies in relation with visual morphological classification. Class 1, 2, 3, and 4, are represented with red triangles, dark blue squares, green filled circles, and light blue diamonds, respectively. The solid black line shows the Elbaz et al. (2011) fit for the MS of star-forming galaxies observed with Herschel, while the dashed lines represent the typical MS width of $\pm$\,0.3\,dex. Top and right histograms
show the normalized distributions of stellar mass and SFR, respectively, and
comparison between the class 1 (red solid lines), 2 (dark blue dashed lines), 3 (green dash-dot-dash lines), and 4 (light blue dotted lines).}
\label{MS_morph}
\end{figure*}

In Fig.~\ref{MS_morph} we represented again the relation between the SFR and stellar mass, as in \cite{Mahoro2017} (see their Fig. 5 and 6), to test the location of all FIR AGN and non-AGN with respect to the main sequence (MS) of star forming galaxies, but now taking into account our visual classification. For MS (solid line) we used again the results of \cite{Elbaz2011}, obtained through the FIR Herschel data, as we did in \cite{Mahoro2017}, while for the width of MS (dashed lines) we used $\pm$ 0.3 dex, found in many previous works to be the typical 1$\sigma$ value \citep[e.g,][]{Elbaz2007, Whitaker2012, Whitaker2014, Shimizu2015}. Table~\ref{tab_MS_stats} summarises the number of galaxies on, above, and below the MS of star formation in relation to morphology. As can be seen, for Class 1, 2, and 4 we find again to be located mainly on the MS of star formation in both AGN and non-AGN, independently on morphology. Class 3 is not inconsistent with this result, though it has poor statistics, especially in case of AGN. In general, similar trends are found in both AGN and non-AGN, where while being on the MS, Class 1 and 2 are mainly located slightly below, and Class 3 slightly above the MS of star formation. These results broaden our knowledge about the large range of properties that green valley galaxies can have if multiwavelength data are taken into account, in comparison to what is known from previous optical studies. 

As mentioned in Sec.~\ref{sec_intro}, previous studies mainly suggested that green valley galaxies are an intermediate population, having properties between those of the red sequence and the blue cloud galaxies \citep[e.g.,][]{Salim2007, Pan2013, Schawinski2014, Salim2014, Lee2015, Smethurst2015, Trayford2016, Ge2018, Coenda2018}, in terms of both their morphologies, stellar masses, luminosities, star formation rates, and star formation histories, etc., and are expected to be located mainly below the MS of star formation \citep[e.g.,][]{Noeske2007, Salim2007, RenziniPeng2015}. Therefore, FIR green valley galaxies selected in this work present an interesting sample for further multiwavelength studies and better understanding of the full and complex picture of galaxy transformation. 

\begin{table*}
\begin{center}
\caption{Number of AGN and non-AGN galaxies per morphological class being located on, above, and below the MS of star formation}
\label{tab_MS_stats}
\begin{tabular}{|c|c|c|c|c|c|c|c|c|}
\hline
&\multicolumn{4}{|c|}{AGN}&\multicolumn{4}{|c|}{non-AGN}\\
\hline
&Class 1&Class 2&Class 3&Class 4&Class 1&Class 2&Class 3&Class 4\\
\hline
Tot. gal.&26&27&2&39&452&1204&87&494\\
\hline
above MS&2 (8\%)&/&1 (50\%)&8 (21\%)&43 (9.5\%)&21 (2\%)&29 (33\%)&86 (17\%)\\
\hline
on MS&19 (73\%)&18 (67\%)&1 (50\%)&27 (69\%)&307 (68\%)&788 (65\%)&57 (66\%)&364 (74\%)\\
\hline
below MS&5 (19\%)&9 (33\%)&/&4 (10\%)&102 (22.5\%)&395 (33\%)&1 (1\%)&44 (9\%)\\
\hline
Total&\multicolumn{4}{|c|}{94}&\multicolumn{4}{|c|}{2240}\\
\hline
\end{tabular}
\end{center}
\end{table*} 

\subsection{Morphology vs. SFR}
\label{sec_discussion_SFRs}

\begin{figure*}
\centering
\begin{minipage}[c]{0.49\textwidth}  
 \includegraphics[width=7.5cm,angle=0]{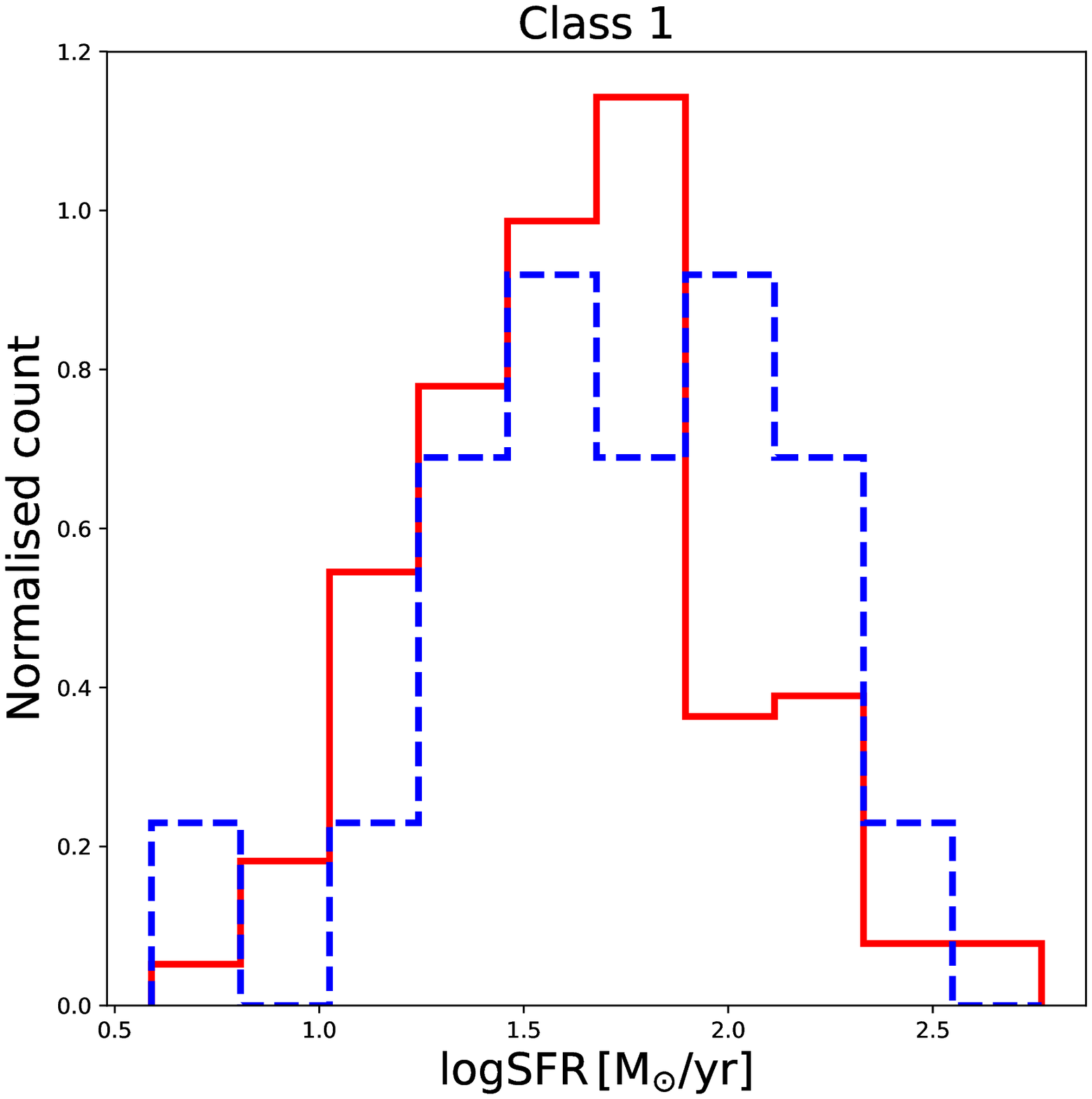}
\end{minipage}
\begin{minipage}[c]{0.49\textwidth}
\includegraphics[width=7.5cm,angle=0]{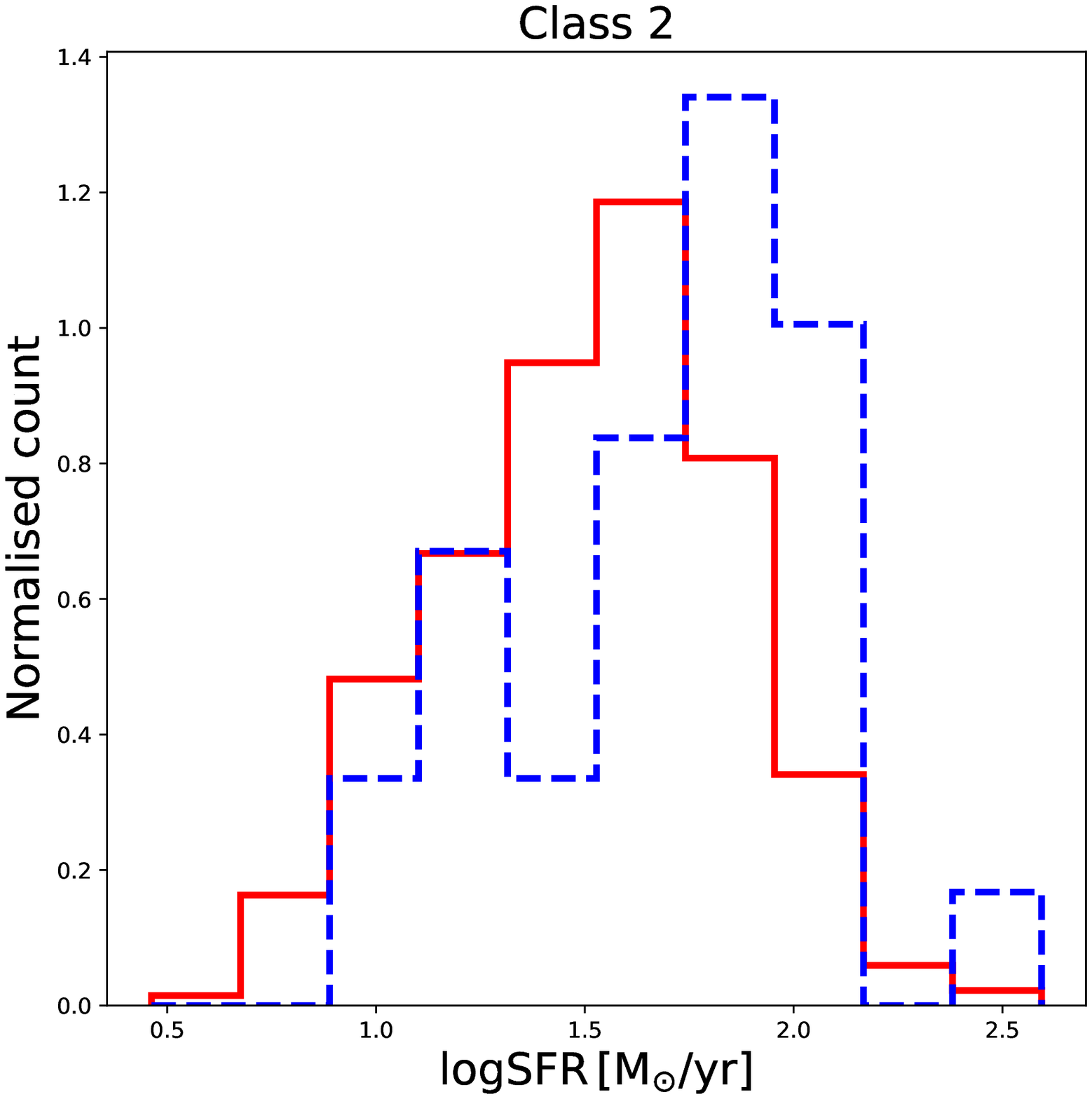}
\end{minipage}
\begin{minipage}[c]{0.49\textwidth}
\includegraphics[width=7.5cm,angle=0]{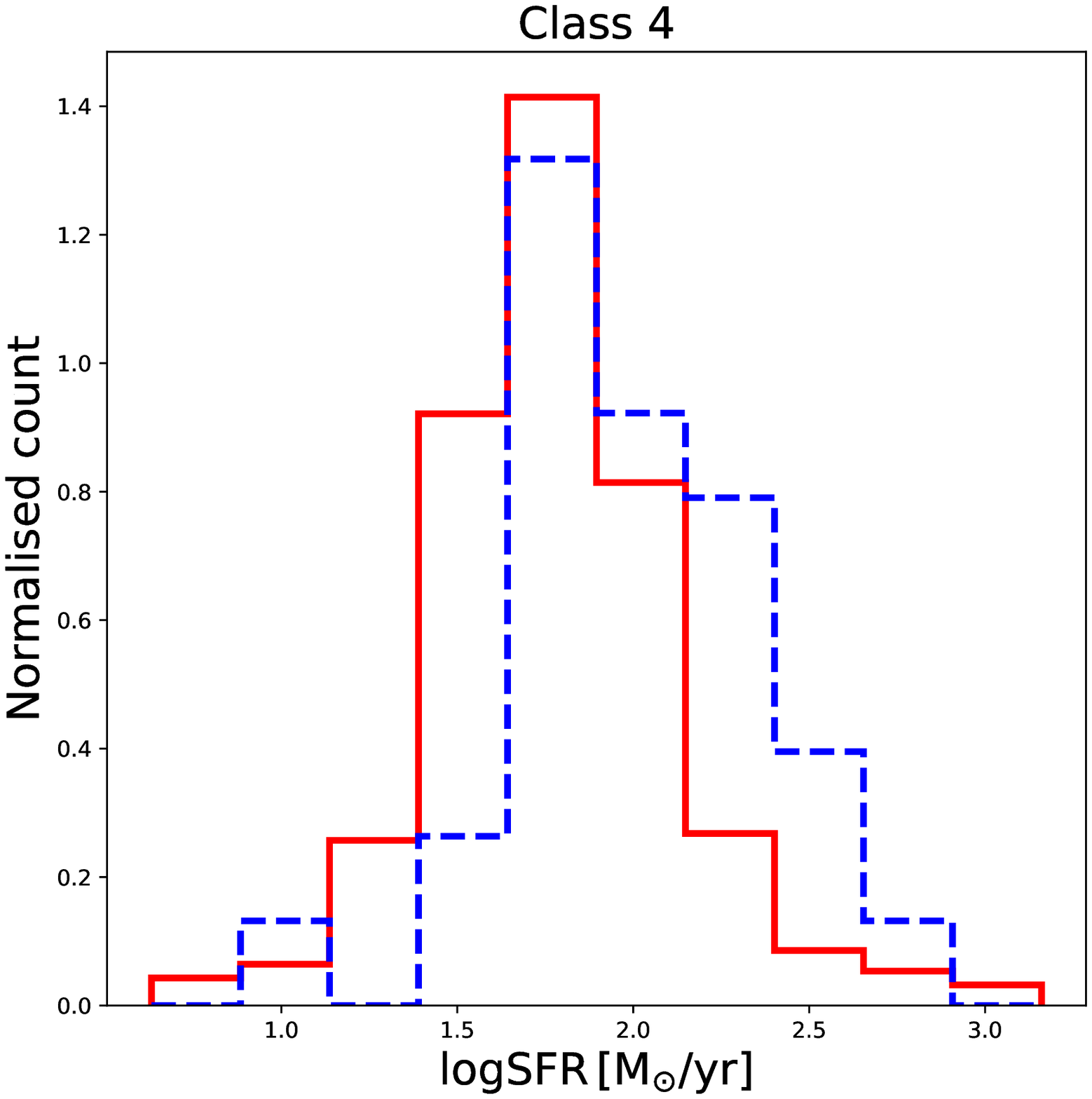}
\end{minipage}
\caption{Normalised distributions of SFR of FIR AGN (blue dashed histograms) and non-AGN (red solid histograms) for a fixed stellar mass range of logM*\,=\,10.6M$_\odot$\,-\,11.6M$_\odot$ in relation to morphology.}
\label{fig_SFR_SameMAssRange}
\end{figure*}

The finding that the majority of X-ray detected FIR green valley AGN are located on the MS of galaxy formation, independently of their morphological type, contradicts previous suggestions that AGN might be responsible for star formation quenching in galaxies  \citep[e.g.][]{Sanchez2004, Nandra2007, Georgakakis2008, Silverman2008, Treister2009, Povic2012, Povic2013b, Leslie2016};  as discussed in detail in \cite{Mahoro2017}. Instead of finding lower star formation, we found higher values of SFRs in the AGN sample in comparison to non-AGN, when observing the same range of stellar mass of logM*\,=\,10.6M$_\odot$\,-\,11.6M$_\odot$, as shown in \cite{Mahoro2017} and described in Sec.~\ref{sec_intro}. For the indicated stellar mass range, we found median SFRs FIR AGN and non-AGN to be 71 and 45\,M$_\odot$/yr, respectively. In this work we compared SFRs of AGN and non-AGN for the same stellar mass range, but also taking into account their morphology. Figure~\ref{fig_SFR_SameMAssRange} shows these comparisons for Class 1 (top left plot), Class 2 (top right plot), and Class 4 (bottom plot), while in Table~\ref{tab_SFRs_SameMassrange} we provide basic statistical comparisons. For Class 3 (irregular galaxies) we do not provide any comparisons due to the small number of sources in the AGN sample (see Table~\ref{tab_Visual_number}). It can be seen from both, Figure~\ref{fig_SFR_SameMAssRange} and Table~\ref{tab_SFRs_SameMassrange}, that in all cases FIR AGN show significantly higher SFRs in comparison to FIR non-AGN within the same stellar mass range, independently on their morphology. As shown in Table~\ref{tab_SFRs_SameMassrange}, 50\% of Class 1, 2, and 4 AGN (non-AGN) have SFRs in the range of 31\,-\,100 (23\,-\,72), 30\,-\,85 (2\,-\,56), and 55\,-\,151 (38\,-\,91), respectively. We run KS test on all plots of Figure~\ref{fig_SFR_SameMAssRange}, finding in all cases that the two distributions do not come from the same one (having the probability factor $<$\,0).    

\begin{table*}
\centering
\caption{SFRs in the same range of mass}
\begin{tabular}{|c|c|c|c|c|c|c|c|c|c|}
\hline
\multirow{2}{*}{} & \multicolumn{3}{c|}{Class 1} & \multicolumn{3}{c|}{Class 2} & \multicolumn{3}{c|}{Class 4} \\ \cline{2-10} 
                  & Q1       & Median      & Q3      & Q1       & Median      & Q3      & Q1      & Median     & Q3      \\ \hline
AGN               & 31    & 48       & 100  & 30    &56     & 85   & 55   & 85     & 151        \\ \hline
Non-AGN           & 23    & 44       & 72   & 2     & 35   & 56   & 38   & 58    & 91         \\ \hline
\end{tabular}
\label{tab_SFRs_SameMassrange}
\end{table*}

The finding in \cite{Mahoro2017} that 82\% of AGN are located either on or above the MS of star formation, and that they show higher SFRs than non-AGN, suggested that for X-ray detected AGN with FIR emission if there is an influence on its star formation in the green valley then positive AGN feedback seems to take place, rather than the negative one. Finding in this work a larger fraction of disturbed morphologies in 38\% of AGN suggests that interactions and mergers play an important role in these active galaxies and could contribute to their higher SFRs. Moreover, it could also be possible that in these galaxies interactions and mergers are responsible for triggering independently both enhanced star formation and AGN activity, as has been shown in previous studies \citep[e.g.,][]{Hopkins08, Lamastra2013, Brassington2015, Hong2015, Knapen2015, Dietrich2018}. However, taking into account high SFRs in the rest of AGN sample (e.g., in 26\% and 25\% of galaxies classified as class 1 and class 2, respectively), and higher obtained SFR values in comparison to non-AGN independently on morphology (as shown above), suggests that interactions and mergers alone cannot explain the results. Therefore, as suggested in \cite{Mahoro2017}, active galaxies analysed in this work (at least those not classified as class 4) could present an interesting observational sample for further studies about AGN positive feedback and their effect on star formation.\\
\indent In line with this, we are performing further analysis using public optical spectroscopic data, accompanied with new spectroscopy from the 11-m Southern African Large Telescope (SALT) data of a smaller AGN sample, in order to understand better the emission line properties of our sample, extinction characteristics, and investigate possible signs of outflows (work in progress). However, beside the positive AGN feedback and influence of mergers and interactions, other factors could also play a role in affecting SFRs of our green valley galaxies in general, and AGN sample in particular. Some of the factors raised in previous works could be related with secular evolution \citep{KormendyKennicutt2004, Bremer2018, Ge2018}, as well as environmental effects \citep[e.g.,][]{Trayford2016, Crossett2017, Coenda2018, Gu2018}. In addition, as mentioned in sec.~\ref{sec_sample_selection}, our galaxies were selected as green valley via their optical colours. However, they are also FIR detections, lie on the main sequence of star formation and have high SFRs, as shown in Fig.~\ref{MS_morph}. It could also be that a certain fraction of our sources are not typical green valley sources and have high amount of dust-obscured star formation not visible in optical due to dust extinction. This possibility has been already observed in many previous studies, especially when comparing the SFRs measured in optical/UV and FIR \citep[e.g.,][]{Wuyts2011, Catalnorrecilla2015, Povic2016, Sklias2017, Mahajan2019}. Therefore, taking into account all mentioned, further studies are needed for understanding the full physics behind the FIR green valley AGN and non-AGN galaxies and their role in galaxy formation and evolution. 

\section{CONCLUSIONS}

This work is a follow-up of the work presented in \cite{Mahoro2017}, where using FIR data, and after removing an AGN contribution, we found for the same mass range higher SFRs in the green valley X-ray detected AGN than in non-AGN, in contrast with previous optical studies. Therefore, in this paper we carried out a morphological study of the same sample of green valley AGN and non-AGN from the COSMOS survey, in order to better understand previously obtained results. We used the public HST/ACS images and classified visually the whole sample of 103 AGN and 2609 non-AGN. In addition, we used all available public catalogues with various morphological parameters measured. We compared in detail our visual and previous automatic classifications. We analysed the standard morphological diagrams of our sample of active and non-active galaxies, and compared their morphological parameters within the same stellar mass range. We analysed SFR vs. stellar mass diagram in relation to morphology, and finally, we compared the SFR distributions of AGN and non-AGN within the same stellar mass range for different morphological types. Our main conclusions are as follows: 
\begin{itemize}
\item We found that the fraction of peculiar galaxies with clear signs of interactions and mergers is significantly higher in AGN (38\%) than non-AGN (19\%) green valley galaxies. On the other hand, non-AGN galaxies from our sample are predominantly spirals (46\%), in comparison to AGN, where only 26\% were classified as spirals (or Class 2). 
\item We confirmed that the standard morphological diagrams follow the same trends when green valley active and non-active galaxies are considered. 
\item We found both FIR AGN and non-AGN green valley galaxies to be located mainly on the MS of star formation, independently on morphology, which contrasts previous results obtained in the optical suggesting that green valley galaxies are mainly located below the MS, and that in addition there is a clear difference in terms of morphology. 
\item Finally, we found that within the same stellar mass range, AGN have significantly higher SFRs in all analysed morphological types.
\end{itemize}
Our findings suggest that in a significant fraction of AGN ($\sim$\,40\%), interactions and mergers play an important role and could be responsible for higher SFRs observed in FIR in comparison to non-AGN sample, and/or for even triggered AGN activity. However, taking into account that in \cite{Mahoro2017} 82\% of AGN are located either on or above the MS of star formation, showing higher SFRs then non-AGN, indicates that in addition to mergers and interactions other factors are also responsible for enhanced star formation in active galaxies, including the AGN positive feedback as one of the possibilities. In follow-up studies we will focus more attention to analyse this aspect, including also extinction characteristics of our sample, using available spectroscopic data from the public archives and the SALT telescope.  

\section*{Acknowledgements}

We thank the anonymous referee for accepting to review this paper, and for giving us constructive and useful comments that improved this paper. This work is based on the research supported in part by the  National  Research Foundation of South Africa (Grant Numbers 110816). AM acknowledges financial support from the Swedish International Development Cooperation Agency (SIDA) through the International Science Programme (ISP) - Uppsala University to University of Rwanda through the Rwanda Astrophysics, Space and Climate Science Research Group (RASCSRG), East African Astrophysics Research Network (EAARN) are gratefully acknowledged. MP acknowledges financial supports from the Ethiopian Space Science and Technology Institute (ESSTI) under the Ethiopian Ministry of Innovation and Technology (MInT), from the Spanish Ministry of Economy and Competitiveness (MINECO) through projects AYA2013-42227-P and AYA2016-76682C3-1-P, and from the State Agency for Research of the Spanish MCIU through the "Center of Excellence Severo Ochoa" award for the Instituto de Astrof\'isica de Andaluc\'ia (SEV-2017-0709). PV acknowledges support from the  National  Research Foundation of South Africa.  We are grateful to the COSMOS survey and all scientists whose data have been used in this research for making them publicly available and free. We are also grateful to the python, Virtual Observatory, and TOPCAT teams for making their packages freely available to the scientific community.



\bibliographystyle{mnras}
\bibliography{Ref_Morph} 

\begin{thebibliography}{}
\makeatletter
\relax
\def\mn@urlcharsother{\let\do\@makeother \do\$\do\&\do\#\do\^\do\_\do\%\do\~}
\def\mn@doi{\begingroup\mn@urlcharsother \@ifnextchar [ {\mn@doi@}
  {\mn@doi@[]}}
\def\mn@doi@[#1]#2{\def\@tempa{#1}\ifx\@tempa\@empty \href
  {http://dx.doi.org/#2} {doi:#2}\else \href {http://dx.doi.org/#2} {#1}\fi
  \endgroup}
\def\mn@eprint#1#2{\mn@eprint@#1:#2::\@nil}
\def\mn@eprint@arXiv#1{\href {http://arxiv.org/abs/#1} {{\tt arXiv:#1}}}
\def\mn@eprint@dblp#1{\href {http://dblp.uni-trier.de/rec/bibtex/#1.xml}
  {dblp:#1}}
\def\mn@eprint@#1:#2:#3:#4\@nil{\def\@tempa {#1}\def\@tempb {#2}\def\@tempc
  {#3}\ifx \@tempc \@empty \let \@tempc \@tempb \let \@tempb \@tempa \fi \ifx
  \@tempb \@empty \def\@tempb {arXiv}\fi \@ifundefined
  {mn@eprint@\@tempb}{\@tempb:\@tempc}{\expandafter \expandafter \csname
  mn@eprint@\@tempb\endcsname \expandafter{\@tempc}}}

\bibitem[\protect\citeauthoryear{{Abraham}, {Tanvir}, {Santiago}, {Ellis},
  {Glazebrook}  \& {van den Bergh}}{{Abraham} et~al.}{1996}]{Abraham1996}
{Abraham} R.~G.,  {Tanvir} N.~R.,  {Santiago} B.~X.,  {Ellis} R.~S.,
  {Glazebrook} K.,   {van den Bergh} S.,  1996, \mn@doi [\mnras]
  {10.1093/mnras/279.3.L47}, \href
  {http://adsabs.harvard.edu/abs/1996MNRAS.279L..47A} {279, L47}

\bibitem[\protect\citeauthoryear{{Abraham}, {van den Bergh}  \&
  {Nair}}{{Abraham} et~al.}{2003}]{Abraham2003}
{Abraham} R.~G.,  {van den Bergh} S.,   {Nair} P.,  2003, \mn@doi [\apj]
  {10.1086/373919}, \href {http://adsabs.harvard.edu/abs/2003ApJ...588..218A}
  {588, 218}

\bibitem[\protect\citeauthoryear{{Alexander}, {Brandt}, {Hornschemeier}  \& {et
  al.,}}{{Alexander} et~al.}{2001}]{Alexander2001}
{Alexander} D.~M.,  {Brandt} W.~N.,  {Hornschemeier} A.~E.,   {et al.,} 2001,
  \mn@doi [\aj] {10.1086/323540}, \href
  {http://adsabs.harvard.edu/abs/2001AJ....122.2156A} {122, 2156}

\bibitem[\protect\citeauthoryear{{Arnouts} \& {Ilbert}}{{Arnouts} \&
  {Ilbert}}{2011}]{Arnouts2011}
{Arnouts} S.,  {Ilbert} O.,  2011, {LePHARE: Photometric Analysis for Redshift
  Estimate}, Astrophysics Source Code Library (\mn@eprint {ascl} {1108.009})

\bibitem[\protect\citeauthoryear{{Baldry}, {Glazebrook}, {Brinkmann},
  {Ivezi{\'c}}, {Lupton}, {Nichol}  \& {Szalay}}{{Baldry}
  et~al.}{2004}]{Baldry2004}
{Baldry} I.~K.,  {Glazebrook} K.,  {Brinkmann} J.,  {Ivezi{\'c}} {\v Z}.,
  {Lupton} R.~H.,  {Nichol} R.~C.,   {Szalay} A.~S.,  2004, \mn@doi [\apj]
  {10.1086/380092}, \href {http://adsabs.harvard.edu/abs/2004ApJ...600..681B}
  {600, 681}

\bibitem[\protect\citeauthoryear{{Baldwin}, {Phillips}  \&
  {Terlevich}}{{Baldwin} et~al.}{1981}]{Baldwin1981}
{Baldwin} J.~A.,  {Phillips} M.~M.,   {Terlevich} R.,  1981, \mn@doi [\pasp]
  {10.1086/130766}, \href {http://adsabs.harvard.edu/abs/1981PASP...93....5B}
  {93, 5}

\bibitem[\protect\citeauthoryear{{Bauer}, {Alexander}, {Brandt}, {Schneider},
  {Treister}, {Hornschemeier}  \& {Garmire}}{{Bauer} et~al.}{2004}]{Bauer2004}
{Bauer} F.~E.,  {Alexander} D.~M.,  {Brandt} W.~N.,  {Schneider} D.~P.,
  {Treister} E.,  {Hornschemeier} A.~E.,   {Garmire} G.~P.,  2004, \mn@doi
  [\aj] {10.1086/424859}, \href
  {http://adsabs.harvard.edu/abs/2004AJ....128.2048B} {128, 2048}

\bibitem[\protect\citeauthoryear{{Belfiore} et~al.,}{{Belfiore}
  et~al.}{2017}]{Belfiore2017}
{Belfiore} F.,  et~al., 2017, \mn@doi [\mnras] {10.1093/mnras/stw3211}, \href
  {http://adsabs.harvard.edu/abs/2017MNRAS.466.2570B} {466, 2570}

\bibitem[\protect\citeauthoryear{{Belfiore} et~al.,}{{Belfiore}
  et~al.}{2018}]{Belfiore2018}
{Belfiore} F.,  et~al., 2018, \mn@doi [\mnras] {10.1093/mnras/sty768}, \href
  {http://adsabs.harvard.edu/abs/2018MNRAS.477.3014B} {477, 3014}

\bibitem[\protect\citeauthoryear{{Bertin} \& {Arnouts}}{{Bertin} \&
  {Arnouts}}{1996}]{Bertin1996}
{Bertin} E.,  {Arnouts} S.,  1996, \mn@doi [\aaps] {10.1051/aas:1996164}, \href
  {http://adsabs.harvard.edu/abs/1996A%26AS..117..393B} {117, 393}

\bibitem[\protect\citeauthoryear{{Beyoro-Amado}, {Povi{\'c}}  \&
  {S{\'a}nchez-Portal}}{{Beyoro-Amado} et~al.}{2018}]{BeyoroAmado2018}
{Beyoro-Amado} Z.,  {Povi{\'c}} M.,   {S{\'a}nchez-Portal} M.,  2018, \mnras,
  submitted

\bibitem[\protect\citeauthoryear{{Brammer} et~al.,}{{Brammer}
  et~al.}{2009}]{Brammer2009}
{Brammer} G.~B.,  et~al., 2009, \mn@doi [\apjl] {10.1088/0004-637X/706/1/L173},
  \href {http://adsabs.harvard.edu/abs/2009ApJ...706L.173B} {706, L173}

\bibitem[\protect\citeauthoryear{{Brassington}, {Zezas}, {Ashby}, {Lanz},
  {Smith}, {Willner}  \& {Klein}}{{Brassington} et~al.}{2015}]{Brassington2015}
{Brassington} N.~J.,  {Zezas} A.,  {Ashby} M.~L.~N.,  {Lanz} L.,  {Smith}
  H.~A.,  {Willner} S.~P.,   {Klein} C.,  2015, \mn@doi [\apjs]
  {10.1088/0067-0049/218/1/6}, \href
  {http://adsabs.harvard.edu/abs/2015ApJS..218....6B} {218, 6}

\bibitem[\protect\citeauthoryear{{Bremer} et~al.,}{{Bremer}
  et~al.}{2018}]{Bremer2018}
{Bremer} M.~N.,  et~al., 2018, \mn@doi [\mnras] {10.1093/mnras/sty124}, \href
  {http://adsabs.harvard.edu/abs/2018MNRAS.476...12B} {476, 12}

\bibitem[\protect\citeauthoryear{{Brusa} et~al.,}{{Brusa}
  et~al.}{2007}]{Brusa2007}
{Brusa} M.,  et~al., 2007, \mn@doi [\apjs] {10.1086/516575}, \href
  {http://adsabs.harvard.edu/abs/2007ApJS..172..353B} {172, 353}

\bibitem[\protect\citeauthoryear{{Bundy}, {Treu}  \& {Ellis}}{{Bundy}
  et~al.}{2007}]{Bundy2007}
{Bundy} K.,  {Treu} T.,   {Ellis} R.~S.,  2007, \mn@doi [\apjl]
  {10.1086/519526}, \href {http://adsabs.harvard.edu/abs/2007ApJ...665L...5B}
  {665, L5}

\bibitem[\protect\citeauthoryear{{Cardamone}, {Urry}, {Schawinski}, {Treister},
  {Brammer}  \& {Gawiser}}{{Cardamone} et~al.}{2010}]{Cardamone2010}
{Cardamone} C.~N.,  {Urry} C.~M.,  {Schawinski} K.,  {Treister} E.,  {Brammer}
  G.,   {Gawiser} E.,  2010, \mn@doi [\apjl] {10.1088/2041-8205/721/1/L38},
  \href {http://adsabs.harvard.edu/abs/2010ApJ...721L..38C} {721, L38}

\bibitem[\protect\citeauthoryear{{Cassata} et~al.,}{{Cassata}
  et~al.}{2007}]{Cassata2007}
{Cassata} P.,  et~al., 2007, \mn@doi [\apjs] {10.1086/516591}, \href
  {http://adsabs.harvard.edu/abs/2007ApJS..172..270C} {172, 270}

\bibitem[\protect\citeauthoryear{{Catal{\'a}n-Torrecilla}
  et~al.,}{{Catal{\'a}n-Torrecilla} et~al.}{2015}]{Catalnorrecilla2015}
{Catal{\'a}n-Torrecilla} C.,  et~al., 2015, \mn@doi [\aap]
  {10.1051/0004-6361/201526023}, \href
  {http://adsabs.harvard.edu/abs/2015A%26A...584A..87C} {584, A87}

\bibitem[\protect\citeauthoryear{{Cibinel} et~al.,}{{Cibinel}
  et~al.}{2013}]{Cibinel2013}
{Cibinel} A.,  et~al., 2013, \mn@doi [\apj] {10.1088/0004-637X/776/2/72}, \href
  {http://adsabs.harvard.edu/abs/2013ApJ...776...72C} {776, 72}

\bibitem[\protect\citeauthoryear{{Civano} et~al.,}{{Civano}
  et~al.}{2012}]{Civano2012}
{Civano} F.,  et~al., 2012, \mn@doi [\apjs] {10.1088/0067-0049/201/2/30}, \href
  {http://adsabs.harvard.edu/abs/2012ApJS..201...30C} {201, 30}

\bibitem[\protect\citeauthoryear{{Coenda}, {Mart{\'{\i}}nez}  \&
  {Muriel}}{{Coenda} et~al.}{2018}]{Coenda2018}
{Coenda} V.,  {Mart{\'{\i}}nez} H.~J.,   {Muriel} H.,  2018, \mn@doi [\mnras]
  {10.1093/mnras/stx2707}, \href
  {http://adsabs.harvard.edu/abs/2018MNRAS.473.5617C} {473, 5617}

\bibitem[\protect\citeauthoryear{{Coil} et~al.,}{{Coil}
  et~al.}{2009}]{Coil2009}
{Coil} A.~L.,  et~al., 2009, \mn@doi [\apj] {10.1088/0004-637X/701/2/1484},
  \href {http://adsabs.harvard.edu/abs/2009ApJ...701.1484C} {701, 1484}

\bibitem[\protect\citeauthoryear{{Conselice}, {Bershady}  \&
  {Jangren}}{{Conselice} et~al.}{2000}]{Conselice2000}
{Conselice} C.~J.,  {Bershady} M.~A.,   {Jangren} A.,  2000, \mn@doi [\apj]
  {10.1086/308300}, \href {http://adsabs.harvard.edu/abs/2000ApJ...529..886C}
  {529, 886}

\bibitem[\protect\citeauthoryear{{Crossett}, {Pimbblet}, {Jones}, {Brown}  \&
  {Stott}}{{Crossett} et~al.}{2017}]{Crossett2017}
{Crossett} J.~P.,  {Pimbblet} K.~A.,  {Jones} D.~H.,  {Brown} M.~J.~I.,
  {Stott} J.~P.,  2017, \mn@doi [\mnras] {10.1093/mnras/stw2228}, \href
  {http://adsabs.harvard.edu/abs/2017MNRAS.464..480C} {464, 480}

\bibitem[\protect\citeauthoryear{{Di Matteo}, {Springel}  \& {Hernquist}}{{Di
  Matteo} et~al.}{2005}]{DiMatteo2005}
{Di Matteo} T.,  {Springel} V.,   {Hernquist} L.,  2005, \mn@doi [\nat]
  {10.1038/nature03335}, \href
  {http://adsabs.harvard.edu/abs/2005Natur.433..604D} {433, 604}

\bibitem[\protect\citeauthoryear{{Dietrich} et~al.,}{{Dietrich}
  et~al.}{2018}]{Dietrich2018}
{Dietrich} J.,  et~al., 2018, preprint, \href
  {http://adsabs.harvard.edu/abs/2018arXiv180104328D} {} (\mn@eprint {arXiv}
  {1801.04328})

\bibitem[\protect\citeauthoryear{{Elbaz} et~al.,}{{Elbaz}
  et~al.}{2007}]{Elbaz2007}
{Elbaz} D.,  et~al., 2007, \mn@doi [\aap] {10.1051/0004-6361:20077525}, \href
  {http://adsabs.harvard.edu/abs/2007A%26A...468...33E} {468, 33}

\bibitem[\protect\citeauthoryear{{Elbaz} et~al.,}{{Elbaz}
  et~al.}{2011}]{Elbaz2011}
{Elbaz} D.,  et~al., 2011, \mn@doi [\aap] {10.1051/0004-6361/201117239}, \href
  {http://adsabs.harvard.edu/abs/2011A%26A...533A.119E} {533, A119}

\bibitem[\protect\citeauthoryear{{Ellison}, {Teimoorinia}, {Rosario}  \&
  {Mendel}}{{Ellison} et~al.}{2016}]{Ellison2016}
{Ellison} S.~L.,  {Teimoorinia} H.,  {Rosario} D.~J.,   {Mendel} J.~T.,  2016,
  \mn@doi [\mnras] {10.1093/mnrasl/slw012}, \href
  {http://adsabs.harvard.edu/abs/2016MNRAS.458L..34E} {458, L34}

\bibitem[\protect\citeauthoryear{{Ge}, {Gu}, {Chen}  \& {Ding}}{{Ge}
  et~al.}{2018}]{Ge2018}
{Ge} X.,  {Gu} Q.-S.,  {Chen} Y.-Y.,   {Ding} N.,  2018, preprint, \href
  {http://adsabs.harvard.edu/abs/2018arXiv180801709G} {} (\mn@eprint {arXiv}
  {1808.01709})

\bibitem[\protect\citeauthoryear{{Georgakakis} et~al.,}{{Georgakakis}
  et~al.}{2008}]{Georgakakis2008}
{Georgakakis} A.,  et~al., 2008, \mn@doi [\mnras]
  {10.1111/j.1365-2966.2008.12962.x}, \href
  {http://adsabs.harvard.edu/abs/2008MNRAS.385.2049G} {385, 2049}

\bibitem[\protect\citeauthoryear{{Gu}, {Fang}, {Yuan}, {Cai}  \& {Wang}}{{Gu}
  et~al.}{2018}]{Gu2018}
{Gu} Y.,  {Fang} G.,  {Yuan} Q.,  {Cai} Z.,   {Wang} T.,  2018, \mn@doi [\apj]
  {10.3847/1538-4357/aaad0b}, \href
  {http://adsabs.harvard.edu/abs/2018ApJ...855...10G} {855, 10}

\bibitem[\protect\citeauthoryear{{Hickox} et~al.,}{{Hickox}
  et~al.}{2009}]{Hickox2009}
{Hickox} R.~C.,  et~al., 2009, \mn@doi [\apj] {10.1088/0004-637X/696/1/891},
  \href {http://adsabs.harvard.edu/abs/2009ApJ...696..891H} {696, 891}

\bibitem[\protect\citeauthoryear{{Hong}, {Im}, {Kim}  \& {Ho}}{{Hong}
  et~al.}{2015}]{Hong2015}
{Hong} J.,  {Im} M.,  {Kim} M.,   {Ho} L.~C.,  2015, \mn@doi [\apj]
  {10.1088/0004-637X/804/1/34}, \href
  {http://adsabs.harvard.edu/abs/2015ApJ...804...34H} {804, 34}

\bibitem[\protect\citeauthoryear{{Hopkins}, {Cox}, {Kere{\v s}}  \&
  {Hernquist}}{{Hopkins} et~al.}{2008}]{Hopkins08}
{Hopkins} P.~F.,  {Cox} T.~J.,  {Kere{\v s}} D.,   {Hernquist} L.,  2008,
  \mn@doi [\apjs] {10.1086/524363}, \href
  {http://adsabs.harvard.edu/abs/2008ApJS..175..390H} {175, 390}

\bibitem[\protect\citeauthoryear{{Huertas-Company}, {Rouan}, {Tasca}, {Soucail}
   \& {Le F{\`e}vre}}{{Huertas-Company} et~al.}{2008}]{HuertasCompany2008}
{Huertas-Company} M.,  {Rouan} D.,  {Tasca} L.,  {Soucail} G.,   {Le F{\`e}vre}
  O.,  2008, \mn@doi [\aap] {10.1051/0004-6361:20078625}, \href
  {http://adsabs.harvard.edu/abs/2008A%26A...478..971H} {478, 971}

\bibitem[\protect\citeauthoryear{{Ilbert} et~al.,}{{Ilbert}
  et~al.}{2006}]{Ilbert2006}
{Ilbert} O.,  et~al., 2006, \mn@doi [\aap] {10.1051/0004-6361:20065138}, \href
  {http://adsabs.harvard.edu/abs/2006A%26A...457..841I} {457, 841}

\bibitem[\protect\citeauthoryear{{Ilbert} et~al.,}{{Ilbert}
  et~al.}{2009}]{Ilbert2009}
{Ilbert} O.,  et~al., 2009, \mn@doi [\apj] {10.1088/0004-637X/690/2/1236},
  \href {http://adsabs.harvard.edu/abs/2009ApJ...690.1236I} {690, 1236}

\bibitem[\protect\citeauthoryear{{Kauffmann} et~al.,}{{Kauffmann}
  et~al.}{2003}]{Kauffmann2003}
{Kauffmann} G.,  et~al., 2003, \mn@doi [\mnras]
  {10.1046/j.1365-8711.2003.06292.x}, \href
  {http://adsabs.harvard.edu/abs/2003MNRAS.341...54K} {341, 54}

\bibitem[\protect\citeauthoryear{{Kirkpatrick}, {Pope}, {Sajina}, {Roebuck},
  {Yan}, {Armus}, {D{\'{\i}}az-Santos}  \& {Stierwalt}}{{Kirkpatrick}
  et~al.}{2015}]{Kirkpatrick2015}
{Kirkpatrick} A.,  {Pope} A.,  {Sajina} A.,  {Roebuck} E.,  {Yan} L.,  {Armus}
  L.,  {D{\'{\i}}az-Santos} T.,   {Stierwalt} S.,  2015, \mn@doi [\apj]
  {10.1088/0004-637X/814/1/9}, \href
  {http://adsabs.harvard.edu/abs/2015ApJ...814....9K} {814, 9}

\bibitem[\protect\citeauthoryear{{Knapen}, {Cisternas}  \&
  {Querejeta}}{{Knapen} et~al.}{2015}]{Knapen2015}
{Knapen} J.~H.,  {Cisternas} M.,   {Querejeta} M.,  2015, \mn@doi [\mnras]
  {10.1093/mnras/stv2135}, \href
  {http://adsabs.harvard.edu/abs/2015MNRAS.454.1742K} {454, 1742}

\bibitem[\protect\citeauthoryear{{Koekemoer} et~al.,}{{Koekemoer}
  et~al.}{2007}]{Koekemoer2007}
{Koekemoer} A.~M.,  et~al., 2007, \mn@doi [\apjs] {10.1086/520086}, \href
  {http://adsabs.harvard.edu/abs/2007ApJS..172..196K} {172, 196}

\bibitem[\protect\citeauthoryear{{Kormendy} \& {Kennicutt}}{{Kormendy} \&
  {Kennicutt}}{2004}]{KormendyKennicutt2004}
{Kormendy} J.,  {Kennicutt} Jr. R.~C.,  2004, \mn@doi [\araa]
  {10.1146/annurev.astro.42.053102.134024}, \href
  {http://adsabs.harvard.edu/abs/2004ARA%26A..42..603K} {42, 603}

\bibitem[\protect\citeauthoryear{{Lamastra}, {Menci}, {Fiore}, {Santini},
  {Bongiorno}  \& {Piconcelli}}{{Lamastra} et~al.}{2013}]{Lamastra2013}
{Lamastra} A.,  {Menci} N.,  {Fiore} F.,  {Santini} P.,  {Bongiorno} A.,
  {Piconcelli} E.,  2013, \mn@doi [\aap] {10.1051/0004-6361/201322667}, \href
  {http://adsabs.harvard.edu/abs/2013A%26A...559A..56L} {559, A56}

\bibitem[\protect\citeauthoryear{{Leauthaud} et~al.,}{{Leauthaud}
  et~al.}{2007}]{Leauthaud2007}
{Leauthaud} A.,  et~al., 2007, \mn@doi [\apjs] {10.1086/516598}, \href
  {http://adsabs.harvard.edu/abs/2007ApJS..172..219L} {172, 219}

\bibitem[\protect\citeauthoryear{{Lee}, {Hwang}, {Lee}, {Ko}, {Sohn}, {Shim}
  \& {Diaferio}}{{Lee} et~al.}{2015}]{Lee2015}
{Lee} G.-H.,  {Hwang} H.~S.,  {Lee} M.~G.,  {Ko} J.,  {Sohn} J.,  {Shim} H.,
  {Diaferio} A.,  2015, \mn@doi [\apj] {10.1088/0004-637X/800/2/80}, \href
  {http://adsabs.harvard.edu/abs/2015ApJ...800...80L} {800, 80}

\bibitem[\protect\citeauthoryear{{Leslie}, {Kewley}, {Sanders}  \&
  {Lee}}{{Leslie} et~al.}{2016}]{Leslie2016}
{Leslie} S.~K.,  {Kewley} L.~J.,  {Sanders} D.~B.,   {Lee} N.,  2016, \mn@doi
  [\mnras] {10.1093/mnrasl/slv135}, \href
  {http://adsabs.harvard.edu/abs/2016MNRAS.455L..82L} {455, L82}

\bibitem[\protect\citeauthoryear{{Lotz}, {Primack}  \& {Madau}}{{Lotz}
  et~al.}{2004}]{Lotz2004}
{Lotz} J.~M.,  {Primack} J.,   {Madau} P.,  2004, \mn@doi [\aj]
  {10.1086/421849}, \href {http://adsabs.harvard.edu/abs/2004AJ....128..163L}
  {128, 163}

\bibitem[\protect\citeauthoryear{{Lotz}, {Jonsson}, {Cox}  \& {Primack}}{{Lotz}
  et~al.}{2010}]{Lotz2010}
{Lotz} J.~M.,  {Jonsson} P.,  {Cox} T.~J.,   {Primack} J.~R.,  2010, \mn@doi
  [\mnras] {10.1111/j.1365-2966.2010.16268.x}, \href
  {http://adsabs.harvard.edu/abs/2010MNRAS.404..575L} {404, 575}

\bibitem[\protect\citeauthoryear{{Lutz} et~al.,}{{Lutz}
  et~al.}{2011}]{Lutz2011}
{Lutz} D.,  et~al., 2011, \mn@doi [\aap] {10.1051/0004-6361/201117107}, \href
  {http://adsabs.harvard.edu/abs/2011A%26A...532A..90L} {532, A90}

\bibitem[\protect\citeauthoryear{{Mahajan}, {Ashby}, {Willner}, {Barmby},
  {Fazio}, {Maragkoudakis}, {Raychaudhury}  \& {Zezas}}{{Mahajan}
  et~al.}{2019}]{Mahajan2019}
{Mahajan} S.,  {Ashby} M.~L.~N.,  {Willner} S.~P.,  {Barmby} P.,  {Fazio}
  G.~G.,  {Maragkoudakis} A.,  {Raychaudhury} S.,   {Zezas} A.,  2019, \mn@doi
  [\mnras] {10.1093/mnras/sty2699}, \href
  {http://adsabs.harvard.edu/abs/2019MNRAS.482..560M} {482, 560}

\bibitem[\protect\citeauthoryear{{Mahoro}, {Povi{\'c}}  \&
  {Nkundabakura}}{{Mahoro} et~al.}{2017}]{Mahoro2017}
{Mahoro} A.,  {Povi{\'c}} M.,   {Nkundabakura} P.,  2017, \mn@doi [\mnras]
  {10.1093/mnras/stx1762}, \href
  {http://adsabs.harvard.edu/abs/2017MNRAS.471.3226M} {471, 3226}

\bibitem[\protect\citeauthoryear{{Massey}, {Stoughton}, {Leauthaud}, {Rhodes},
  {Koekemoer}, {Ellis}  \& {Shaghoulian}}{{Massey} et~al.}{2010}]{Massey2010}
{Massey} R.,  {Stoughton} C.,  {Leauthaud} A.,  {Rhodes} J.,  {Koekemoer} A.,
  {Ellis} R.,   {Shaghoulian} E.,  2010, \mn@doi [\mnras]
  {10.1111/j.1365-2966.2009.15638.x}, \href
  {http://adsabs.harvard.edu/abs/2010MNRAS.401..371M} {401, 371}

\bibitem[\protect\citeauthoryear{{Mendez}, {Coil}, {Lotz}, {Salim}, {Moustakas}
   \& {Simard}}{{Mendez} et~al.}{2011}]{Mendez2011}
{Mendez} A.~J.,  {Coil} A.~L.,  {Lotz} J.,  {Salim} S.,  {Moustakas} J.,
  {Simard} L.,  2011, \mn@doi [\apj] {10.1088/0004-637X/736/2/110}, \href
  {http://adsabs.harvard.edu/abs/2011ApJ...736..110M} {736, 110}

\bibitem[\protect\citeauthoryear{{Nandra} et~al.,}{{Nandra}
  et~al.}{2007}]{Nandra2007}
{Nandra} K.,  et~al., 2007, \mn@doi [\apjl] {10.1086/517918}, \href
  {http://adsabs.harvard.edu/abs/2007ApJ...660L..11N} {660, L11}

\bibitem[\protect\citeauthoryear{{Nkundabakura}, {Mahoro}  \&
  {Povi{\'c}}}{{Nkundabakura} et~al.}{2016}]{Nkundabakura2016}
{Nkundabakura} P.,  {Mahoro} A.,   {Povi{\'c}} M.,  2016, \mn@doi [Rwanda
  Journal- Series D] {10.4314/rj.vlilS.3D}, 1, 34

\bibitem[\protect\citeauthoryear{{Noeske} et~al.,}{{Noeske}
  et~al.}{2007}]{Noeske2007}
{Noeske} K.~G.,  et~al., 2007, \mn@doi [\apjl] {10.1086/517926}, \href
  {http://adsabs.harvard.edu/abs/2007ApJ...660L..43N} {660, L43}

\bibitem[\protect\citeauthoryear{{Nogueira-Cavalcante}, {Gon{\c c}alves},
  {Men{\'e}ndez-Delmestre}  \& {Sheth}}{{Nogueira-Cavalcante}
  et~al.}{2018}]{NogueiraCavalcante2018}
{Nogueira-Cavalcante} J.~P.,  {Gon{\c c}alves} T.~S.,  {Men{\'e}ndez-Delmestre}
  K.,   {Sheth} K.,  2018, \mn@doi [\mnras] {10.1093/mnras/stx2399}, \href
  {http://adsabs.harvard.edu/abs/2018MNRAS.473.1346N} {473, 1346}

\bibitem[\protect\citeauthoryear{{Pan}, {Kong}  \& {Fan}}{{Pan}
  et~al.}{2013}]{Pan2013}
{Pan} Z.,  {Kong} X.,   {Fan} L.,  2013, \mn@doi [\apj]
  {10.1088/0004-637X/776/1/14}, \href
  {http://adsabs.harvard.edu/abs/2013ApJ...776...14P} {776, 14}

\bibitem[\protect\citeauthoryear{{Pintos-Castro} et~al.,}{{Pintos-Castro}
  et~al.}{2016}]{Pintos2016}
{Pintos-Castro} I.,  et~al., 2016, \mn@doi [\aap]
  {10.1051/0004-6361/201526744}, \href
  {http://adsabs.harvard.edu/abs/2016A%26A...592A.108P} {592, A108}

\bibitem[\protect\citeauthoryear{{Povi{\'c}} et~al.,}{{Povi{\'c}}
  et~al.}{2012}]{Povic2012}
{Povi{\'c}} M.,  et~al., 2012, \mn@doi [\aap] {10.1051/0004-6361/201117314},
  \href {http://adsabs.harvard.edu/abs/2012A%26A...541A.118P} {541, A118}

\bibitem[\protect\citeauthoryear{{Povi{\'c}} et~al.,}{{Povi{\'c}}
  et~al.}{2013a}]{Povic2013a}
{Povi{\'c}} M.,  et~al., 2013a, \mn@doi [\mnras] {10.1093/mnras/stt1538}, \href
  {http://adsabs.harvard.edu/abs/2013MNRAS.435.3444P} {435, 3444}

\bibitem[\protect\citeauthoryear{{Povi{\'c}}, {S{\'a}nchez-Portal}, {P{\'e}rez
  Garc{\'{\i}}a}  \& {Otelo Group}}{{Povi{\'c}} et~al.}{2013b}]{Povic2013b}
{Povi{\'c}} M.,  {S{\'a}nchez-Portal} M.,  {P{\'e}rez Garc{\'{\i}}a} A.~M.,
  {Otelo Group} 2013b, in {Sun} W.-H.,  {Xu} C.~K.,  {Scoville} N.~Z.,
  {Sanders} D.~B.,  eds,  Astronomical Society of the Pacific Conference Series
  Vol. 477, Galaxy Mergers in an Evolving Universe. p.~177

\bibitem[\protect\citeauthoryear{{Povi{\'c}} et~al.,}{{Povi{\'c}}
  et~al.}{2015}]{Povic2015}
{Povi{\'c}} M.,  et~al., 2015, \mn@doi [\mnras] {10.1093/mnras/stv1663}, \href
  {http://adsabs.harvard.edu/abs/2015MNRAS.453.1644P} {453, 1644}

\bibitem[\protect\citeauthoryear{{Povi{\'c}}, {M{\'a}rquez}, {Netzer},
  {Masegosa}, {Nordon}, {P{\'e}rez}  \& {Schoenell}}{{Povi{\'c}}
  et~al.}{2016}]{Povic2016}
{Povi{\'c}} M.,  {M{\'a}rquez} I.,  {Netzer} H.,  {Masegosa} J.,  {Nordon} R.,
  {P{\'e}rez} E.,   {Schoenell} W.,  2016, \mn@doi [\mnras]
  {10.1093/mnras/stw1842}, \href
  {http://adsabs.harvard.edu/abs/2016MNRAS.462.2878P} {462, 2878}

\bibitem[\protect\citeauthoryear{{Renzini} \& {Peng}}{{Renzini} \&
  {Peng}}{2015}]{RenziniPeng2015}
{Renzini} A.,  {Peng} Y.-j.,  2015, \mn@doi [\apjl]
  {10.1088/2041-8205/801/2/L29}, \href
  {http://adsabs.harvard.edu/abs/2015ApJ...801L..29R} {801, L29}

\bibitem[\protect\citeauthoryear{{Rieke} et~al.,}{{Rieke}
  et~al.}{2004}]{Rieke2004}
{Rieke} G.~H.,  et~al., 2004, \mn@doi [\apjs] {10.1086/422717}, \href
  {http://adsabs.harvard.edu/abs/2004ApJS..154...25R} {154, 25}

\bibitem[\protect\citeauthoryear{{Salim}}{{Salim}}{2014}]{Salim2014}
{Salim} S.,  2014, \mn@doi [Serbian Astronomical Journal]
  {10.2298/SAJ1489001S}, \href
  {http://adsabs.harvard.edu/abs/2014SerAJ.189....1S} {189, 1}

\bibitem[\protect\citeauthoryear{{Salim} et~al.,}{{Salim}
  et~al.}{2007}]{Salim2007}
{Salim} S.,  et~al., 2007, \mn@doi [\apjs] {10.1086/519218}, \href
  {http://adsabs.harvard.edu/abs/2007ApJS..173..267S} {173, 267}

\bibitem[\protect\citeauthoryear{{Salpeter}}{{Salpeter}}{1955}]{Salpeter1955}
{Salpeter} E.~E.,  1955, \mn@doi [\apj] {10.1086/145971}, \href
  {http://adsabs.harvard.edu/abs/1955ApJ...121..161S} {121, 161}

\bibitem[\protect\citeauthoryear{{Salvato} et~al.,}{{Salvato}
  et~al.}{2011}]{Salvato2011}
{Salvato} M.,  et~al., 2011, \mn@doi [\apj] {10.1088/0004-637X/742/2/61}, \href
  {http://adsabs.harvard.edu/abs/2011ApJ...742...61S} {742, 61}

\bibitem[\protect\citeauthoryear{{S{\'a}nchez} et~al.,}{{S{\'a}nchez}
  et~al.}{2004}]{Sanchez2004}
{S{\'a}nchez} S.~F.,  et~al., 2004, \mn@doi [\apj] {10.1086/423234}, \href
  {http://adsabs.harvard.edu/abs/2004ApJ...614..586S} {614, 586}

\bibitem[\protect\citeauthoryear{{Scarlata} et~al.,}{{Scarlata}
  et~al.}{2007}]{Scarlata2007}
{Scarlata} C.,  et~al., 2007, \mn@doi [\apjs] {10.1086/516582}, \href
  {http://adsabs.harvard.edu/abs/2007ApJS..172..406S} {172, 406}

\bibitem[\protect\citeauthoryear{{Schawinski} et~al.,}{{Schawinski}
  et~al.}{2014}]{Schawinski2014}
{Schawinski} K.,  et~al., 2014, \mn@doi [\mnras] {10.1093/mnras/stu327}, \href
  {http://adsabs.harvard.edu/abs/2014MNRAS.440..889S} {440, 889}

\bibitem[\protect\citeauthoryear{{Scoville} et~al.,}{{Scoville}
  et~al.}{2007}]{Scoville2007}
{Scoville} N.,  et~al., 2007, \mn@doi [\apjs] {10.1086/516585}, \href
  {http://adsabs.harvard.edu/abs/2007ApJS..172....1S} {172, 1}

\bibitem[\protect\citeauthoryear{{Shimizu}, {Mushotzky}, {Mel{\'e}ndez}, {Koss}
   \& {Rosario}}{{Shimizu} et~al.}{2015}]{Shimizu2015}
{Shimizu} T.~T.,  {Mushotzky} R.~F.,  {Mel{\'e}ndez} M.,  {Koss} M.,
  {Rosario} D.~J.,  2015, \mn@doi [\mnras] {10.1093/mnras/stv1407}, \href
  {http://adsabs.harvard.edu/abs/2015MNRAS.452.1841S} {452, 1841}

\bibitem[\protect\citeauthoryear{{Silverman} et~al.,}{{Silverman}
  et~al.}{2008}]{Silverman2008}
{Silverman} J.~D.,  et~al., 2008, \mn@doi [\apj] {10.1086/527283}, \href
  {http://adsabs.harvard.edu/abs/2008ApJ...675.1025S} {675, 1025}

\bibitem[\protect\citeauthoryear{{Sklias}, {Schaerer}, {Elbaz}, {Pannella},
  {Schreiber}  \& {Cava}}{{Sklias} et~al.}{2017}]{Sklias2017}
{Sklias} P.,  {Schaerer} D.,  {Elbaz} D.,  {Pannella} M.,  {Schreiber} C.,
  {Cava} A.,  2017, \mn@doi [\aap] {10.1051/0004-6361/201628330}, \href
  {http://adsabs.harvard.edu/abs/2017A%26A...605A..29S} {605, A29}

\bibitem[\protect\citeauthoryear{{Smethurst} et~al.,}{{Smethurst}
  et~al.}{2015}]{Smethurst2015}
{Smethurst} R.~J.,  et~al., 2015, \mn@doi [\mnras] {10.1093/mnras/stv161},
  \href {http://adsabs.harvard.edu/abs/2015MNRAS.450..435S} {450, 435}

\bibitem[\protect\citeauthoryear{{Tasca} et~al.,}{{Tasca}
  et~al.}{2009}]{Tasca2009}
{Tasca} L.~A.~M.,  et~al., 2009, \mn@doi [\aap] {10.1051/0004-6361/200912213},
  \href {http://adsabs.harvard.edu/abs/2009A%26A...503..379T} {503, 379}

\bibitem[\protect\citeauthoryear{{Trayford}, {Theuns}, {Bower}, {Crain},
  {Lagos}, {Schaller}  \& {Schaye}}{{Trayford} et~al.}{2016}]{Trayford2016}
{Trayford} J.~W.,  {Theuns} T.,  {Bower} R.~G.,  {Crain} R.~A.,  {Lagos}
  C.~d.~P.,  {Schaller} M.,   {Schaye} J.,  2016, \mn@doi [\mnras]
  {10.1093/mnras/stw1230}, \href
  {http://adsabs.harvard.edu/abs/2016MNRAS.460.3925T} {460, 3925}

\bibitem[\protect\citeauthoryear{{Treister}, {Virani}, {Gawiser}  \& {et
  al.,}}{{Treister} et~al.}{2009}]{Treister2009}
{Treister} E.,  {Virani} S.,  {Gawiser} E.,   {et al.,} 2009, \mn@doi [\apj]
  {10.1088/0004-637X/693/2/1713}, \href
  {http://adsabs.harvard.edu/abs/2009ApJ...693.1713T} {693, 1713}

\bibitem[\protect\citeauthoryear{{Trump} et~al.,}{{Trump}
  et~al.}{2009}]{Trump2009}
{Trump} J.~R.,  et~al., 2009, \mn@doi [\apj] {10.1088/0004-637X/696/2/1195},
  \href {http://adsabs.harvard.edu/abs/2009ApJ...696.1195T} {696, 1195}

\bibitem[\protect\citeauthoryear{{Walker} et~al.,}{{Walker}
  et~al.}{2013}]{Walker2013}
{Walker} L.~M.,  et~al., 2013, \mn@doi [\apj] {10.1088/0004-637X/775/2/129},
  \href {http://adsabs.harvard.edu/abs/2013ApJ...775..129W} {775, 129}

\bibitem[\protect\citeauthoryear{{Whitaker}, {van Dokkum}, {Brammer}  \&
  {Franx}}{{Whitaker} et~al.}{2012}]{Whitaker2012}
{Whitaker} K.~E.,  {van Dokkum} P.~G.,  {Brammer} G.,   {Franx} M.,  2012,
  \mn@doi [\apjl] {10.1088/2041-8205/754/2/L29}, \href
  {http://adsabs.harvard.edu/abs/2012ApJ...754L..29W} {754, L29}

\bibitem[\protect\citeauthoryear{{Whitaker} et~al.,}{{Whitaker}
  et~al.}{2014}]{Whitaker2014}
{Whitaker} K.~E.,  et~al., 2014, \mn@doi [\apj] {10.1088/0004-637X/795/2/104},
  \href {http://adsabs.harvard.edu/abs/2014ApJ...795..104W} {795, 104}

\bibitem[\protect\citeauthoryear{{Willmer} et~al.,}{{Willmer}
  et~al.}{2006}]{Willmer07}
{Willmer} C.~N.~A.,  et~al., 2006, \mn@doi [\apj] {10.1086/505455}, \href
  {http://adsabs.harvard.edu/abs/2006ApJ...647..853W} {647, 853}

\bibitem[\protect\citeauthoryear{{Wuyts} et~al.,}{{Wuyts}
  et~al.}{2011}]{Wuyts2011}
{Wuyts} S.,  et~al., 2011, \mn@doi [\apj] {10.1088/0004-637X/738/1/106}, \href
  {http://adsabs.harvard.edu/abs/2011ApJ...738..106W} {738, 106}

\makeatother
\end{thebibliography}






\bsp	
\label{lastpage}
\end{document}